\documentclass[aps,prd,onecolumn,showpacs,11pt,superscriptaddress,preprintnumbers,nofootinbib]{revtex4-2}
\usepackage[T1]{fontenc}
\usepackage{amsmath,amssymb,mathtools,bm}\allowdisplaybreaks[1]
\usepackage{mathrsfs}
\usepackage{graphicx}
\usepackage{physics, esint}
\usepackage{xcolor}
\usepackage[colorlinks, linkcolor=blue!80!black, citecolor=green!50!black, urlcolor=magenta]{hyperref}
\usepackage[normalem]{ulem}
\usepackage{adjustbox}
\usepackage{tikz}
\usetikzlibrary{arrows.meta}
\usepackage{enumitem}

\usepackage{soul}
\usepackage{multirow}
\usepackage{orcidlink}

\renewcommand{\sec}[1]{Sect.~\ref{#1}}
\newcommand{\fig}[1]{Fig.~\ref{#1}}
\newcommand{\figs}[2]{Figs.~\ref{#1} and~\ref{#2}}
\newcommand{\eq}[1]{Eq.~\eqref{#1}}
\newcommand{\eqs}[2]{Eqs.~\eqref{#1} and~\eqref{#2}}

\newcommand{\refcite}[1]{Ref.~\cite{#1}}
\newcommand{\refs}[1]{Refs.~\cite{#1}}

\newcommand{\pp}[1]{\left(#1\right)}
\newcommand{\bb}[1]{\left[#1\right]}
\newcommand{\cc}[1]{\left\{#1\right\}}
\newcommand{\vv}[1]{\left\langle #1 \right\rangle}
\newcommand{\bigpp}[1]{\big(#1\big)}
\newcommand{\bigbb}[1]{\big[#1\big]}
\newcommand{\bigcc}[1]{\big\{#1\big\}}
\newcommand{\bigvv}[1]{\big\langle #1 \big\rangle}
\newcommand{\Bigpp}[1]{\Big(#1\Big)}
\newcommand{\Bigbb}[1]{\Big[#1\Big]}
\newcommand{\Bigcc}[1]{\Big\{#1\Big\}}
\newcommand{\Bigvv}[1]{\Big\langle #1 \Big\rangle}
\newcommand{\biggpp}[1]{\bigg(#1\bigg)}
\newcommand{\biggbb}[1]{\bigg[#1\bigg]}
\newcommand{\biggcc}[1]{\bigg\{#1\bigg\}}

\newcommand{\beq}[1][]{\begin{equation}\label{#1}}
\newcommand{\eeq}{\end{equation}}
\newcommand{\bse}[1][]{\begin{subequations}\label{#1}}
\newcommand{\ese}{\end{subequations}}
\newcommand{\nn}{\nonumber}

\renewcommand{\slash}[1]{#1 \hspace{-0.45em} / }
\newcommand{\Slash}[1]{#1 \hspace{-0.66em} / }

\newcommand{\Lag}{\mathcal{L}}
\newcommand{\C}{\mathscr{C}}
\renewcommand{\l}{\ell}

\renewcommand{\O}{\mathcal{O}}
\newcommand{\wt}[1]{\widetilde{#1}}

\newcommand{\LQCD}{ \Lambda_{\rm QCD} }
\newcommand{\M}{\mathcal{M}}

\def\x{\hat{x}}
\def\z{\hat{z}}

\newcommand{\TeV}{{\rm TeV}}
\newcommand{\GeV}{{\rm GeV}}

\newcommand{\fb}{{\rm fb}}
\newcommand{\MS}{{\overline{\rm MS}}}

\begin{document}
%%%%%%%%%%%%%%%%%%%%%%%%%%%%%%%%%%%%%%%%%%%%%%%%%%%%%%%%

%================================================================================
\title{QCD corrections to light-quark dipole sensitivity via transverse spin asymmetries at the Electron-Ion Collider}
%================================================================================

%--------
\author{Sally Dawson\,\orcidlink{0000-0002-5598-695X}}
\email{dawson@bnl.gov}
\affiliation{High Energy Theory Group, Physics Department, Brookhaven National Laboratory, Upton, New York 11973, USA}
%--------

%--------
\author{Zhite Yu\,\orcidlink{0000-0003-1503-5364}}
\thanks{Corresponding author: \href{mailto:zyu1@bnl.gov}{zyu1@bnl.gov}}
\affiliation{High Energy Theory Group, Physics Department, Brookhaven National Laboratory, Upton, New York 11973, USA}
%--------

\date{\today}
%================================================================

%================================================================
\begin{abstract}
We consider the effects of next-to-leading order (NLO) QCD corrections to transverse spin asymmetries (TSAs) in deep inelastic scattering at the future Electron-Ion Collider (EIC) with polarized electron and proton beams that depend on light quark dipole operators. The TSAs are robust under the inclusion of the NLO QCD effects and provide sensitivity to new physics that is complementary to limits extracted from the Drell-Yan measurements at the LHC and from neutron electric dipole moments. The major uncertainty on our results is from the poor existing determination of the transversity parton distribution function for the down quark. Interesting relations from parity and time reversal invariance, along with chiral symmetry, constrain the analytic form of the TSAs.
\end{abstract}
%================================================================

\maketitle
\tableofcontents

%================================================================================
\section{Introduction}
%================================================================================

Since the discovery of the Higgs boson completed the framework of the Standard Model (SM) of particle physics,
it has withstood all tests from high-energy collider experiments,
and no clear deviations from the SM predictions have been found to date.  
The theoretical predictions of the SM have been verified in many very different channels and energies 
and at accuracies ranging from the percent level to ${\cal{O}}(10\%)$.  
This general agreement between theory and experiment could indicate that 
any new physics lies well above the energy scale that is currently directly within our reach at the LHC.

The Standard Model Effective Field Theory (SMEFT) provides a model-independent framework 
to quantify signals of new physics beyond the SM that correspond to high energies well above the weak scale.
Treating the SM as a weakly interacting low-energy effective theory, 
it parameterizes the effects of new physics at the SM scale in terms of 
a Lagrangian built from an infinite tower of higher-dimensional operators 
that includes only the SM field content and satisfies 
the SU(3)$\times$SU(2)$\times$U(1) gauge invariance of the SM,
\begin{align}
    \Lag_{\rm SMEFT} = \Lag_{\rm SM}
        + \sum_i \frac{C_i^{(d)} \, O_i^{(d)}}{\Lambda^{d-4}}
    \, ,
\end{align}
where $C_i^{(d)}$ represent unknown Wilson coefficients corresponding to the $d$-dimensional operators $O_i^{(d)}$. 
All information about the new physics resides in the $C_i^{(d)}$.  
The effects of the higher dimension operators are suppressed by powers of the new physics scale $\Lambda$ 
that is often taken to be of the TeV scale.  
Including only dimension-6 operators is typically a good numerical approximation, 
and one we follow here (and so we drop the superscript $(d)$ from here on in.) 
The difficulty with the SMEFT approach is, of course, that 
there are many operators whose Wilson coefficients are {\it a priori} independent. 
At the $1/\Lambda^{2}$ level, there are already 2499 distinct operators~\cite{Grzadkowski:2010es} when flavor is included.  
Even neglecting flavor, there are $59$ baryon number conserving dimension-6 operators. 

A useful strategy is to concentrate on subsets of SMEFT operators 
that offer {\it qualitatively} different phenomena from the SM interactions, 
such as baryon number or $CP$ violation,
so that their effects can be separated experimentally from those of other operators.
Of particular interest in this paper are the dimension-6 dipole operators 
involving the structure $\bar \psi \sigma_{\mu\nu} \psi F^{\mu\nu}$, 
where $\psi$ is a light fermion whose mass can be neglected  
compared to the hard scales of the physical processes.
The SM contains a chiral symmetry that is broken only by the Yukawa couplings of the fermions 
and it is useful to classify SMEFT operators relative to this chiral symmetry.   
Dipole operators break the chiral symmetry of the SM 
and their Wilson coefficients mix only among themselves 
under renormalization group (RG) evolution when Yukawa couplings are neglected~\cite{Alonso:2013hga},
forming a self-contained subset that describes unique new physics effects.

It is exactly because of this chiral-odd property that 
light-fermion dipole operators are poorly probed experimentally.
To contribute a non-vanishing result in a cross section, 
a dipole interaction needs to pair with another chiral-odd insertion.
This can be either itself or another dipole interaction, 
and the result will be suppressed by $1 / \Lambda^4$,  
instead of the naive $1 / \Lambda^2$ from a single dipole interaction. 
Similarly, a mass insertion or Yukawa interaction in an amplitude can 
flip the chirality and allow the dipole operator to contribute at $\mathcal{O}(1 / \Lambda^2)$.  
Such effects, however, are suppressed by the fermion mass.
The consequence of this chiral-odd property of the dipole interactions is that 
recent global SMEFT fits find rather poor constraints~\cite{Escribano:1993xr, Alioli:2018ljm, daSilvaAlmeida:2019cbr, Boughezal:2021tih, Cao:2021trr, Gauld_2025, Armadillo:2026mvp} 
or omit light fermion dipoles from the fit~\cite{Celada:2024mcf, deBlas:2025xhe, Ellis:2020unq}. 

A possibility to observe dipole interactions that evades both the $1/\Lambda^4$ and light mass suppressions
is to measure the single transverse spin asymmetry (TSA), which originates from 
the interference of left- and right-handed fermion states and
so is by definition a chiral-odd probe.
While the concept of transverse spin is not new, 
the study in the literature has focused on 
the hadron structure and hadron production 
(see, e.g., \refs{Barone:2001sp, DAlesio:2007bjf, Barone:2010zz, GrossePerdekamp:2015xdx, Anselmino:2020vlp} for reviews),
and has only recently been applied to enhance effects of the SMEFT dipole operators~\cite{Boughezal:2023ooo, Wen:2023xxc, Wang:2024zns, Wen:2024cfu, Wen:2024nff, Huang:2025ljp}.%%%
\footnote{In the same spirit, measurement of the transverse spin asymmetry has also been proposed as a mechanism to provide linear sensitivity to light fermion Yukawa couplings in the SM~\cite{Boughezal:2024yjk, Cao:2025wfg}.}
%%%

By rotational symmetry, a single transverse spin does not affect the total cross section 
but is associated with an observable azimuthal asymmetry.
This can be studied for both initial- and final-state fermions. 
An initial-state transversely polarized fermion beam can 
produce an overall azimuthally asymmetric event distribution, 
which can be observed for any specific particle in the final state.
This can be applied 
directly to electrons or muons in possible future colliders~\cite{Wen:2023xxc, Boughezal:2024yjk, AlAli:2021let, Accettura:2023ked, Grzadkowski:2000hm}
or indirectly to light quarks which acquire their transverse spins 
from transversely polarized proton beams via the transversity parton distribution function (PDF)~\cite{Boughezal:2023ooo} 
or from unpolarized proton beams via correlations with energy flows in the forward direction~\cite{Huang:2025ljp}.
%%%
On the other hand, when produced from a hard scattering, 
a final-state transversely polarized fermion can have an azimuthal asymmetry in its decay products,
which can be observed for the tau lepton~\cite{Bernabeu:1990na, Alemany:1991ki, Ananthanarayan:2002fh}
and also for quarks via their fragmentation products~\cite{Vidal:1998jc, Collins:1992kk}.

In either of these two cases, the transverse spin dependence of the differential cross section
gives a linear sensitivity to the dipole coupling at $\mathcal{O}(1 / \Lambda^2)$
without a suppression from light fermion masses. 
The proposed electron-ion collider (EIC) will provide beams of both 
longitudinally and transversely polarized electrons and protons 
and so will offer the opportunity to use transverse spin observables to probe dipole operators.
Previous studies of single spin asymmetries at the EIC have found that 
sensitivity to the light fermion dipole operators could significantly
improve on existing SMEFT bounds~\cite{Boughezal:2023ooo, Wen:2024cfu}.

The difficulty, however, in extracting limits on dipole operators from transverse spin observables at the EIC 
is that the magnitude of the azimuthal asymmetry induced by the dimension-6 dipoles is generally small.
Therefore, a high precision is needed for such observables in both theory and experiment.
Here, we take the first step in this program and calculate the one-loop correction due to  
gluon exchange to the light quark dipole-induced azimuthal asymmetry in polarized deep inelastic electron-proton scattering (DIS)~\cite{Boughezal:2023ooo}.
This observable is the simplest of its kind experimentally 
because we only need to measure the final-state electron, without constructing any hadronic states. 
Nevertheless, our calculations can readily be extended to other observables.

The calculation is done in dimensional regularization (DR) and 
the most nontrivial part of the calculation concerns the treatment of $\gamma_5$ in the SMEFT. 
This issue is still under intensive discussion in the literature~\cite{Jegerlehner:2000dz, Belusca-Maito:2023wah, DiNoi:2025arz, Fuentes-Martin:2025meq}, 
yet our work provides a new context for the $\gamma_5$ problem in which it enters via the transverse spin projector to the {\it cross section}, 
in addition to its presence in the parity-odd interaction vertices in the {\it amplitude}. 
We find it most convenient to carry out our calculation using  
Kreimer's scheme for treating the $\gamma_5$'s occurring in traces~\cite{Kreimer:1989ke, Korner:1991sx, Kreimer:1993bh},
with which we show the cancellation of soft and collinear divergences 
and obtain the one-loop hard scattering coefficients. 

In this paper, we focus on the phenomenological consequences of the SMEFT QCD corrections 
to polarized transverse spin asymmetries in DIS. 
We find that while as large as $20\%$ corrections can be induced to the tree-level results,
the impacts are numerically very similar for the spin-independent and spin-dependent cross sections 
and so cancel to a large extent in their ratios when forming the asymmetries.
This makes single transverse spin asymmetries not just infrared safe 
but also perturbatively robust, and thus they can serve as 
precision observables in constraining the dipole operators.

The rest of the paper is organized as follows. 
\sec{sec:dip} defines our notation for the dipole operators. 
In \sec{sec:obs}, we discuss the observables needed to construct the DIS transverse spin asymmetry 
and derive constraints on the amplitudes from parity, time reversal, and chiral symmetry.  
The NLO calculation is summarized in \sec{sec:nlores} and 
phenomenological results at the EIC are given in \sec{sec:pheno}.  
We give our conclusion in \sec{sec:con}.
Appendix~\ref{app:sec:summary-xsec} presents analytical results for the different contributions 
to the spin polarized differential cross sections involving the dipole operators. 

%================================================================================
\section{Electroweak dipole operators}
\label{sec:dip}
%================================================================================

We use the Warsaw basis for the dimension-6 SMEFT operators~\cite{Grzadkowski:2010es}.
Restricting to the dipole interactions of the first-generation quarks 
and neglecting the flavor mixing with higher generations, 
we consider the dimension-6 Lagrangian,
\begin{align}
    \Lag_{\rm SMEFT} = \Lag_{\rm SM}
        + \sum_i \biggpp{ 
                \frac{C_i}{\Lambda^2} \, O_i 
                + {\rm h.c.}
            },
\end{align}
where we include the four electroweak dipole operators in the sum over $i$,%%%
\footnote{We have omitted the color dipole operators. 
Although they may contribute at one-loop to the TSA through QCD corrections, 
their contribution is separate from that of the electroweak dipoles. 
There are no meaningful experimental limits on the contribution 
from the real parts of the light quark color dipole operators,  
while limits on the imaginary parts can be derived from the neutron EDM, 
as discussed in \sec{ssec:others}.}
%%%
\begin{align}
	O_{uW} &= ( \bar{Q}_L \, \sigma^{\mu\nu} u_R ) W^b_{\mu\nu} \sigma^b \wt{H}, 
    \nn\\
	O_{uB} 	&= ( \bar{Q}_L \, \sigma^{\mu\nu} u_R ) B_{\mu\nu} \wt{H}, 
    \nn\\
	O_{dW} &= ( \bar{Q}_L \, \sigma^{\mu\nu} d_R ) W^b_{\mu\nu} \sigma^b H, 
    \nn\\
	O_{dB} &= ( \bar{Q}_L \, \sigma^{\mu\nu} d_R ) B_{\mu\nu} H,
\label{eq:dipoles}
\end{align}
and $C_i$ are their Wilson coefficients, suppressed by the new physics scale $\Lambda$. 
We note that the dipole operators are not Hermitian.
We denote $Q_L = (\psi_{uL}, \psi_{dL})^T$ as the quark SU(2) doublets, $\psi_{uR}$ and $\psi_{dR}$ the singlets, 
and $H = (\phi^+, \phi^0)^T$ and $\wt{H} = i \sigma^2 H^* = [ (\phi^0)^*, -(\phi^+)^* ]^T$,
the Higgs doublet. 
Since we will only consider the first generation fermions, we omit flavor indices.
$B_{\mu\nu}$ and $W^b_{\mu\nu}$ are the field strength tensors of the U(1) and SU(2) gauge fields, respectively,
and $b = 1, 2, 3$ labels the SU(2) generators $\sigma^b / 2$.

We work to the next-to-leading order (NLO) of QCD 
but to leading order (LO) in the electroweak SU(2)$\times$U(1) sector. 
This means that we can carry out the calculation in the unitary gauge, 
using as degrees of freedom the photon $\gamma$, $W^{\pm}$, $Z$, and Higgs boson, 
without the inclusion of Goldstone bosons.
The Lagrangian can then be reduced to the broken phase by 
$H \to (0, 1)^T (v + h) / \sqrt{2}$
and 
$\wt{H} \to (1, 0)^T (v + h) / \sqrt{2}$, 
involving the physical Higgs boson, $h$, and the vacuum expectation value, $v$.
The relevant dipole interactions for our calculation of neutral-current DIS are
\begin{align}
    \Lag_{\rm SMEFT}  &\supset \Lag_{\rm SM}
    - \frac{v + h}{2 v^2} \Bigbb{
        e A_{\mu\nu} \,
        \bar{\psi}_u \, \sigma^{\mu\nu} \bigpp{ \Re \Gamma_{\gamma}^u + i \, \Im \Gamma_{\gamma}^u \, \gamma_5 } \psi_u \nn\\
        &\hspace{8em} 
        + g_Z Z_{\mu\nu} \,
        \bar{\psi}_u \, \sigma^{\mu\nu} \bigpp{ \Re \Gamma_Z^u + i \, \Im \Gamma_Z^u \, \gamma_5 } \psi_u
        + (u \to d)
    }.
\label{eq:Lag-AZ}
\end{align}
Here $A_{\mu\nu} = \partial_{\mu} A_{\nu} - \partial_{\nu} A_{\mu}$ 
and $Z_{\mu\nu} = \partial_{\mu} Z_{\nu} - \partial_{\nu} Z_{\mu}$
are the field strength tensors of the photon and $Z$ boson, respectively,
and we have redefined the dipole couplings for this physical basis, 
\begin{align}
    e \, \Gamma_{\gamma}^u
    &= \frac{\sqrt{2} v^2}{\Lambda^2} \pp{
        -s_W \, C_{uW} - c_W \, C_{uB}
    },
    \nn\\
    e \, \Gamma_{\gamma}^d
    &= \frac{\sqrt{2} v^2}{\Lambda^2} \pp{
        s_W \, C_{dW} - c_W \, C_{dB}
    },
    \nn\\
    g_Z \, \Gamma_{Z}^u
    &= \frac{\sqrt{2} v^2}{\Lambda^2} \pp{
        -c_W \, C_{uW} + s_W \, C_{uB}
    },
    \nn\\
    g_Z \, \Gamma_{Z}^d
    &= \frac{\sqrt{2} v^2}{\Lambda^2} \pp{
        c_W \, C_{dW} + s_W \, C_{dB}
    }.
\label{eq:dip-wcoefs-conver}
\end{align}
We use $(g_1, g_2, g_3)$ for the gauge couplings of the SM gauge groups, U(1), SU(2), and SU(3), respectively,
$(c_W, s_W) = (\cos\theta_W, \sin\theta_W)$ for the weak mixing angle $\theta_W = \arctan(g_1 / g_2)$,
and $(e, g_Z) = (g_2 s_W, g_2 / (2 c_W))$ for the photon and $Z$ couplings, respectively.
Note that we have rescaled the dipole couplings in \eq{eq:dip-wcoefs-conver} 
by factors of $e$, $g_Z$, and $v^2$ to have a better correspondence to the SM couplings;
the factor $v$ will formally play the role of the high energy scale suppression in our calculation.

To fix the remaining notation, we list the relevant Feynman rules for quark couplings to the photon and $Z$,
\begin{align}
    \adjincludegraphics[valign=c, scale=0.8]{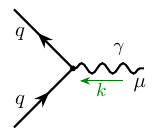}
	& = - i e \biggcc{
        e_q \gamma^{\mu}
		+ \frac{i \sigma^{\mu \nu} k_{\nu}}{v}
		\bigbb{ \Re(\Gamma^{q}_{\gamma}) 
        + i \Im(\Gamma^{q}_{\gamma}) \gamma_5 }
    },
	\nn\\
    \adjincludegraphics[valign=c, scale=0.8]{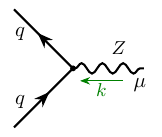}	
    & = - i g_Z \biggcc{
        \gamma^{\mu} (g_V^q - g_A^q \gamma_5) 
		+ \frac{i  \sigma^{\mu \nu} k_{\nu}}{v}
		\bigbb{ \Re(\Gamma^{q}_Z) 
        + i \Im(\Gamma^{q}_Z) \gamma_5}
    },
\label{eq:feyn-rules}
\end{align}
where $q = u, d$, and $(e_q, g_V^q, g_A^q)$ are defined as
\beq
    g_V^q = \tau^3_q - 2 e_q s_W^2, 
    \quad
	g_A^q = \tau^3_q,
    \quad
    (e_u, e_d) = (2/3, -1/3), 
    \quad
    (\tau^3_u, \tau^3_d) = (1/2, -1/2).
\eeq
The SM part of the Feynman rules in \eq{eq:feyn-rules} applies also to the electron, 
with $e_e = -1$ and $\tau_e^3 = -1/2$.
Then, we define the currents associated with $\gamma$ and $Z$,
\begin{align}
	J_{\gamma}^{\mu} 
	&\equiv e_e \, \bar \psi_e \gamma^{\mu} \psi_e + 
		\sum_{q = u, d} \biggcc{ 
		e_q \, \bar \psi_q \gamma^{\mu} \psi_q 
		+ \frac{1}{v} \, \partial_{\nu} \Bigpp{ 
			\bar \psi_q \, \sigma^{\mu\nu} \bigbb{ \Re(\Gamma_{\gamma}^q) + i \Im(\Gamma_{\gamma}^q) \gamma_5 } \psi_q 
		}
	}, \nn\\
	J_{Z}^{\mu} 
	&\equiv \bar \psi_e \gamma^{\mu} (g_V^e - g_A^e \gamma_5) \psi_e + 
		\sum_{q = u, d} \biggcc{ 
		\bar \psi_q \gamma^{\mu} (g_V^q - g_A^q \gamma_5) \psi_q
		+ \frac{1}{v} \, \partial_{\nu} \Bigpp{ 
			\bar \psi_q \, \sigma^{\mu\nu} \bigbb{ \Re(\Gamma_{Z}^q) + i \Im(\Gamma_{Z}^q) \gamma_5 } \psi_q 
		}
	},
\label{eq:currents}
\end{align}
which are both Hermitian, and
where we have only retained the electron $(\psi_e)$ and quark terms that are relevant to this study. 

We are interested in the one-loop QCD effects involving dipole operators in polarized DIS. 
The Wilson coefficients $C_i$ corresponding to the dipole operators in \eq{eq:dipoles} 
all have the same renormalization group (RG) equation~\cite{Alonso:2013hga},
\beq[eq:RGE-C]
    \frac{d C_i}{d\ln \mu}
    = \frac{\alpha_s C_F}{2\pi} C_i,
\eeq
where $\alpha_s \equiv g_3^2 / (4\pi)$, $\mu$ is the RG scale,%%%
\footnote{We take the renormalization and factorization scales to be equal throughout the paper.}
%%%
and $C_F = 4/3$ is the SU(3) group factor.
This means that the $C_i$ grow with increasing energy scale.
The RG evolutions of SM parameters are unaffected by the dipole couplings.
Among them we include only the RG for $\alpha_s$,
\beq[eq:RGE-als]
    \frac{d \alpha_s}{d\ln \mu}
    = -\frac{\alpha_s^2}{2\pi}\beta_0 ,
\eeq
where $\beta_0 = 11 - 2 n_f / 3$ depends on the active quark number $n_f$.
Combining \eqs{eq:RGE-C}{eq:RGE-als}, we obtain the RG solution of the dipole couplings,
\beq[eq:run-C]
    C_i(\mu)
    = C_i(Q_0) \bb{
        \frac{\alpha_s(\mu)}{\alpha_s(Q_0)}
    }^{-\frac{C_F}{\beta_0}},
\eeq
in any regime where $n_f$ stays constant.
This RG evolution applies to the dipole couplings 
in the physical basis [Eq.~\eqref{eq:dip-wcoefs-conver}] as well.

%================================================================================
\section{Observables, factorization, and symmetry constraints}
\label{sec:obs}
%================================================================================

\begin{figure}[htbp]
	\centering
	\includegraphics[scale=0.5]{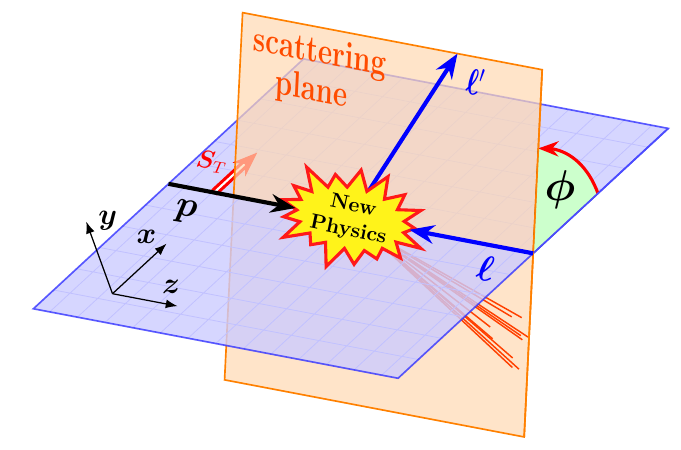}
	\caption{Lab-frame kinematics of DIS with a transversely polarized proton.}
\label{fig:geo}
\end{figure}

The process that we study is DIS of an electron $e$ and proton $P$,
\beq[eq:dis]
	P(p) + e(\l) \to e(\l') + X,
\eeq
where the quantities inside parentheses are the momenta, and 
we take the proton to move along the $+z$ direction
and the electron along $-z$.
This process is described by the two Lorentz invariants,
\beq[eq:kin-xQ]
	Q^2 = -q^2 \equiv -(\l - \l')^2, 
	\qquad
	x_B \equiv \frac{Q^2}{2p \cdot q},
\eeq
and the azimuthal angle $\phi$ of the final-state electron in the lab frame, 
as shown in \fig{fig:geo}. 
We also define $\bm{\l}_T = (\l_T \cos\phi, \l_T \sin\phi)$ as the transverse momentum of the scattered electron,
and the DIS inelasticity variable,
\beq[eq:kin-y]
	y = \frac{p \cdot q}{p \cdot \l} = \frac{Q^2}{x_B s},
\eeq
with $s = (p + \l)^2$ the $e$-$P$ center-of-mass energy squared.
Electron and proton masses are neglected throughout.
Then, combining \eqs{eq:kin-xQ}{eq:kin-y}, one may easily derive 
\beq[eq:lT-expr]
	\l_T \equiv |\bm{\l}_T|
    = Q \sqrt{1 - y}.
\eeq

As introduced in \refcite{Boughezal:2023ooo}, when the proton has a transverse spin 
$\bm{S}_T = (S_T \cos\phi_S, S_T \sin\phi_S)$
with magnitude $S_T \equiv |\bm{S}_T|$ along the azimuthal direction $\phi_S$ in the lab frame,
the azimuthal $\phi$ distribution of the final-state electron will have a nontrivial modulation sensitive to the quark dipole operators in \eq{eq:dipoles} at $\order{1 / \Lambda^2}$.
This modulation is the main focus of this paper;
our aim is to include the one-loop QCD corrections to the results in \refcite{Boughezal:2023ooo}.
We do not consider the transverse spin of the electron beam as in \refcite{Boughezal:2023ooo}, 
which does not generate effects that are sensitive to the quark dipole operators.
Instead, we include its longitudinal polarization $\lambda_e$, 
which will enhance the photon channel signal~\cite{Wen:2024cfu, Huang:2025ljp} that dominates at the EIC energy.

%----------------------------------------------------------------
% Fig: geometry
%------------------------------------------------
%----------------------------------------------------------------

%----------------------------------------------------------------
\subsection{Setup and factorization}
%----------------------------------------------------------------

\begin{figure}[htbp]
	\centering
	\includegraphics[scale=1]{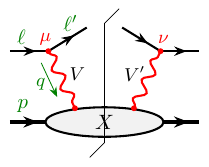}
	\caption{Cut diagram for the contribution of the $(VV')$ channel to the DIS cross section.}
\label{fig:dis}
\end{figure}

We work at LO in the electroweak (EW) interactions.
Then, a generic amplitude squared contributing to the DIS cross section is shown in \fig{fig:dis} as a cut diagram.
To the left of the cut is the amplitude of $e$-$P$ scattering to $e$ plus any set of particles $X$,
initiated via an exchange of $V$ ($= \gamma$ or $Z$) carrying a highly virtual spacelike momentum $q$,%%%
\footnote{We adopt the standard notation where $q$ refers to both the momentum exchange $q = \l - \l'$ 
and the quark flavor label $(u, d)$. The distinction should be clear from the context.}
%%%
and to the right is a conjugate amplitude into the same final state, 
but mediated by $V'$ ($= \gamma$ or $Z$).

%----------------------------------------------------------------
% Fig: Cut diagram
%------------------------------------------------
%----------------------------------------------------------------

The DIS cross section is given by the sum over the four combinations of $(VV')$ and all final states $X$,
\begin{align}
	\frac{d\sigma}{dx_B \, dQ^2 \, d\phi}
	&= \frac{y^2}{(4\pi)^3 Q^2} \sum_{V, V'} \sum_X \Bigvv{ \M_X^V \bigpp{ \M_X^{V'} }^*} \,
		(2\pi)^4 \delta^{(4)}(p + \l - \l' - p_X)
	\nn\\
	&\equiv \sum_{V, V'} \frac{N_{VV'}}{2\pi} \biggbb{ 
		\frac{y^2}{Q^2} \, L^{\mu\nu}_{VV'} W_{VV', \mu\nu}
	},
\label{eq:xsec-LW}
\end{align}
where $\M_X^V$ is the amplitude for the final state $X$ mediated by $V$, 
and $\vv{\cdot}$ refers to a spin sum for final states and average for initial states (weighted by density matrices).
In the second line, we have decomposed each channel into 
a leptonic tensor $L^{\mu\nu}_{VV'}$ and a hadronic tensor $W_{VV', \mu\nu}$.
The coupling constants and propagators associated with $V$ and $V'$ 
have been included in the normalization factors $N_{VV'}$, 
defined as
\begin{align}
	N_{\gamma\gamma} = \frac{2\pi \alpha_e^2}{Q^4}, 
	\quad
	N_{ZZ} = \frac{2\pi \alpha_Z^2}{(Q^2 + m_Z^2)^2},
	\quad
	N_{\gamma Z} = N_{Z \gamma} = \frac{2\pi \alpha_e \alpha_Z}{Q^2 (Q^2 + m_Z^2)},
\label{eq:norm-NI}
\end{align}
with $(\alpha_e, \alpha_Z) = (e^2, g_Z^2)/4\pi$
and $m_Z$ is the mass of the $Z$ boson.

The leptonic tensor $L^{\mu\nu}_{VV'}$ is formally defined as
\begin{align}
	L^{\mu\nu}_{VV'}
	= \frac{1}{2} \sum_{\alpha, \bar \alpha, \alpha'}
		\rho^e_{\alpha \bar\alpha}(\lambda_e) 
		\vv{ e(\l', \alpha') | \, J_{V}^{\mu}(0) \, | e(\l, \alpha) }
		\vv{ e(\l', \alpha') | \, J_{V'}^{\nu}(0) \, | e(\l, \bar \alpha) }^* \big|_{\text{LO EW}},
\label{eq:leptonic-tensor}
\end{align}
where the local currents $J_V^{\mu}(0)$ are defined in \eq{eq:currents}.
The matrix elements are evaluated at LO in the EW interactions, 
which are simply given at tree level.
The $\alpha$'s represent electron helicities, among which the initial-state ones
are averaged with the density matrix $\rho^e$.
The latter has only two nonzero diagonal elements,
\beq
	\rho^e_{\pm\pm} = \frac{1 \pm \lambda_e}{2},
	\quad
	\rho^e_{\pm\mp} = 0,
\eeq
where $\lambda_e = \rho^e_{++} - \rho^e_{--}$ is the longitudinal polarization degree of the electron beam.

The leptonic tensor $L^{\mu\nu}_{VV'}$ is constrained by: 
(1) current conservation associated with the massless electron, 
$q_{\mu} L^{\mu\nu}_{VV'} = L^{\mu\nu}_{VV'} q_{\nu} = 0$,
and 
(2) tree-level transversality for the vector current of $V$ or $V'$,
$\l_{\mu} L^{\mu\nu}_{VV'} = L^{\mu\nu}_{VV'} \l_{\nu} = 0$,
which can be easily seen by transforming into the frame 
where $V(q)$ is radiated collinear to $\l$.
These properties allow the leptonic tensor to be decomposed into 
two scalar form factors $(L^{VV'}_1, L^{VV'}_2)$,
\beq[eq:lepton-L-param]
	L_{VV'}^{\mu\nu}
	\equiv \pp{ -\frac{Q^2}{2} \wt{g}^{\mu\nu} + 2 \wt{\l}^{\mu} \wt{\l}^{\nu} } L^{VV'}_1 
		- i \, \epsilon^{\mu \nu \l q} \, L^{VV'}_2 ,
\eeq
for each channel $(VV')$,
where we define the gauge-invariant structures, 
\beq[eq:gi-notat]
	\wt{g}^{\mu\nu} = g^{\mu\nu} - \frac{q^{\mu} q^{\nu}}{q^2},
	\quad
	\wt{k}^{\mu} = \wt{g}^{\mu\alpha} k_{\alpha}.
\eeq
with the latter defined for any four vector $k$.
We use the notation
\beq	
	\epsilon^{\mu \nu \l q} \equiv \epsilon^{\mu \nu \rho \sigma} \l_{\rho} q_{\sigma},
\eeq
with the Levi-Civita tensor convention $\epsilon_{0123} = -\epsilon^{0123} = 1$.
%%%
By Hermiticity,  
$L^{VV'}_i = L^{V'V}_i$ are real functions ($i = 1, 2$).
From parity invariance, 
$L^{VV'}_1$ must be even in $(\lambda_e, g_A^e)$
while $L^{VV'}_2$ is odd.

For example, the leptonic tensor of the $\gamma\gamma$ channel explicitly reads 
\begin{align}
	L_{\gamma\gamma}^{\mu\nu}
	&= \frac{1}{2} \Tr\bb{ \frac{\slash{\l}}{2} \, (1 - \lambda_e \gamma_5) \, 
		(e_e \gamma^{\nu}) \, \Slash{\l'} \, (e_e \gamma^{\mu}) }
	\nn\\
	&= -\frac{Q^2}{2} \wt{g}^{\mu\nu} + 2 \wt{\l}^{\mu} \wt{\l}^{\nu}
			- i \lambda_e \, \epsilon^{\mu\nu \l q},
\label{eq:AA-lepton-trace-tree}
\end{align} 
which gives 
\beq[eq:AA-lepton-coefs]
	L^{\gamma\gamma}_1 = 1,
	\qquad
	L^{\gamma\gamma}_2 = \lambda_e.
\eeq
Results for other channels are given in Appendix~\ref{app:sec:summary-xsec}.

The hadronic tensor $W^{\mu\nu}_{VV'}$ is defined as 
\begin{align}
	W^{\mu\nu}_{VV'}(p, q, \bm{S}_T)
	&= \frac{1}{2\pi} \sum_{\lambda, \lambda'} 
		\rho_{\lambda \lambda'}(\bm{S}_T) 
		\int d^4 r \, e^{-i q \cdot r} \, 
		\vv{ p(p, \lambda') | \,
			J_{V'}^{\nu}(0) \, J_V^{\mu}(r) \,
		|p(p, \lambda) },
\label{eq:hadronic-tensor}
\end{align}
where $r$ is the space-time coordinate.
\eq{eq:hadronic-tensor} has absorbed the $\delta$-function and the sum over $X$ in \eq{eq:xsec-LW} 
to convert the amplitude squared into the forward matrix element of the current correlation,
which is again understood to be defined to LO in the EW interactions, but to all orders in QCD. 
The proton helicities $\lambda$ and $\lambda'$ are averaged with the density matrix,
which is related to the transverse spin vector $\bm{S}_T$ in the lab frame by
\beq[eq:den-mtx]
	\rho_{\lambda \lambda'}(\bm{S}_T) 
	= \frac{1}{2} \begin{pmatrix}
		1 & S_T \, e^{-i \phi_S} \\
		S_T \, e^{i \phi_S} & 1
		\end{pmatrix}_{\lambda \lambda'}.
\eeq

In the DIS limit, with $Q$ much greater than the proton mass,
the hadronic tensor can be factorized into PDFs 
that are convoluted with hard scattering coefficients,
\begin{align}
	W^{\mu\nu}_{VV'}(p, q, \bm{S}_T)
	&= \sum_a \int_{x_B}^1 \frac{dx}{x}
		f_a(x, \mu) \,
		\hat{W}_{VV', a}^{\mu\nu}\biggpp{ x p, q, \bm{s}_{T, a}(x, \mu), \frac{Q}{\mu}, \alpha_s(\mu), C_i(\mu) }
		+ \order{\frac{\LQCD}{Q}},
\label{eq:W-fact}
\end{align}
where $\mu$ is the factorization scale, $f_a$ is the unpolarized PDF of parton $a$, 
which is summed over quarks, antiquarks, and gluons. 
The $\hat{W}$ is the ``partonic tensor'' given analogously to \eq{eq:hadronic-tensor} 
but for on-shell parton states and  
after subtraction of collinear divergences to render it infrared safe. 
For $a = q$ or $\bar{q}$, 
$\hat{W}$ is given by 
\begin{align}
	&\hat{W}_{VV', a}^{\mu\nu}\biggpp{ k, q, \bm{s}_{T, a}, \frac{Q}{\mu}, \alpha_s(\mu), C_i(\mu) }
	\nn\\
    &\qquad
    = \frac{1}{2\pi} \sum_{\lambda, \lambda'} 
		\rho_{\lambda \lambda'}\bigpp{ \bm{s}_{T, a} }
		\int d^4 r \, e^{-i q \cdot r} \, 
		\vv{ a(k, \lambda') | \,
			J_{V'}^{\nu}(0) \, J_V^{\mu}(r) \,
		|a(k, \lambda) } \big|_{\text{col. sub.}},
\label{eq:partonic-tensor-q}
\end{align}
which is understood within perturbation theory,
and where $k = x p$ is the lightlike parton momentum 
which defines the ``partonic Bjorken-${\hat x}$'',
\beq[eq:xhat-def]
	\x = \frac{Q^2}{2 k \cdot q} = \frac{x_B}{x}.
\eeq
The spin density matrix takes the same form as \eq{eq:den-mtx},
with the parton's transverse spin related to that of the proton by
\beq[eq:sT-h/f]
	\bm{s}_{T, a}
	\equiv \bm{s}_{T, a}(x, \mu)
	= \frac{h_a(x, \mu)}{f_a(x, \mu)} \bm{S}_T.
\eeq
Here $h_a(x, \mu)$ is the transversity PDF~\cite{Ralston:1979ys, Bukhvostov:1984rns, Jaffe:1991kp, Barone:2001sp, Barone:2010zz, Collins:2011zzd}, 
physically given by the difference in number densities of partons 
polarized in the same and in the opposite directions of the fully transversely polarized proton.
Since gluons do not have transversity PDFs, we have, for $a = g$, 
\begin{align}
	&\hat{W}_{VV', g}^{\mu\nu}\biggpp{ k, q, \frac{Q}{\mu}, \alpha_s(\mu), C_i(\mu) }
	\nn\\
    &\qquad
    = \frac{1}{2\pi} \sum_{\lambda} \frac{1}{2}
		\int d^4 r \, e^{-i q \cdot r}
		\vv{ g(k, \lambda) | \,
			J_{V'}^{\nu}(0) \, J_V^{\mu}(r) \,
		|g(k, \lambda) } \big|_{\text{col. sub.}},
\label{eq:partonic-tensor-g}
\end{align}
which is the same as in the unpolarized DIS case.
The $C_i(\mu)$ arguments on the left-hand sides of \eqs{eq:partonic-tensor-q}{eq:partonic-tensor-g}
are the dipole couplings, entering at LO in the EW interactions only through the currents.

The correction terms suppressed by $\LQCD / Q$ in \eq{eq:W-fact}
arise from proton mass dependence and higher-twist effects.
As estimated in \refcite{Boughezal:2023ooo}, 
they give negligible contributions to the transverse spin asymmetry observables
so will be dropped in the following.

%----------------------------------------------------------------
\subsection{Partonic tensor decomposition}
\label{ssec:tensor}
%----------------------------------------------------------------

Before the actual perturbative calculation of the partonic tensors, 
it is useful to decompose them  into an independent gauge-invariant tensor basis.

In our case, only the quark (and antiquark) tensors involve the spin dependence,
which enters \eq{eq:partonic-tensor-q} linearly through the density matrix.
The helicity form of \eq{eq:partonic-tensor-q} can be rewritten equivalently 
in terms of a covariant Feynman diagram calculation by replacing the quark spin average $\slash{k} / 2$
with
\beq[eq:q-spin-avg]
	\sum_{\lambda, \lambda'} 
		\rho_{\lambda \lambda'}(\bm{s}_{T,a}) \, u(k, \lambda) \, \bar{u}(k, \lambda')
	= \frac{\slash{k}}{2} (1 - \slash{s}_a \gamma_5),
\eeq
where $s_a$ is the spin four-vector that takes the form in the lab frame,
\beq[eq:smu]
	s_a^{\mu} = (0, \bm{s}_{T,a}, 0),
\eeq
and {\it formally} transforms covariantly in the same way as $k$ under a Lorentz transformation.
Since its appearance in \eq{eq:q-spin-avg} is next to $\slash{k}$,
it is equivalent to the gauge-invariant form,
\beq
	s_{a\perp}^{\mu} \equiv s_a^{\mu} - \frac{s_a \cdot q }{ k \cdot q } k^{\mu},
\eeq
simply because $\slash{k}\slash{k} = 0$.

The tensor basis is thus made of $\cc{ k, q, s_{a\perp}, g_{\mu\nu}, \epsilon_{\mu\nu\rho\sigma} }$, 
which leads to seven independent gauge-invariant structures for the partonic tensors.
Three of them are associated with unpolarized partons,
\begin{align}
	\tau_L^{\mu\nu} 
	&= \frac{Q^2}{(k \cdot q)^2} \wt{k}^{\mu} \wt{k}^{\nu}, 
	\nn\\
	\tau_T^{\mu\nu}
	&= - \wt{g}^{\mu\nu} + \tau_L^{\mu\nu},
	\nn\\
	\tau_3^{\mu\nu} 
	&= \frac{i}{k \cdot q} \epsilon^{\mu\nu k q},
\label{eq:gi-struct-U}
\end{align}
where we have used the notation of \eq{eq:gi-notat}.
The other four structures are associated with the transverse spin,
\begin{align}
	\Delta\tau_1^{\mu\nu}
	&= \frac{i \, Q}{(k \cdot q)^2} \pp{ 
		\epsilon^{\mu k q s_a} \wt{k}^{\nu} 
		- \epsilon^{\nu k q s_a} \wt{k}^{\mu} 
	}, 
	\nn\\
	\Delta\tau_2^{\mu\nu}
	&= \frac{Q}{k \cdot q} \pp{ 
		s_{a\perp}^{\mu} \wt{k}^{\nu} 
		+ s_{a\perp}^{\nu} \wt{k}^{\mu} 
	},
	\nn\\
	\Delta\wt{\tau}_1^{\mu\nu}
	&= \frac{-i \, Q}{k \cdot q} \pp{ 
		s_{a\perp}^{\mu} \wt{k}^{\nu} 
		- s_{a\perp}^{\nu} \wt{k}^{\mu} 
	},
	\nn\\
	\Delta\wt{\tau}_2^{\mu\nu}
	&= \frac{Q}{(k \cdot q)^2} \pp{ 
		\epsilon^{\mu k q s_a} \wt{k}^{\nu} 
		+ \epsilon^{\nu k q s_a} \wt{k}^{\mu} 
	}.
\label{eq:gi-struct-T}
\end{align}
These structures are orthogonal to each other and are normalized as
\begin{align}
	\tau_L^{\mu\nu} \tau_{L, \mu\nu} &= 1, 
	\quad
	\tau_T^{\mu\nu} \tau_{T, \mu\nu} 
	= \tau_3^{\mu\nu} \tau_{3, \mu\nu}^* = 2,
	\nn\\
	\Delta\tau_i^{\mu\nu} \Delta\tau_{i, \mu\nu}^* 
	&= \Delta\wt{\tau}_i^{\mu\nu} \Delta\wt{\tau}_{i, \mu\nu}^{\,*}
	= 2 s_a^{\mu} s_{a \mu}
	= -2 \bm{s}_{T, a}^2,
	\quad
	(i = 1, 2).
\end{align}

Next, we make use of the property $L^{VV'}_{\mu\nu} = L^{V'V}_{\mu\nu}$ of the leptonic tensor 
to organize the sum over channels in \eq{eq:xsec-LW} as
\begin{align}
	\sum_{V, V'} L_{VV'} \cdot W_{VV'}
	= L_{\gamma} \cdot W_{\gamma} + L_{Z} \cdot W_{Z} 
		+ L_{\rm int} \cdot W_{\rm int} 
	\equiv \sum_{I = \gamma, Z, {\rm int}} L_I \cdot W_I,
\end{align}
with the channel-symmetric notation,
\begin{align}
	(L_V^{\mu\nu}, W_V^{\mu\nu}) 
    &\equiv (L_{VV}^{\mu\nu}, W_{VV}^{\mu\nu}),
	\quad
	L_{\rm int}^{\mu\nu} \equiv L_{\gamma Z}^{\mu\nu},
	\quad
	W_{\rm int}^{\mu\nu} \equiv W_{\gamma Z}^{\mu\nu} + W_{Z \gamma}^{\mu\nu},
    \nn \\
    N_V & \equiv N_{VV}, 
    \quad 
    N_{\rm int} \equiv N_{\gamma Z} = N_{Z\gamma}.
\label{eq:ch-notat}
\end{align}
The Hermiticity property implies 
\beq
	\pp{ W_I^{\mu\nu} }^* = W_I^{\nu\mu} ,
	\qquad
	(I = \gamma, Z, {\rm int})
\eeq
which also applies to the partonic tensor $\hat W_{I, a}^{\mu\nu}$,
so that when expanded onto the basis of \eqs{eq:gi-struct-U}{eq:gi-struct-T},
all its scalar coefficients are real functions.
This is why an $i$ factor has been assigned to each antisymmetric structure
in $(\tau_3, \Delta\tau_1, \Delta\wt{\tau}_1)$.

When contracted with the leptonic tensor in \eq{eq:lepton-L-param},
the gauge-invariant structures in \eqs{eq:gi-struct-U}{eq:gi-struct-T} give 
\bse[eq:L-tau-contract]\begin{align}
	\frac{y^2}{Q^2} \, L_{I}^{\mu\nu} \cdot \bigcc{
		\tau_T, \, \tau_L, \, \tau_3
	}_{\mu\nu}
	&=  \Bigcc{
		L^I_1 \, \bigpp{ 1 + (1 - y)^2 }, \, 
		L^I_1 \, \bigpp{ 2(1 - y) }, \,
		L^I_2 \, \bigpp{ y (2 - y) }
	}, \\
	\frac{y^2}{Q^2} \, L_{I}^{\mu\nu} \cdot \bigcc{
		\Delta\tau_1, \, \Delta\tau_2
	}_{\mu\nu}
	&= \frac{2\bm{s}_{T, a} \cdot \bm{\l}_T}{Q} \, 
	\Bigcc{
		L^I_2 \cdot y , \, 
		-L^I_1 \, \bigpp{ 2 - y }
	}, 
	\label{eq:L-tau-T1} \\
	\frac{y^2}{Q^2} \, L_{I}^{\mu\nu} \cdot \bigcc{
		\Delta\wt{\tau}_1, \, \Delta\wt{\tau}_2
	}_{\mu\nu}
	&= -\frac{2(\bm{s}_{T, a} \times \bm{\l}_T)^z}{Q} \, 
	\Bigcc{
		L^I_2 \cdot y , \, 
		-L^I_1 \, \bigpp{ 2 - y }
	},
	\label{eq:L-tau-T2} 
\end{align}\ese
where $\bm{s}_{T, a} \cdot \bm{\l}_T$ and $(\bm{s}_{T, a} \times \bm{\l}_T)^z$ are evaluated in the lab frame.
These factors come from the Lorentz invariant expressions $q \cdot s_a$ and $\epsilon^{k q \l s_a}$
and yield the explicit form of the azimuthal dependence in the cross section,
\begin{align}
	\bm{s}_{T, a} \cdot \bm{\l}_T = s_{T, a} \, \l_T \, \cos(\Delta\phi), 
	\qquad
	(\bm{s}_{T, a} \times \bm{\l}_T)^z
	= s_{T, a} \, \l_T \, \sin(\Delta\phi),
\label{eq:dot-cross}
\end{align}
with $\Delta\phi \equiv \phi - \phi_S$,
where $\phi$ is defined below \eq{eq:kin-xQ} and 
$\phi_S$ is defined below \eq{eq:lT-expr} or in \eq{eq:den-mtx}.

%-----------------------------------------
\subsection{Implications from time reversal and chiral symmetry}
%-----------------------------------------

The symmetry between \eqs{eq:L-tau-T1}{eq:L-tau-T2} suggests 
a corresponding relation between the scalar  coefficients of $\Delta \tau_i$ and $\Delta \wt \tau_i$
in the decomposition of the partonic tensors. 
In this subsection, we derive such a relation.
The essential two steps are to 
extend the Christ-Lee theorem~\cite{Christ:1966zz} to incorporate dipole currents
and then to use chiral symmetry.

Let us first define the spin unaveraged version of \eq{eq:partonic-tensor-q},
for given quark helicity states $(\lambda,\lambda^\prime)$,
\begin{align}
	\hat{W}_{I, \lambda\lambda'}^{\mu\nu}(k, q)
	&= \frac{1}{2\pi} 
		\int d^4 r \, e^{-i q \cdot r} \, 
		\vv{ k, \lambda' | \,
			J_I^{\nu}(0) \, J_I^{\mu}(r) \,
		|k, \lambda } ,
\label{eq:partonic-tensor-q-hel}
\end{align}
where we have suppressed the flavor label, which stays unchanged in this analysis, 
as well as the dependence on $C_i$ and $\alpha_s$.
We use the channel notation of \eq{eq:ch-notat},
with the current correlator symmetrized with respect to the vector bosons,
i.e.,  
$J_I^{\nu}(0) \, J_I^{\mu}(r) = J_\gamma^{\nu}(0) \, J_\gamma^{\mu}(r)$
or $J_Z^{\nu}(0) \, J_Z^{\mu}(r)$ for $I = (\gamma)$ or $(Z)$,
and $J_I^{\nu}(0) \, J_I^{\mu}(r) = J_\gamma^{\nu}(0) \, J_Z^{\mu}(r) + J_Z^{\nu}(0) \, J_\gamma^{\mu}(r)$
for $I = ({\rm int})$.
At this point, \eq{eq:partonic-tensor-q-hel} denotes the
matrix element without collinear subtraction,
so does not contain the factorization scale dependence $Q/\mu$, to which we will come back later.

We work in the lab frame, where $q = \l - \l'$ has an azimuthal angle $\phi + \pi$.
This $\phi$ dependence can be factored out 
by inserting a rotation operator ${\cal{U}}[R_z(-\phi)]\equiv \exp\bigpp{ i J_z \phi }$
around the $z$ direction, with $J_z$ the corresponding angular momentum operator,
\begin{align}
	\hat{W}_{I, \lambda\lambda'}^{\mu\nu}(k, q)
	&= \frac{1}{2\pi} 
		\int d^4 r \, e^{-i q \cdot r} \, 
		\vv{ k, \lambda' | \, e^{-i J_z \phi} e^{i J_z \phi} \,
			J_I^{\nu}(0) \, J_I^{\mu}(r) \,
		e^{-i J_z \phi} e^{i J_z \phi} \, |k, \lambda } 
	\nn\\
	&= \frac{1}{2\pi} \,
		e^{i (\lambda - \lambda') \phi} \,
		\bb{ R_z(\phi) }^{\nu}{}_{\sigma} \,
		\bb{ R_z(\phi) }^{\mu}{}_{\rho} \,
		\int d^4 r \, e^{-i q \cdot r} \, 
		\vv{ k, \lambda' | \,
			J_I^{\sigma}(0) \, J_I^{\rho}\bigpp{ R_z(-\phi) r } \,
		|k, \lambda } 
	\nn\\
	&= e^{i (\lambda - \lambda') \phi} \,
		\bb{ R_z(\phi) }^{\mu}{}_{\rho} \,
		\bb{ R_z(\phi) }^{\nu}{}_{\sigma} \,
		\hat{W}_{I, \lambda\lambda'}^{\rho\sigma}(k, q_{\star}),
\label{eq:partonic-tensor-rot}
\end{align}
where%%%
\footnote{While we generally write ``$\pm$'' for helicity labels $\lambda$, 
when their specific values are used, we have $\lambda = \pm 1/2$ for fermions.}
%%%
$q_{\star}^{\mu} = \bb{ R_z(-\phi) }^{\mu}{}_{\nu} \, q^{\nu}$ rotates $q$ to the azimuthal plane of $\pi$.
When contracted with the leptonic tensor in \eq{eq:leptonic-tensor}, 
the rotation matrix $R_z(\phi)$ rotates the lepton momentum $\l'$ to have azimuthal angle $0$,
\beq
	L_I^{\mu\nu}(\l, q) \,
		\bb{ R_z(\phi) }_{\mu}{}^{\rho}
		\bb{ R_z(\phi) }_{\nu}{}^{\sigma}
	= L_I^{\rho\sigma}(\l, q_{\star}).
\eeq
Then the contraction of $L$ and $W$ in the quark helicity basis becomes
\begin{align}
	L_{I, \mu\nu}(\l, q) \, \hat{W}_{I, \lambda\lambda'}^{\mu\nu}(k, q)
	&= e^{i (\lambda - \lambda') \phi} \,
		L_{I, \mu\nu}(\l, q_{\star}) \,
		\hat{W}_{I, \lambda\lambda'}^{\mu\nu}( k, q_{\star} )
	\nn\\
	&\equiv
		e^{i (\lambda - \lambda') \phi} \,
		[L\hat W]^I_{\lambda\lambda'}( k, \l, q_{\star} ).
\label{eq:partonic-tensor-LW}
\end{align}
In the last factor $[L\hat W]$,
all momenta have been oriented to be in the $x$-$z$ plane, 
so it has no $\phi$ dependence.

\eq{eq:partonic-tensor-LW} is a general result that depends only on the rotation property.
It does not rely on the separation of leptonic and partonic tensors.
Nevertheless, as in \refcite{Christ:1966zz}, 
the separation enables us to examine their symmetry properties individually,
which may not hold for their products beyond the LO EW expansion.
In particular, under time reversal $\hat{T}$, the currents in \eq{eq:currents} transform as 
\beq[eq:current-T]
	\hat{T} J_V^{\mu}(r) \hat{T}^{-1} = J_{V, \mu}(-\bar r) \big|_{\Gamma_V^q \to \Gamma_V^{q*} },
	\quad
	(V = \gamma, Z),
\eeq
with the notation $\bar k^{\mu} = k_{\mu}$ for any four-vector $k$. 
Then the partonic tensor satisfies 
\begin{align}
	\hat{W}_{I, \lambda\lambda'}^{\mu\nu}( k, q_{\star}, \Gamma_j )
	&= \int d^4 r \, e^{-i q_{\star} \cdot r} \, 
		\vv{ k, \lambda' | \, \hat{T}^{\dag} \hat{T} \, 
			J_I^{\nu}(0) \, J_I^{\mu}( r ) \,
			\hat{T}^{-1} \hat{T} \,
		|k, \lambda } ^*
	\nn\\
	&= (-1)^{\lambda + \lambda' - 1} \int d^4 r \, e^{-i q_{\star} \cdot r} \, 
		\vv{ \bar k, \lambda | \, 
			J_{I, \mu}( - \bar r ) \, J_{I, \nu}(0) \, 
		|\bar k, \lambda' }
		\big|_{\Gamma_j \to \Gamma_j^* }
	\nn\\
	&= (-1)^{\lambda + \lambda' - 1} \,
		\hat{W}_{I, \nu\mu, \lambda'\lambda} \pp{ \bar k, \bar{q}_{\star}, \Gamma_j^* },
\end{align}
where we have indicated the dependence on dipole couplings
$\Gamma_j \in \{ \Gamma_{\gamma}^q, \Gamma_{Z}^q \}$
and in the second step implicitly used the fact that the channel index $I$ is permutation invariant.
Similarly, for the leptonic tensor, we have
\beq
	L_{I}^{\mu\nu}( \l, q_{\star} )
	= L_{I, \nu\mu}\pp{ \bar \l, \bar{q}_{\star} },
\eeq
which can be trivially checked with \eq{eq:lepton-L-param}.
Their contraction, $[L \hat W]$, then satisfies
\begin{align}
	[L \hat W]^{I}_{\lambda\lambda'}( k, \l, q_{\star}, \Gamma_j )
	= (-1)^{\lambda + \lambda' - 1} \, 
		[L \hat W]^{I}_{\lambda'\lambda}( \bar k, \bar \l, \bar{q}_{\star}, \Gamma_j^{*} ).
\label{eq:LW-T}
\end{align}
Since $k$, $\l$, and $q_{\star}$ all lie in the $x$-$z$ plane, 
we can perform a rotation $R_y(\pi)$ in \eq{eq:LW-T}
to remove the ``bars''.
This rotation introduces the same phase 
$(-1)^{\lambda + \lambda' - 1}$~\cite{Yu:2023shd} 
that cancels the phase in \eq{eq:LW-T}.
\begin{align}
	[L \hat W]^{I}_{\lambda\lambda'}( k, \l, q_{\star}, \Gamma_j )
	= [L \hat W]^{I}_{\lambda'\lambda}( k, \l, q_{\star}, \Gamma_j^{*} ).
\label{eq:LW-TRy}
\end{align}
Clearly, this property does not necessarily apply beyond
fully inclusive DIS with single vector boson exchange.
 
Combining \eq{eq:LW-TRy} with the obvious Hermiticity property,
\begin{align}
	[L \hat W]^{I}_{\lambda\lambda'}( k, \l, q_{\star}, \Gamma_j )
	&= \bigcc{ [L \hat W]^{I}_{\lambda'\lambda}( k, \l, q_{\star}, \Gamma_j ) }^*,
\end{align}
we then establish the important relation,
\beq[eq:LW-real]
	[L \hat W]^{I}_{\lambda\lambda'}( k, \l, q_{\star}, \Gamma_j )
	= \cc{ [L \hat W]^{I}_{\lambda\lambda'}( k, \l, q_{\star}, \Gamma_j^{*} ) }^*.
\eeq
The above reasoning did not use any special property of partons,
so \eq{eq:LW-real} applies to the hadronic tensor $W$ as well as to the partonic $\hat W$.
In the case of the SM, this was used to show~\cite{Christ:1966zz} 
that no imaginary part exists in $[L \hat W]^{\gamma}_{\lambda\lambda'}$,
so that the single transverse spin asymmetry does not exist 
in inclusive DIS in the one-photon exchange approximation.
In our case, however, the existence of dipole couplings 
allows for a non-zero single transverse spin asymmetry. 

In perturbation theory with massless quarks, 
the QCD interactions underlying the partonic tensor $\hat W$ satisfy an additional chiral symmetry.
The only mechanism to flip the quark helicity is by means of the dipole interaction,
which, by \eq{eq:dipoles}, flips right-handed chiral fermions to left-handed via $\Gamma_j$ 
and left to right via $\Gamma_j^{*}$.
Keeping up to linear dependence on $\Gamma_j$ at $\order{1 / \Lambda^2}$,
we see that the helicity-diagonal elements 
$[L \hat W]^{I}_{\pm \pm}( k, \l, q_{\star} )$ 
do not depend on dipole couplings and 
so must be real by \eq{eq:LW-real}.
The off-diagonal elements, on the other hand, carry linear dipole dependence and can be parametrized as
\beq[eq:dip-param]
	[L \hat W]^{I}_{+-}( k, \l, q_{\star}, \Gamma_j ) 
	= \cc{ [L \hat W]^{I}_{-+}( k, \l, q_{\star}, \Gamma_j ) }^*
	= \sum_j t^I_j \, \Gamma_j,
\eeq
where the sum over $j$ depends on the channel $I$,
e.g., the coefficients $t^I_j$ vanish for $\Gamma_Z^{q}$ in the $\gamma$ channel, etc. 
Then, \eq{eq:LW-real} constrains the coefficients $t^I_j$ to be real.
Any possible imaginary part is carried by the dipole couplings,
and no phase can arise from loop corrections to arbitrary order 
in perturbation theory since the argument results from symmetry considerations.

With these results established, let us contract \eq{eq:partonic-tensor-LW} 
with the helicity density matrix in \eq{eq:partonic-tensor-q}, 
\begin{align}
	L_I^{\mu\nu}(\l, q) \, \hat{W}_{I, \mu\nu}(k, q, \Gamma_j)
	&= \sum_{\lambda, \lambda'} 
		\rho_{\lambda \lambda'}\bigpp{ \bm{s}_{T} } \,
		e^{i (\lambda - \lambda') \phi} \,
		[L\hat W]^I_{\lambda\lambda'}( k, \l, q_{\star}, \Gamma_j )
	\nn\\
	&= \frac{1}{2} \, \Bigcc{
			[L\hat W]^I_{++}( k, \l, q_{\star} ) + [L\hat W]^I_{--}( k, \l, q_{\star} )
		}
    \nn\\
    &\quad
		+ s_{T} \, \Re\Bigcc{ e^{i \Delta \phi} \, [L\hat W]^I_{+-}( k, \l, q_{\star}, \Gamma_j ) },
\end{align}
where the last term carries the $\phi$ dependence 
and is reduced to
\begin{align}
	\Re\Bigcc{ e^{i \Delta \phi} \, [L\hat W]^I_{+-}( k, \l, q_{\star}, \Gamma_j ) }
	&= \sum_j t_j^I \, \Re\Bigcc{ \Gamma_j \, e^{i \Delta \phi} }
	\nn\\
    &= \sum_j t_j^I \, \Bigbb{ \Re(\Gamma_j) \cos(\Delta \phi) - \Im(\Gamma_j) \sin(\Delta \phi) }.
\label{eq:gamma-phi}
\end{align}
This result is achieved in the helicity basis,
which enables the use of chiral symmetry
and is complementary to the covariant tensor method in \sec{ssec:tensor}.

Comparing \eq{eq:gamma-phi} with \eqs{eq:L-tau-contract}{eq:dot-cross},
we find that in the tensor decomposition of $\hat{W}_I^{\mu\nu}$,
the $\Delta\tau_1^{\mu\nu}$ and $\Delta\tau_2^{\mu\nu}$ structures 
must only be associated with the real parts of the dipole couplings,
and $\Delta\wt \tau_1^{\mu\nu}$ and $\Delta\wt \tau_2^{\mu\nu}$ only with the imaginary parts,
and their coefficients must be correspondingly equal. 
Therefore, for each channel and parton flavor $a$,  
we can decompose the partonic tensor as 
\begin{align}
	\hat W_{I, a}^{\mu\nu}
	&= U^{I, a}_T \, \tau_T^{\mu\nu} + U^{I, a}_L \, \tau_L^{\mu\nu} + U^{I, a}_3 \, \tau_3^{\mu\nu}
	\nn\\
	&\hspace{1.2em}
	  + \frac{Q}{v} \sum_j \bb{
	  	\Re(\Gamma_j) 
	  		\pp{ T^{I, a}_{1, j} \, \Delta \tau_1^{\mu\nu} + T^{I, a}_{2, j} \, \Delta \tau_2^{\mu\nu} } 
		+ \Im(\Gamma_j) 
	  		\pp{ \wt{T}^{I, a}_{1, j} \, \Delta \wt{\tau}_1^{\mu\nu} + \wt{T}^{I, a}_{2, j} \, \Delta \wt{\tau}_2^{\mu\nu} }
	},
\label{eq:w-decomp}
\end{align}
where the scalar coefficients are functions of $(\x, Q/\mu, \alpha_s(\mu))$,
and where we must have
\beq[eq:consistency]
	T^{I, a}_{1, j} = \wt T^{I, a}_{1, j},
	\quad
	T^{I, a}_{2, j} = \wt T^{I, a}_{2, j}.
\eeq
This property holds starting from the lowest order and 
can be preserved by the collinear subtracted cross section order by order in $\alpha_s$
as long as the different structures are factorized into the same transversity PDF in \eq{eq:sT-h/f},
so \eqs{eq:w-decomp}{eq:consistency} apply to $\hat W_{I, a}^{\mu\nu}$ 
whether collinear subtraction is included or not. 
We will see this explicitly in \sec{sssec:subtraction}.

Contracting \eq{eq:w-decomp} with the leptonic tensor 
and using \eqs{eq:L-tau-contract}{eq:consistency}, 
we get the master formula,
\begin{align}
	\frac{y^2}{Q^2} \, L_I \cdot \hat W_{I, a}
	&= L^I_1 \, \Bigbb{ U^{I, a}_T \, \bigpp{ 1 + (1 - y)^2 }
			+ U^{I, a}_L \, \bigpp{ 2(1 - y) }
		}
		+ L^I_2 \, U^{I, a}_3 \, \bigpp{ y (2 - y) }
	\nn\\
	&\hspace{1.2em}
		+ \frac{2 s_{T, a} \l_T}{v} \sum_j \Bigbb{
			\Re(\Gamma_j) \cos(\Delta \phi) 
			- \Im(\Gamma_j) \sin(\Delta \phi)
		} \Bigbb{
			L^I_2 \, T^{I, a}_{1, j} \cdot y
			- L^I_1 \, T^{I, a}_{2, j} \, \bigpp{ 2 - y }
		}.
\label{eq:LW-generic}
\end{align}
Note that all the partonic quantities here, 
including the coefficients $U$ and $T$ and the spin $s_{T, a}$,
depend on $\x$ as defined in \eq{eq:xhat-def}.
They will need to be convoluted with the PDF $f_a(x, \mu)$ in \eq{eq:W-fact}.

Since the above reasoning explicitly used chiral symmetry, 
which may not be preserved by a particular $\gamma_5$ scheme in DR,
\eqs{eq:w-decomp}{eq:consistency} serve as a consistency condition 
for checking the practical calculation as well as to construct counterterms 
to restore chiral symmetry when it is broken, which we will use below.

%----------------------------------------------------------------
\subsection{Implications from parity}
%----------------------------------------------------------------

Having seen that time reversal $\hat T$ constrains the 
relative phases between $(T^{I, a}_{1, j}, T^{I, a}_{2, j})$ and $(\wt T^{I, a}_{1, j}, \wt T^{I, a}_{2, j})$,
we now examine the consequence of parity symmetry $\hat P$,
which sets further constraints on these coefficients channel by channel.

It is convenient to directly look at the effect of the combined operator $\hat P \hat T$,
which acts on a spin-dependent matrix element as in \eq{eq:partonic-tensor-q} to give
\begin{align}
	\sum_{\lambda, \lambda'} 
		\rho_{\lambda \lambda'}( \bm{s} )
		\vv{ k, \lambda' | \,
			O \,
		| k, \lambda }
	&= \sum_{\lambda, \lambda'} 
		\rho_{\lambda \lambda'}( \bm{s} ) \,
		\bigvv{ k, \lambda' | \, (\hat P \hat T)^{\dag}
			\bigbb{ \hat P \hat T
			\, O \,
			(\hat P \hat T)^{-1}
			}
			(\hat P \hat T) \,
		| k, \lambda }^*
	\nn\\
	&= \sum_{\lambda, \lambda'} (-1)^{-\lambda -\lambda' - 1} \,
		\rho_{-\lambda', -\lambda}( \bm{s} ) \,
		\bigvv{ k, \lambda' | \, 
			\bigbb{ \hat P \hat T
			\, O \,
			(\hat P \hat T)^{-1}
			}^{\dag} \,
		| k, \lambda }
	\nn\\
	&= \sum_{\lambda, \lambda'}
		\rho_{\lambda \lambda'}( -\bm{s} ) \,
		\bigvv{ k, \lambda' | \, 
			\bigbb{ \hat P \hat T
			\, O \,
			(\hat P \hat T)^{-1}
			}^{\dag} \,
		| k, \lambda },
\end{align}
which preserves the momenta but flips the spin.
The currents in \eq{eq:currents} transform under ${\hat P} {\hat T}$ as
\beq[eq:current-PT]
	\hat P \hat T
	\, J_V^{\mu}(r) \,
	(\hat P \hat T)^{-1}
	= J_V^{\mu}(-r) \big|_{g_A^f \to -g_A^f},
	\quad
	(V = \gamma, Z),
\eeq
where, unlike \eq{eq:current-T}, the dipole couplings are ${\hat P}{\hat T}$-even 
but the axial fermion-$Z$ couplings $g_A^f$ are ${\hat P} {\hat T}$-odd.
Then, in the covariant notation using \eqs{eq:q-spin-avg}{eq:smu}, 
the partonic tensor in \eq{eq:partonic-tensor-q} satisfies
\beq[eq:W-PT]
	\hat{W}_{I, a}^{\mu\nu}\bigpp{ k, q, s_a, g_A^q }
	= \hat{W}_{I, a}^{\nu\mu}\bigpp{ k, q, -s_a, -g_A^q },
\eeq
where only relevant arguments are retained and the dependence on $g_A^q$ is indicated. 

We now apply \eq{eq:W-PT} to the tensor expansion in \eq{eq:w-decomp}.
For the three spin-independent terms, the symmetric tensor structures $\tau_T$ and $\tau_L$
must have coefficients $U_T$ and $U_L$ that are even in $g_A^q$,
while the coefficient of the antisymmetric $\tau_3$ must be odd.
This reproduces the well-known result that 
the $\gamma$ channel only generates the DIS structure functions $U_T$ and $U_L$.%%%
\footnote{These are conventionally denoted as $F_T$, $F_L$, and $F_3$. 
But we use $U$ due to a non-standard $1/(2\pi)$ normalization and 
to avoid confusion with the Fisher information in \sec{ssec:bound}.}
%%%
The $U_3$ structure function is present for both the $Z$ and interference channels 
and is linearly dependent on $g_A^q$.
This is also true for the gluon tensor, \eq{eq:partonic-tensor-g}.
Since the $g_A^q$ coupling has opposite charge conjugation parity 
to the vector $g_V^q$ component with which it interferes in the $U_3$ structure,
the $U_3$ contribution vanishes due to a cancellation between quark and antiquark contributions.

The four spin-dependent terms are generated only by the $(u, d, \bar{u}, \bar{d})$ initial states. 
Since all the tensor structures in \eq{eq:gi-struct-T} are linear in $s_a^{\mu}$,
\eq{eq:W-PT} requires that the coefficients of the anti-symmetric structures, 
$\Delta\tau_1^{\mu\nu}$ and $\Delta\wt\tau_1^{\mu\nu}$,
are even in $g_A^q$,
and that those of the symmetric ones, 
$\Delta\tau_2^{\mu\nu}$ and $\Delta\wt\tau_2^{\mu\nu}$,
are odd in $g_A^q$.
Hence, the $\gamma$ channel only allows the $T_1$ and $\wt T_1$ structures,
whereas the $Z$ and interference channels also produce $T_2$ and $\wt T_2$ structures
that are linear in $g_A^q$.

Again, the constraints following \eq{eq:W-PT} can be preserved order by order,
and apply to both the collinear-subtracted and to the unsubtracted versions of the partonic tensors.

%================================================================================
\section{NLO calculation}
\label{sec:nlores}
%================================================================================

Now, we switch to the details of the one-loop calculation.
Most of the steps are standard and will not be elaborated on.
We  use DR to regulate  ultraviolet divergences and infrared divergences 
in both the virtual loops and the phase space of the real emissions.
The new ingredients are the transverse spin projector in \eq{eq:q-spin-avg}
and the issue of $\gamma_5$ in both the projector and the dipole vertices.
We specifically illustrate the issues with the $\gamma$ channel,
and collect results for the $Z$ and interference channels in Appendix~\ref{app:sec:summary-xsec}.

%----------------------------------------------------------------
\subsection{Tree-level calculation for photon channel}
%----------------------------------------------------------------

%----------------------------------------------------------------
% Fig: tree level
%------------------------------------------------
\begin{figure}[htbp]
	\centering
	\begin{tabular}{cc}
		\includegraphics[scale=1]{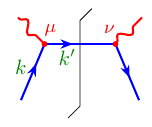} &
		\includegraphics[scale=1]{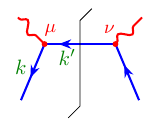} \\
		(a) & (b)
	\end{tabular}
	\caption{Tree-level diagram for the partonic tensor $\hat W_{\gamma, a}^{\mu\nu}$ of the $\gamma$ channel
		for $a=$ quark (a) and antiquark (b).}
\label{fig:tree}
\end{figure}
%----------------------------------------------------------------

We start with the tree-level diagrams shown in \fig{fig:tree}.
The Feynman rule for the quark channel reads 
\begin{align}
	\hat W_{\gamma, q}^{(0)\mu\nu}
	&= \frac{1}{2\pi} \, \Tr\bb{ 
			\frac{\slash{k}}{2} (1 - \slash{s}_q \gamma_5) \,
			\bar{\Gamma}^{q \nu} \,
			  (\slash{k} + \slash{q})  \,
			\Gamma^{q \mu}
		} \cdot
		(2\pi) \delta^+\bigpp{ (k + q)^2 },
\label{eq:AA-Wq0-tensor}
\end{align}
where the superscript ``$(0)$'' refers to LO,
and the photon vertices are given by \eq{eq:feyn-rules}
\begin{align}
	\Gamma^{q \mu}
	&= e_{q} \gamma^{\mu}
		+ \frac{i \sigma^{\mu \alpha} q_{\alpha}}{v}
		\bb{ \Re(\Gamma^{q}_{\gamma}) + i \Im(\Gamma^{q}_{\gamma}) \gamma_5 },
	\nn\\
	\bar \Gamma^{q \mu} 
	&\equiv \gamma^0 (\Gamma^{q \mu})^{\dag} \gamma^0
	= e_{q} \gamma^{\mu}
		- \frac{i \sigma^{\mu \alpha} q_{\alpha}}{v} 
		\bb{ \Re(\Gamma^{q}_{\gamma}) + i \Im(\Gamma^{q}_{\gamma}) \gamma_5 }.
\end{align}
The $\delta$-function constrains the cut quark line to be on shell, 
giving a $\delta(1 - \x)$ factor that sets $\x$ to 1.
The result of explicit calculation agrees with the decomposition of \eq{eq:w-decomp} and gives
\begin{align}
	\hat W_{\gamma, q}^{(0)\mu\nu}
	&= U^{(0) \gamma, q}_{T} \, \tau_T^{\mu\nu}
	+ \frac{Q}{v} \bb{
	  	\Re(\Gamma_{\gamma}^q) 
	  		\, T^{(0) \gamma, q}_{1} \, \Delta \tau_1^{\mu\nu}(s_q)
		+ \Im(\Gamma_{\gamma}^q) 
	  		\, \wt{T}^{(0) \gamma, q}_{1} \, \Delta \wt{\tau}_1^{\mu\nu}(s_q)
	},
\label{eq:AA-Wq0-decomp}
\end{align}
where the polarized tensor structures $\Delta \tau_1$ and $\Delta \wt \tau_1$ are constructed with 
the quark spin vector $s_q^{\mu}$, as explicitly indicated.
\eq{eq:AA-Wq0-decomp} has only three nonzero gauge-invariant structures,
\begin{align}
	U^{(0) \gamma, q}_{T} 
	&= e_q^2 \, \delta(1 - \x),
	\nn\\
	T^{(0) \gamma, q}_{1} = \wt T^{(0) \gamma, q}_{1}
	&= e_q \, \delta(1 - \x),
\label{eq:AA-Wq0-coefs}
\end{align}
which is consistent with \eq{eq:consistency}.
$U^{\gamma, q}_{L}$ is generated at one loop,
but $U^{\gamma, q}_{3}$, $T^{\gamma, q}_{2}$, and $\wt T^{\gamma, q}_{2}$ vanish to all orders,
as expected from \eq{eq:W-PT}.
Since every element in \eq{eq:AA-Wq0-tensor} satisfies charge conjugation symmetry,
the result for antiquark is the same,
except with $s_q \to s_{\bar q}$.

The contraction with the leptonic tensor in \eq{eq:AA-lepton-trace-tree}, 
with the use of \eq{eq:LW-generic}, gives
\begin{align}
	\frac{y^2}{Q^2} \, L_{\gamma} \cdot \hat W_{\gamma, q}^{(0)}
	&= \delta(1 - \x) 
		\cc{ e_q^2 \pp{ 1 + (1 - y)^2 }
			+ \lambda_e \, s_{T, q}(x) \, e_q \, \frac{2 y \l_T}{v} 
			\Re\bigpp{ \Gamma_{\gamma}^q \, e^{i \Delta \phi} }
		}.
\end{align}
Applying this to \eq{eq:xsec-LW}, 
using the factorization for the hadronic tensor in \eq{eq:W-fact}
and \eq{eq:sT-h/f} for $s_{T, q}(x)$,
and including antiquark contributions,
gives the LO cross section for the $\gamma$ channel,
\begin{align}
	\frac{d\sigma_{\gamma}^{(0)}}{dx_B \, dQ^2 \, d\phi}
	= \frac{N_{\gamma}}{2\pi} \, \Sigma^{\gamma (0)}_{UU}(x_B, Q^2) \, \Bigbb{
		1 + \lambda_e S_T \, A^{\gamma (0)}_{LT}(x_B, Q^2, \Delta \phi)
	},
\label{eq:xsec-AA-0}
\end{align}
which contains an unpolarized term that includes all active quark flavors, 
\begin{align}
	\Sigma^{\gamma (0)}_{UU}(x_B, Q^2)
	= \pp{ 1 + (1 - y)^2 } \cdot \sum_q e_q^2 \, 
		\bigbb{ f_q(x_B, \mu) + f_{\bar q}(x_B, \mu) },
\label{eq:UU-AA-0}
\end{align}
and a double spin asymmetry (DSA) $A^{\gamma (0)}_{LT}$ which only depends on $u$ and $d$ quarks
via their photon dipole couplings,
\begin{align}
	\bb{ \Sigma^{\gamma (0)}_{UU} \cdot A^{\gamma (0)}_{LT} }(x_B, Q^2, \Delta \phi)
	= \frac{2 y \l_T }{v} 
		\sum_{q = u, d} e_q \, \Re\bigpp{ \Gamma_{\gamma}^q \, e^{i \Delta \phi} } \,
		\bigbb{ h_q(x_B, \mu) + h_{\bar q}(x_B, \mu) }.
\label{eq:LT-AA-0}
\end{align}
As argued below \eq{eq:W-PT}, parity constrains 
the nonzero polarized tensor structures in \eq{eq:AA-Wq0-decomp} to be antisymmetric, 
so they require the antisymmetric leptonic structure $L_2^{\gamma}$ 
induced by the longitudinal electron beam polarization $\lambda_e$ to survive in the cross section.
Since the $\gamma$ channel dominates over the interference and $Z$ channels at the EIC energy,
the DSA will greatly enhance the sensitivity to the photon dipole couplings.

%----------------------------------------------------------------
\subsection{One-loop correction to photon channel: quark channel}
%----------------------------------------------------------------

%--------------------------------------------
\subsubsection{Virtual gluon correction}
\label{sssec:virt}
%--------------------------------------------

%----------------------------------------------------------------
% Fig: 1 loop virtual
%------------------------------------------------
\begin{figure}[htbp]
	\centering
	\begin{tabular}{ccc}
		\includegraphics[scale=0.8]{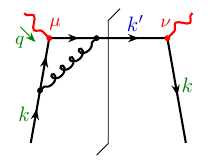} &
		\includegraphics[scale=0.8]{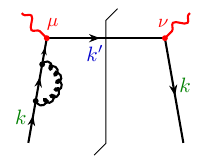} & 
		\includegraphics[scale=0.8]{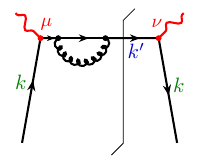}\\
		(a) & (b) & (c)
	\end{tabular}
	\caption{Virtual gluon corrections to the partonic tensor $\hat W_{\gamma, a}^{\mu\nu}$ of the $\gamma$ channel,
		each implicitly accompanied by its UV counterterm diagram and mirror diagram with respect to the cut line.}
\label{fig:virt}
\end{figure}
%----------------------------------------------------------------

The one-loop diagrams for the virtual gluon corrections are shown in \fig{fig:virt}, 
where (a) is the photon vertex that involves both the SM and dipole couplings,
and (b)(c) are external leg gluon contributions that are independent of the vertices.
The loop corrections to the SM photon vertex and to the dipole vertex are different,
but due to chiral symmetry, 
the QCD correction to each corresponds to a multiplicative factor to the tree-level vertex.
This property holds whether the latter involves a $\gamma_5$ or not.

Specifically for \fig{fig:virt}(a), let us denote the irreducible SM quark-photon vertex by
\beq[eq:loop-v-v]
	\bar{u}(k') \, \Gamma_{{v}}^{\mu}(k, k') \, u(k)
	= \bar{u}(k') \bb{ 1 + \delta_1 F_v(q^2) + \order{\alpha_s^2} } \gamma^{\mu} \, u(k).
\eeq
On the right-hand side, the term ``$1$'' corresponds to the tree-level vertex.
The $\delta_1 F_v(q^2)$ term refers to the renormalized one-loop gluon correction,
given by~\cite{Altarelli:1979ub}
\begin{align}
	\delta_1 F_v(q^2)
	&= - \frac{\alpha_s C_F}{2\pi} \, S_{\epsilon} \bb{
		\pp{ \frac{\mu^2 }{Q^2} }^{\epsilon} \, 
		\frac{\Gamma(\epsilon) \, \Gamma^3(1 - \epsilon)}{\epsilon \, \Gamma(2 - 2\epsilon) }
			\pp{ 1 + \epsilon^2 - \frac{\epsilon}{2} }
		+ \frac{1}{2\epsilon}
	} \nn\\
	&= - \frac{\alpha_s C_F}{2\pi} \, S_{\epsilon} \bb{
		\pp{ \frac{\mu^2 }{Q^2} }^{\epsilon} \, 
		\pp{ \frac{1}{\epsilon^2} + \frac{3}{2\epsilon} + 4 + \O(\epsilon) }
		+ \frac{1}{2\epsilon}
	} ,
\label{eq:virt-fV}
\end{align}
where we use DR to work in $d = 4 - 2\epsilon$ dimensions 
and the first line is exact to all orders in $\epsilon$.
The $1 / (2\epsilon)$ term is the UV counterterm,
$\mu$ is the renormalization scale,
and $S_{\epsilon}$ implements the $\MS$ subtraction scheme~\cite{Collins:2011zzd},
\beq
	S_{\epsilon} \equiv \frac{ \pp{ 4\pi }^{\epsilon} }{ \Gamma(1 - \epsilon) }
	= 1 + \epsilon \, \ln(4\pi \, e^{-\gamma_E}) + \order{\epsilon^2},
\eeq
with $\gamma_E$ the Euler constant.
%%%
For the dipole contribution, on the other hand, we denote it by
\beq[eq:loop-v-t]
	\bar{u}(k') \, \Gamma_t^{\mu}(k, k') \, u(k)
	= \bar{u}(k') \bb{ 1 + \delta_1 F_t(q^2) + \order{\alpha_s^2} } \sigma^{\mu q} \, u(k),
\eeq
where $\sigma^{\mu q}\equiv\sigma^{\mu\nu}q_\nu$ and the one-loop correction $\delta_1 F_t(q^2)$ is
\begin{align}
	\delta_1 F_t(q^2)
	&= - \frac{\alpha_s C_F}{2\pi} \, S_{\epsilon} \,
		\pp{ \frac{\mu^2 }{Q^2} }^{\epsilon} \, 
		\frac{\Gamma(\epsilon) \, \Gamma^3(1 - \epsilon)}{\epsilon \, \Gamma(2 - 2\epsilon) }
	\nn\\
	&= - \frac{\alpha_s C_F}{2\pi} \, S_{\epsilon} \,
		\pp{ \frac{\mu^2 }{Q^2} }^{\epsilon} \, 
		\pp{ \frac{1}{\epsilon^2} + \frac{2}{\epsilon} + 4 + \O(\epsilon) },
\label{eq:virt-fT}
\end{align}
which has no UV divergence so there is no counterterm. 
Replacing the $\gamma^{\mu}$ in \eq{eq:loop-v-v} by $\gamma^{\mu} \gamma_5$
or the $\sigma^{\mu q}$ in \eq{eq:loop-v-t} by $\sigma^{\mu q} \gamma_5$
gives the same correction factor $\delta_1 F_v(q^2)$ or $\delta_1 F_t(q^2)$.

The external leg corrections are included via the LSZ reduction formula,
obtained by vanishing scaleless integrals plus UV counterterms.
The sum of \fig{fig:virt}(b) and (c) amounts to multiplying the tree-level diagram by
\beq[eq:virt-leg]
	\delta_1 Z_2
	= \frac{\alpha_s C_F}{2\pi} \frac{S_{\epsilon}}{2\epsilon}.
\eeq
When applied to the SM vertex, \eq{eq:virt-leg} combines with \eq{eq:virt-fV} to cancel the UV counterterm.
Since the dipole vertex correction does not have a UV divergence, 
the combination of \eqs{eq:virt-leg}{eq:virt-fT} gives rise to an additional logarithmic term, 
\begin{align}
	\bb{ \delta_1 F_t(q^2) + \delta_1 Z_2 }
	- \bb{ \delta_1 F_v(q^2) + \delta_1 Z_2 }
	&= - \frac{\alpha_s C_F}{2\pi} \, S_{\epsilon} \bb{
		\pp{ \frac{\mu^2 }{Q^2} }^{\epsilon} \, 
		\pp{ \frac{1}{2\epsilon} + \O(\epsilon) }
		- \frac{1}{2\epsilon}
	}
	\nn\\
	&= - \frac{\alpha_s C_F}{2\pi} \ln\pp{ \frac{\mu}{Q} }
	+ \O(\epsilon).
\label{eq:fT-log}
\end{align}
This is a signature of the renormalization of the dipole coupling.
While the dipole vertex counterterm vanishes as a whole, 
it means that the coupling counterterm cancels that of the wavefunction renormalization factor,
unlike the vector vertex, for which the coupling is unrenormalized at $\order{\alpha_s}$.
The resultant log compensates the $\mu$ dependence of the dipole coupling as in \eq{eq:RGE-C}.

Combining with the conjugate diagrams, 
the effects of virtual gluon corrections are given, 
in terms of structure functions defined in \eq{eq:w-decomp}, 
by 
\beq[eq:AA-UT-1V]
	U^{(1V) \gamma, q}_{T}
	= 2\bb{ \delta_1 F_v(q^2) + \delta_1 Z_2 } \cdot U^{(0) \gamma, q}_{T}
\eeq
for the unpolarized case, and 
\beq[eq:AA-T1-1V]
	T^{(1V) \gamma, q}_{1} 
	= \wt T^{(1V) \gamma, q}_{1}
	= \Bigcc{ \bb{ \delta_1 F_t(q^2) + \delta_1 Z_2 }
		+ \bb{ \delta_1 F_v(q^2) + \delta_1 Z_2 } 
	}
	\cdot T^{(0) \gamma, q}_{1}
\eeq
for the polarized contributions, 
where the superscripts ${(1V)}$ denote the one-loop virtual corrections.
The partonic tensor has the same decomposition as \eq{eq:AA-Wq0-decomp},
without new structure functions.
All the $\epsilon$ poles that remain are now of infrared nature,
since the counterterms have removed all UV poles.

%--------------------------------------------
\subsubsection{Real gluon corrections}
\label{sssec:real}
%--------------------------------------------

%----------------------------------------------------------------
% Fig: 1-loop real
%------------------------------------------------
\begin{figure}[htbp]
	\centering
	\begin{tabular}{cccc}
		\includegraphics[scale=0.8]{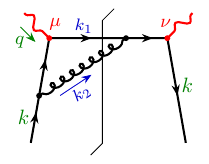} &
		\includegraphics[scale=0.8]{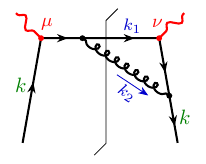} & 
		\includegraphics[scale=0.8]{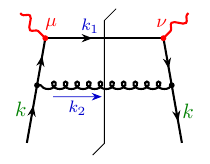} & 
		\includegraphics[scale=0.8]{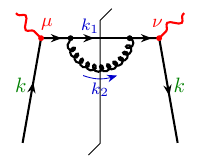} \\
		(a) & (b) & (c) & (d)
	\end{tabular}
	\caption{Real gluon corrections to the partonic tensor 
    $\hat W_{\gamma, a}^{\mu\nu}$ of the $\gamma$ channel for $a = q$.}
\label{fig:real}
\end{figure}
%----------------------------------------------------------------

We have evaded the $\gamma_5$ issue in the virtual corrections 
because we could focus ``locally'' on the vertex loop
which is sandwiched between on-shell spinors as in \eqs{eq:loop-v-v}{eq:loop-v-t}.
The loop integration reduces to the same spinor structure as the tree level,
and so the resultant traces can be evaluated in a trivial way.
A naive anticommuting scheme is sufficient.

However, for the real corrections, 
which correspond to the $2\to2$ scattering,
\beq[eq:real-2to2]
	q(k) + \gamma^*(q) \to q(k_1) + g(k_2),
\eeq
as shown in \fig{fig:real},
the phase space of $(k_1, k_2)$ lives in $d$ dimensions,
which regulates soft and collinear divergences,
so we need DR to act on the whole fermion trace,
with $\slash{k}_1$ inserted for the final state quark and \eq{eq:q-spin-avg} for the initial state quark.
This ``cut fermion loop'' of the cut diagram is then sensitive to the treatment of $\gamma_5$.
We use the Kreimer scheme~\cite{Kreimer:1989ke, Korner:1991sx, Kreimer:1993bh},
where $\gamma_5$ and $\epsilon_{\mu\nu\rho\sigma}$ are defined as quantities in four dimensions, 
with anticommutation relations and Hermiticity preserved, 
\beq
	\cc{\gamma_{\mu}, \gamma_5} = 0, \quad
	\gamma_5^{\dag} = \gamma_5, \quad
	\gamma_5 = \frac{i}{4!} \epsilon_{\mu\nu\rho\sigma} \gamma^{\mu} \gamma^{\nu} \gamma^{\rho} \gamma^{\sigma}, \quad
	\epsilon_{0123} = -\epsilon^{0123} = 1,
\eeq
and $\epsilon_{\mu\nu\rho\sigma}$ is zero if any of its indices take values beyond $\{0, 1, 2, 3 \}$.
The cyclicity property for $\gamma$-traces is not satisfied,
and a consistent ``reading point'' needs to be chosen for all the fermion loops.
%%%
We start every fermion trace of \fig{fig:real} from the spinor projector, 
as done for the tree-level trace in \eq{eq:AA-Wq0-tensor}.
For example, the fermion trace in \fig{fig:real}(a) reads
$$
	\Tr\biggbb{ 
		\frac{\slash{k}(1 - \slash{s}_q \gamma_5)}{2} \,
		\bar \Gamma^{q \nu} \,
		( \slash{k} + \slash{q} ) \, 
		\gamma_{\alpha} \, \slash{k}_1 \,
		\Gamma^{q \mu} \, ( \slash{k} - \slash{k}_2 ) \,
		\gamma^{\alpha} 
	}\, .
$$
Rules for other Feynman diagrams are the same as usual.

The phase space integral in $d $ dimensions is given by
\beq[eq:phase-space-def]
	\int d\Pi_2
	= \int \frac{d^d k_1}{(2\pi)^d} \frac{d^d k_2}{(2\pi)^d} \, 
		(2\pi) \delta^+(k_1^2) \, 
		(2\pi) \delta^+(k_2^2) \,
		(2\pi)^d \delta^{(d)}(k + q - k_1 - k_2).
\eeq
The $\delta$-functions reduce the $2d$ phase space  to $(d-2)$ kinematic degrees of freedom.
Among them two are in the physical four dimensions, 
which are conveniently chosen as the polar angle $\theta_1$ and the azimuthal angle $\phi_1$ of $k_1$
in the center-of-mass (c.m.) frame of $k$ and $q$,
where $\theta_1$ can be written in an invariant form as
\beq[eq:def-z-kin]
	\z = \frac{k \cdot k_1}{k \cdot q}
	= \frac{1 - \cos\theta_1}{2}.
\eeq
This also defines the three partonic Mandelstam variables,
\begin{align}
	\hat s &= (k + q)^2 = \frac{(1 - \x) \, Q^2}{\x},
	\nn\\
	\hat t &= (k - k_1)^2 = - \frac{\z \, Q^2}{\x},
	\nn\\
	\hat u &= (k - k_2)^2 = - \frac{(1-\z) \, Q^2}{\x},
\label{eq:stu-def}
\end{align}
where $\x$ is the same as in \eq{eq:xhat-def}.
%%%
The remaining $(d - 4) = -2\epsilon$ degrees of freedom are unphysical.
They circulate completely within the (cut) loop where $k_1$ and $k_2$ flow
and are orthogonal to all the external momenta.
Nevertheless, they do constitute part of the phase space, 
which is constrained by the magnitude of the transverse momentum,
$$
	k_{1T} = \frac{ \sqrt{ \hat s} }{ 2 } \sin\theta_1 
	= \sqrt{ \z (1 - \z) \hat s }
	= Q \sqrt{ \frac{\z}{\x} (1 - \x) (1 - \z) }
	\,,
$$
and will give a suppression factor when $k_{1T}$ vanishes (with $\epsilon < 0$),
corresponding to soft or collinear configurations.
Hence, integrating them out reduces \eq{eq:phase-space-def} to 
\begin{align}
	\int d\Pi_2
	&= \frac{ S_{\epsilon} }{16\pi^2 } \pp{ \frac{1}{Q^2} }^{\epsilon} 
		\int_0^1 d\z \int_0^{2\pi} d\phi_1 \,
		\frac{\x^\epsilon \z^{-\epsilon}}{(1 - \x)^\epsilon (1 - \z)^\epsilon}
	= \frac{ S_{\epsilon} }{8\pi } \pp{ \frac{1}{Q^2} }^{\epsilon} 
		\frac{\x^\epsilon }{(1 - \x)^\epsilon}
		\int_0^1 d\z \,
		\frac{\z^{-\epsilon}}{(1 - \z)^\epsilon},
\label{eq:phase-space}
\end{align}
where in the second step we also integrated out $\phi_1$
since no external momenta will be sensitive to it.
%%%
The real gluon is collinear to the final-state quark at $\x = 1$ 
or to the initial-state quark at $\z = 1$,
and becomes soft when these two overlap at $\x = \z = 1$.
These singularities are regulated with $\epsilon < 0$. 

Our calculational procedure is as follows:
\begin{itemize}
% step 1
\item For each trace, anti-commute all $\gamma_5$, which may occur at the spin projector or the $\gamma$ vertices,
	to the rightmost position.
	At the same time, simplify the $\gamma$-matrix products by applying identities
	such as 
	$\slash{a}^2 = a^2$,
	and
	$\gamma_5^2 = 1$ and contracting all indices in terms such as $\gamma^{\alpha} \cdots \gamma_{\alpha}$, 
    but never use the cyclicity property.
	
	This step reduces all traces involving $\gamma_5$ to contain only 4 or 6 gamma matrices in our calculation.

% step 2
\item Replace $k_2$ by $k + q - k_1$ in the integrands,
	and express the latter as explicit tensor functions of $k_1$, 
	\beq
		D(\hat s, \hat t, \hat u) \pp{ A_0^{\mu\nu} 
			+ A_1^{\mu\nu,\alpha}  k_{1 \alpha} 
			+ A_2^{\mu\nu,\alpha\beta} k_{1 \alpha} k_{1 \beta} 
			+ A_3^{\mu\nu,\alpha\beta\gamma} k_{1 \alpha} k_{1 \beta} k_{1 \gamma} 
			+ \cdots
		},
	\eeq
	where $D$ is a scalar function of the Mandelstam variables in \eq{eq:stu-def},
	and $A_i$ are tensors depending on $(k, q, s_a, g_{\rho\sigma})$ and unevaluated traces. 

% step 3
\item By Lorentz invariance of the phase space, we replace the $k_{1 \alpha}$ tensors at the integrand level as
    \begin{align}
        k_{1 \alpha} 
        & \to a_1 \, k_{\alpha} + b_1 \, q_{\alpha}, \nn\\
        k_{1 \alpha} k_{1 \beta}
        & \to a_2 \, g_{\alpha\beta} + b_2 \, k_{\alpha} k_{\beta} + c_2 \, q_{\alpha} q_{\beta} 
            + d_2 \, (k_{\alpha} q_{\beta} + k_{\beta} q_{\alpha} ), \nn\\
        k_{1 \alpha} k_{1 \beta} k_{1 \gamma}
        & \to a_3 \, k_{\alpha} k_{\beta} k_{\gamma} 
            + b_3 \, (k_{\alpha} k_{\beta} q_{\gamma} + k_{\alpha} q_{\beta} k_{\gamma} + q_{\alpha} k_{\beta} k_{\gamma})
            + c_3 \, (k_{\alpha} q_{\beta} q_{\gamma} + q_{\alpha} k_{\beta} q_{\gamma} + q_{\alpha} q_{\beta} k_{\gamma})
            \nn\\
        &\quad
            + d_3 \, q_{\alpha} q_{\beta} q_{\gamma} 
            + e_3 \, (g_{\alpha\beta} k_{\gamma} + g_{\alpha\gamma} k_{\beta} + g_{\beta\gamma} k_{\alpha})
            + f_3 \, (g_{\alpha\beta} q_{\gamma} + g_{\alpha\gamma} q_{\beta} + g_{\beta\gamma} q_{\alpha}),
    \label{eq:1R-k1-replace}
    \end{align}
    using the fact that $k$ and $q$ are the only available external vectors from the phase space integral,
    and where $a_i, b_i, c_i, d_i, e_i, f_i$ are scalar functions that are solved in $d$ dimensions
    by mandating equivalence of both sides under contraction with the tensor bases on the right-hand sides.
    Since the diagrams in \fig{fig:real} involve up to only three fermion propagators, 
    \eq{eq:1R-k1-replace} is sufficient for our purpose.

    Incidentally, the tensor reduction implicitly removes the $\phi_1$ dependence from the integrand.

% step 4
\item After the last step, all the unevaluated traces  contain only four gamma matrices, 
	which can be evaluated explicitly using 
	\beq
		\Tr\pp{\gamma^{\mu} \gamma^{\nu} \gamma^{\rho} \gamma^{\sigma} \gamma_5}
		= 4i \epsilon^{\mu\nu\rho\sigma} 
		\,.
	\eeq
	
% step 5
\item Then the integrands will only involve scalar functions of the Mandelstam variables,
	which can be directly integrated in \eq{eq:phase-space} with $\epsilon < 0$.
	All tensors are outside the integrals and are
	made of $(k, q, s_a, g_{\mu\nu}, \epsilon_{\mu\nu\rho\sigma})$. 
\end{itemize}

The $\z$ integral of the sum of the four diagrams in \fig{fig:real}
satisfies gauge invariance, and gives the decomposition of \eq{eq:w-decomp},
\begin{align}
	\hat W_{\gamma, q}^{(1R)\mu\nu}
	&= U^{(1R) \gamma, q}_{T} \, \tau_T^{\mu\nu}
		+ U^{(1R) \gamma, q}_{L} \, \tau_L^{\mu\nu}
	+ \frac{Q}{v} \bb{
	  	\Re(\Gamma_{\gamma}^q) 
	  		\, T^{(1R) \gamma, q}_{1} \, \Delta \tau_1^{\mu\nu}(s_q)
		+ \Im(\Gamma_{\gamma}^q) 
	  		\, \wt{T}^{(1R) \gamma, q}_{1} \, \Delta \wt{\tau}_1^{\mu\nu}(s_q)
	}.
\label{eq:AA-Wq1R-decomp}
\end{align}
The $U^{(1R) \gamma, q}_{T}$, $T^{(1R) \gamma, q}_{1}$, and $\wt{T}^{(1R) \gamma, q}_{1}$
are corrections to the LO structure functions in \eq{eq:AA-Wq0-coefs}
given by
\begin{align}
	U^{(1R) \gamma, q}_{T}
	&= e_q^2 \, \frac{\alpha_s C_F}{2\pi} \, S_{\epsilon} \pp{ \frac{\mu^2 }{Q^2} }^{\epsilon} \frac{\x^\epsilon }{(1 - \x)^{1+\epsilon}}
		\frac{\Gamma(1 - \epsilon) \Gamma(-\epsilon)}{\Gamma(2 - 2\epsilon)}
    \nn\\
    &\hspace{10em} \times
		\biggcc{ (1 - \epsilon) \pp{ 2 - \frac{\epsilon}{2} }
			- (1 - \x) \Bigbb{ 
				1 + \epsilon
				+ \bigpp{ \epsilon^2 + (1 - \epsilon)^2 } \x
			}
		}
	\nn\\
	&= e_q^2 \, \frac{\alpha_s C_F}{2\pi} \, S_{\epsilon} \pp{ \frac{\mu^2 }{Q^2} }^{\epsilon} 
		\bigg\{
			\frac{2 \, \delta(1 - \x)}{\epsilon^2} + \frac{1}{\epsilon} \bb{ \frac{3}{2} \, \delta(1 - \x) + 1 + \x - \frac{2}{(1 - \x)_+} }
			\nn\\
		&\hspace{10em}
            + \pp{ \frac{7}{2} - \frac{\pi^2}{3} } \delta(1 - \x)
			+ 2 \pp{ \frac{\ln(1 - \x)}{1 - \x} }_+ 
			- \frac{3}{2} \frac{1}{(1 - \x)_+} 
            \nn\\
		&\hspace{10em}
			- \frac{1 + \x^2}{1 - \x} \ln \x 
			- (1 + \x) \ln (1 - \x) + 3 
			+ \order{\epsilon}
		\bigg\},
	\nn\\
	T^{(1R) \gamma, q}_{1}
	&= \wt T^{(1R) \gamma, q}_{1} \nn\\
	&= e_q \, \frac{\alpha_s C_F}{2\pi} \, S_{\epsilon} \pp{ \frac{\mu^2 }{Q^2} }^{\epsilon} \frac{\x^\epsilon }{(1 - \x)^{1+\epsilon}}
		\frac{\Gamma(1 - \epsilon) \Gamma(-\epsilon)}{\Gamma(2 - 2\epsilon)}
    \nn\\
    &\hspace{10em} \times
		\biggbb{ (1 - \epsilon) \pp{ 2 - \frac{\epsilon}{2} }
			- (1 - \x) \Bigpp{ 
				2 - \frac{3}{2} \epsilon + \epsilon^2
			}
		}
	\nn\\
	&= e_q \, \frac{\alpha_s C_F}{2\pi} \, S_{\epsilon} \pp{ \frac{\mu^2 }{Q^2} }^{\epsilon} 
		\bigg\{
			\frac{2 \, \delta(1 - \x)}{\epsilon^2} 
			+ \frac{1}{\epsilon} \bb{ \frac{3}{2} \, \delta(1 - \x) + 2 - \frac{2}{(1 - \x)_+} }
			\nn\\
		&\hspace{10em}
            + \pp{ \frac{7}{2} - \frac{\pi^2}{3} } \delta(1 - \x)
			+ 2 \pp{ \frac{\ln(1 - \x)}{1 - \x} }_+ 
			- \frac{3}{2} \frac{1}{(1 - \x)_+} 
            \nn\\
		&\hspace{10em}
			- \frac{2 \x}{1 - \x} \ln \x 
			- 2 \ln (1 - \x) + \frac{5}{2} 
			+ \order{\epsilon}
		\bigg\},
\label{eq:AA-UT-T1-1R}
\end{align}
where \eq{eq:consistency} is satisfied to every order of $\epsilon$, 
strongly indicating the correct treatment of $\gamma_5$ in the calculation.
In the second step of each equation, we expand up to $\order{\epsilon}$ in the bracket
using the well-known identity for $\x^{\epsilon} / (1 - \x)^{\epsilon}$~\cite{Altarelli:1979ub}.
We have kept the prefactor $S_{\epsilon} (\mu^2 / Q^2 )^{\epsilon}$ unchanged 
to be consistent with the virtual corrections in \eqs{eq:virt-fV}{eq:virt-fT}.
The $\tau_L$ structure first appears at this order and has no $\epsilon$ poles,
so we directly give its result in four dimensions~\cite{Altarelli:1978id, Bardeen:1978yd}, 
\beq[eq:AA-UL-1R]
	U^{(1R) \gamma, q}_{L}
	= e_q^2 \, \frac{\alpha_s C_F}{2\pi} \pp{ 2 \x }.
\eeq

%--------------------------------------------
\subsubsection{Combination of virtual and real corrections}
%--------------------------------------------
\eq{eq:AA-UL-1R} is the complete result for $U^{\gamma, q}_{L}$ up to NLO.
For the other three structure functions in \eq{eq:AA-UT-T1-1R}, 
we need to combine them with the virtual corrections in \sec{sssec:virt}.
For the unpolarized structure function, we have 
\begin{align}
	U^{(1B) \gamma, q}_{T}
	&\equiv U^{(1V) \gamma, q}_{T} + U^{(1R) \gamma, q}_{T}
	\nn\\
	&= e_q^2 \, \frac{\alpha_s C_F}{2\pi} \, S_{\epsilon} \pp{ \frac{\mu^2 }{Q^2} }^{\epsilon} 
		\bigg\{
			- \frac{P_{qq}(\x)}{\epsilon}
			- \pp{ \frac{9}{2} + \frac{\pi^2}{3} } \delta(1 - \x)
			+ 2 \pp{ \frac{\ln(1 - \x)}{1 - \x} }_+ 
			- \frac{3}{2} \frac{1}{(1 - \x)_+}
			\nn\\
		&\hspace{6em}
			- \frac{1 + \x^2}{1 - \x} \ln \x 
			- (1 + \x) \ln (1 - \x) + 3 
			+ \order{\epsilon}
		\bigg\},
\label{eq:AA-UT-1}
\end{align}
where the superscript ``$(1B)$'' refers to one-loop bare quantities before collinear subtraction.
The $1 / \epsilon^2$ poles associated with soft divergences cancel,
and the $1 / \epsilon$ poles organize themselves into the unpolarized quark splitting kernel~\cite{Altarelli:1977zs},
\beq[eq:split-Pqq]
	P_{qq}(\x)
	= \frac{3}{2} \delta(1 - \x) - 1 - \x + \frac{2}{(1 - \x)_+}.
\eeq
Similarly, the one-loop polarized structure functions are
\begin{align}
	T^{(1B) \gamma, q}_{1}
	&= \wt T^{(1B) \gamma, q}_{1}
	= T^{(1R) \gamma, q}_{1} + T^{(1V) \gamma, q}_{1}
	\nn\\
	&= e_q \, \frac{\alpha_s C_F}{2\pi} \, S_{\epsilon} \pp{ \frac{\mu^2 }{Q^2} }^{\epsilon} 
		\bigg\{
			- \frac{\delta P_{qq}(\x)}{\epsilon}
			- \pp{ \frac{9}{2} + \frac{\pi^2}{3} } \delta(1 - \x)
			+ 2 \pp{ \frac{\ln(1 - \x)}{1 - \x} }_+ 
			- \frac{3}{2} \frac{1}{(1 - \x)_+}
			\nn\\
		&\hspace{6em}
			- \frac{2 \x}{1 - \x} \ln \x 
			- 2 \ln (1 - \x) + \frac{5}{2}  
			- \frac{\delta(1 - \x)}{2\epsilon} \biggbb{
				1 - \pp{ \frac{\mu^2 }{Q^2} }^{-\epsilon} 
			}
			+ \order{\epsilon}
		\bigg\},
\label{eq:AA-T1-1}
\end{align}
where the $1 / \epsilon^2$ poles also cancel,
but the $1 / \epsilon$ poles are now organized into the transversity quark splitting kernel~\cite{Artru:1989zv},
\beq[eq:split-tPqq]
	\delta P_{qq}(\x)
	= \frac{3}{2} \delta(1 - \x) - 2 + \frac{2}{(1 - \x)_+}.
\eeq
As mentioned in \eq{eq:fT-log}, the last term in \eq{eq:AA-T1-1}
will give an extra log as opposed to \eq{eq:AA-UT-1}.
This is proportional to the LO structure function $T^{(0) \gamma, q}_{1}$
and serves to cancel the $\mu$ dependence from the dipole coupling. 

%--------------------------------------------
\subsubsection{Subtraction of collinear divergence}
\label{sssec:subtraction}
%--------------------------------------------

The collinear subtraction term (as indicated by the subscript in \eq{eq:partonic-tensor-q}) 
for the one-loop partonic tensor $\hat W_{\gamma, q}^{(1)\mu\nu}$
is obtained by factorizing the one-loop diagrams of \figs{fig:virt}{fig:real}
into perturbative PDFs,
taking a form analogous to \eq{eq:partonic-tensor-q},
\begin{align}
	\hat W_{\gamma, q}^{(1C)\mu\nu}(k, q, s_q)
	= \int_{\x}^1 \frac{d x'}{x' } \, f^{(1)}_{q/q}(x') \,
		\hat W_{\gamma, q}^{(0)\mu\nu}\bigpp{ x' k, q, s'_q(x') }.
\label{eq:AA-col-sub-tensor}
\end{align}
The $x'$ is the momentum fraction carried by the quark parton of the initiating quark of momentum $k$,
with the one-loop quark-in-quark PDF $f^{(1)}_{q/q}(x')$ counterterm given in the $\MS$ scheme by
\begin{align}
	f_{q/q}^{(1)}(x')
	&= -\frac{S_{\epsilon}}{\epsilon} \frac{\alpha_s C_F}{2\pi} P_{qq}(x').
\end{align}
The ``hard coefficient'' $\hat W_{\gamma, q}^{(0)\mu\nu}\bigpp{ x' k, q, s'_q(x') }$ 
is the same as the LO partonic tensor, \eq{eq:AA-Wq0-decomp},
but with  $\x$ replaced by $\x / x'$ and  $s_q(x)$ by 
$$
	s^{\prime \mu}_q(x') = \frac{ h^{(1)}_{q/q}(x', \mu) }{ f^{(1)}_{q/q}(x', \mu) } \, s^{\mu}_q(x),
$$
where $h^{(1)}_{q/q}(x', \mu)$ is the one-loop quark-in-quark transversity PDF, 
\begin{align}
	h_{q/q}^{(1)}(x')
	&= -\frac{S_{\epsilon}}{\epsilon} \frac{\alpha_s C_F}{2\pi} \delta P_{qq}(x').
\end{align}

Since we have normalized the tensors in \eqs{eq:gi-struct-U}{eq:gi-struct-T}
to be scale invariant with respect to $k$,
\eq{eq:AA-col-sub-tensor} directly implies the collinear subtraction terms
for the gauge-invariant scalar coefficients,
\begin{align}
	U^{(1C) \gamma, q}_{T}( \x, Q^2 )
	&= \int_{\x}^1 \frac{d x'}{x'} \,
		f_{q/q}^{(1)}(x') \,
		U^{(0) \gamma, q}_{T}\pp{ \frac{\x}{x'}, Q^2 }
	= e_q^2 \, f_{q/q}^{(1)}(\x),
	\nn\\
	T^{(1C) \gamma, q}_{1}( \x, Q^2 )
	&= \wt T^{(1C) \gamma, q}_{1}( \x, Q^2 )
	= \int_{\x}^1 \frac{d x'}{x'} \,
		h_{q/q}^{(1)}(x') \,
		T^{(0) \gamma, q}_{1}\pp{ \frac{\x}{x'}, Q^2 }
	= e_q \, h_{q/q}^{(1)}(\x)
	\,,
\label{eq:AA-col-sub-coefs}
\end{align}
where we have made use of \eq{eq:AA-Wq0-coefs}.

Subtracting these collinear divergences from \eqs{eq:AA-UT-1}{eq:AA-T1-1}
exactly removes the $1 / \epsilon$ poles.
But note that the extra factor $(\mu^2 / Q^2)^{\epsilon}$ leaves finite logarithmic terms,
indicating dependence on the factorization scale $\mu$.
After that, we may take $\epsilon \to 0$ to get the infrared-safe hard scalar coefficients,
\begin{align}
	U^{(1) \gamma, q}_{T}
	&= \lim_{\epsilon \to 0}
		\bb{ U^{(1B) \gamma, q}_{T} 
			- U^{(1C) \gamma, q}_{T}
		}
	\nn\\
	&= e_q^2 \, \frac{\alpha_s C_F}{2\pi} \,
		\bigg\{
			- P_{qq}(\x) \, \ln \biggpp{ \frac{\mu^2 }{Q^2} }
			- \pp{ \frac{9}{2} + \frac{\pi^2}{3} } \delta(1 - \x)
			\nn\\
		&\hspace{6em}
			+ 2 \pp{ \frac{\ln(1 - \x)}{1 - \x} }_+ 
            - \frac{3}{2} \frac{1}{(1 - \x)_+}
			- \frac{1 + \x^2}{1 - \x} \ln \x 
			- (1 + \x) \ln (1 - \x) + 3 
		\bigg\},
	\nn\\
	T^{(1) \gamma, q}_{1}
	&= \wt T^{(1) \gamma, q}_{1}
	= \lim_{\epsilon \to 0}
		\bb{ T^{(1B) \gamma, q}_{1} 
			- T^{(1C) \gamma, q}_{1}
		}
	\nn\\
	&= e_q \, \frac{\alpha_s C_F}{2\pi} \,
		\bigg\{
			- \bb{ \delta P_{qq}(\x) + \frac{1}{2} \delta(1 - \x) } \, \ln \biggpp{ \frac{\mu^2 }{Q^2} }
			- \pp{ \frac{9}{2} + \frac{\pi^2}{3} } \delta(1 - \x)
			\nn\\
		&\hspace{6em}
            + 2 \pp{ \frac{\ln(1 - \x)}{1 - \x} }_+ 
			- \frac{3}{2} \frac{1}{(1 - \x)_+}
			- \frac{2 \x}{1 - \x} \ln \x 
			- 2 \ln (1 - \x) + \frac{5}{2}  
		\bigg\}.
\label{eq:AA-Uq-Tq-sub}
\end{align}
These directly add to the LO coefficients in \eq{eq:AA-Wq0-coefs}.
Clearly, the subtraction terms preserve the condition in \eq{eq:consistency}
and can be extended to all orders.

The hard coefficients for antiquarks are the same.
For the gluon contribution, since it does not involve dipole vertices,
the calculation is the same as unpolarized DIS and will not be repeated.
The combined cross section up to NLO, organized as \eq{eq:xsec-AA-0}, 
is given in Appendix~\ref{app:sec:summary-xsec}.

%================================================================================
\section{Sensitivity study at EIC}
\label{sec:pheno}
%================================================================================

In this section, we study the phenomenology of the TSA at the EIC. 
We use the c.m.\ energy $\sqrt{s} = 105~\GeV$ with integrated luminosity $\mathcal{L} = 100~\fb^{-1}$,
and choose $\lambda_e = S_T = 0.7$ for both the electron's longitudinal polarization and proton's transverse polarization.
The $G_{\mu}$ scheme is used for EW inputs, 
with the $W$ mass $m_W = 80.377~\GeV$,
$Z$ mass $m_Z = 91.1876~\GeV$,
and Fermi constant of the muon, $G_{\mu} = 1.1663787 \times 10^{-5}~\GeV^{-2} = 1 / (\sqrt 2 v^2)$.
Below, we first examine the effects of loop corrections to the unpolarized and polarized cross sections,
and then study the estimated bounds that EIC data can set on the dipole couplings,
which will also be compared to other constraints.

%----------------------------------------------------------------
\subsection{TSA and NLO impacts}
\label{ssec:nlo}
\begin{figure}[htbp]
	\centering
	\includegraphics[scale=0.7]{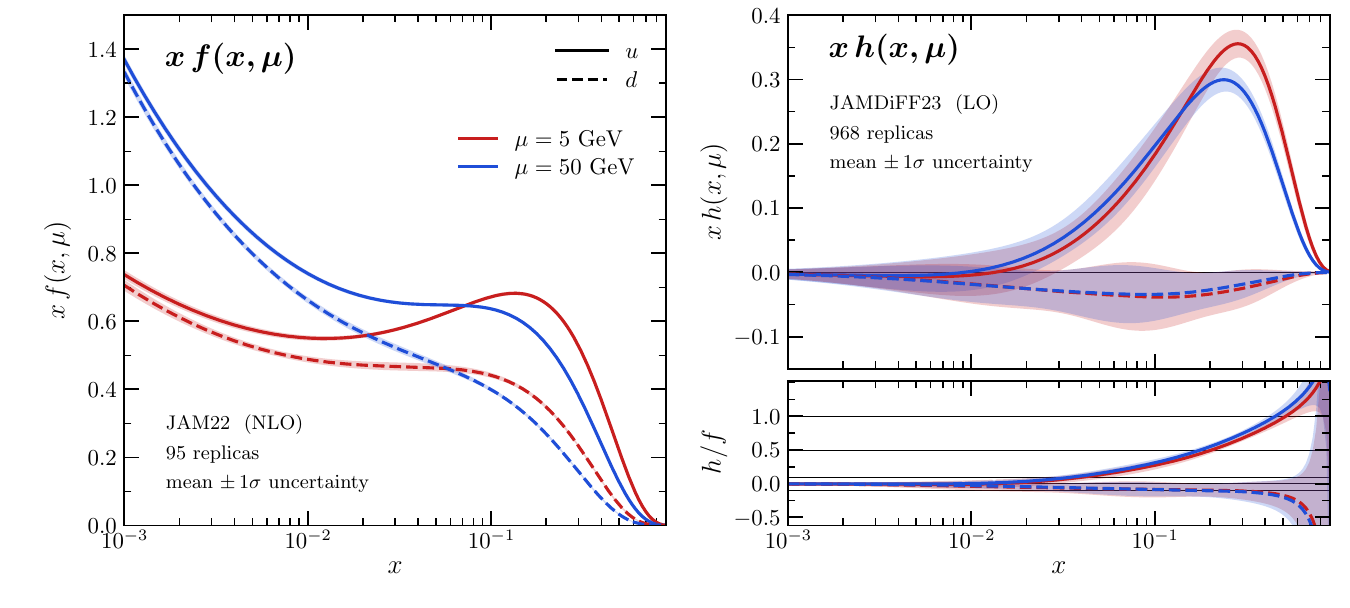}
	\caption{(a) Unpolarized and (b) transversity PDFs for $u$ and $d$ quarks, at scale $\mu = 5$ and $50~\GeV$ for each. 
		The shading refers to the $1\sigma$ uncertainty band for each of the PDF replica sets.
        The uncertainty band for the unpolarized PDFs is not visible on the plot.  
		The lower right panel shows the ratio of $h$ to $f$, with thin grid lines shown for $h / f = \pm 0.1, 0.5, 1$.}
\label{fig:pdf}
\end{figure}

%----------------------------------------------------------------
We start by looking at the PDFs,
for which we use the fit JAM22 (at NLO)~\cite{Cocuzza:2022jye} for the unpolarized PDF
and JAMDiFF (at LO)~\cite{Cocuzza:2023oam, Cocuzza:2023vqs} for the transversity PDF.
In these fits, the PDF uncertainties are treated in a Bayesian analysis 
and are modeled by a set of replica PDFs.
These are shown in \fig{fig:pdf} for the $u$ and $d$ quarks,
in terms of central values, given by the replica set averages,
surrounded by uncertainty bands that are given by 
the standard deviations point by point in the parton momentum fraction $x$.

Clearly, the absolute magnitude of $h_u$ is much bigger than $h_d$, with opposite sign.
They are mostly constrained in the valence quark region, 
with appreciably nonzero values only at $x \gtrsim 0.04$. 
Their uncertainties are generally much bigger than those of the unpolarized PDFs,
and the uncertainty of $h_d$ is also much bigger than that of $h_u$.%%%
\footnote{It appears that the positivity bound $|h_q| \leq |f_q|$ is violated at $x \gtrsim 0.5$.
This is because the positivity constraint is enforced in the JAM fit as a Bayesian penalty prior, not a hard cutoff.
The experimental data used in their fit do not constrain the large-$x$ region, 
leaving it highly sensitive to the lattice QCD tensor charge calculations~\cite{Cocuzza:2023oam, Cocuzza:2023vqs}. 
As a conservative choice, we cut $x_B < 0.5$ to reduce this ambiguity.}
%%%

%----------------------------------------------------------------
% Fig: PDFs
%------------------------------------------------
%----------------------------------------------------------------

\begin{figure}[htbp]
	\centering
	\includegraphics[scale=0.42]{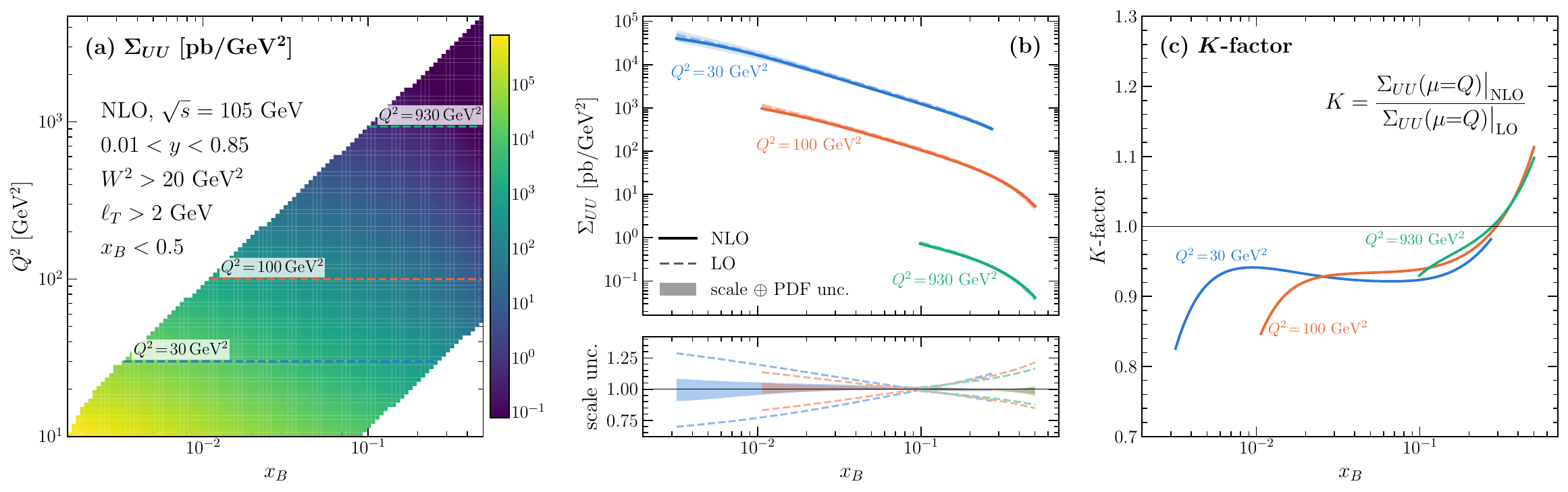}
	\caption{(a) Unpolarized DIS cross section at NLO in the $(x_B, Q^2)$ space.
		(b) the cross section as functions of $x_B$ for three chosen $Q^2$ values indicated by the dashed lines in (a).
		The upper panel compares LO (dashed lines) and NLO (solid lines) results, 
        with bands including both scale and PDF uncertainties
		(invisible on the log scale).
		The lower panel gives the relative scale uncertainties for the three $Q^2$ curves, 
		comparing LO (dashed lines) and NLO (color bands).
		(c) gives the $K$ factors corresponding to the three curves in (b), 
        evaluated at $\mu = Q$ for the central PDF replica.}
\label{fig:unp-xsec}
\end{figure}

Let us write the cross section as
\begin{align}
	\frac{d\sigma}{dx_B \, dQ^2 \, d\phi}
	& = \frac{1}{2\pi} \, \Bigbb{ 
		\Sigma_{UU}(x_B, Q^2)
		+ \lambda_e \, \Sigma_{LU}(x_B, Q^2)
		+ S_T \, \Sigma_{UT}(x_B, Q^2, \phi)
		+ \lambda_e S_T \, \Sigma_{LT}(x_B, Q^2, \phi)
	},
\label{eq:xsec-pol}
\end{align}
where we take $\phi_S = 0$ so that $\Delta \phi = \phi$.
Each term on the right-hand side is a sum of contributions 
from different channels summarized in Appendix~\ref{app:sec:summary-xsec}.
The $\Sigma_{UU}(x_B, Q^2)$ term is the unpolarized cross section $d \sigma / d x_B \, dQ^2$,
given by the convolution of unpolarized PDFs $f_i(x, \mu)$ with 
hard coefficients from all the $\gamma$, $Z$, and interference channels.
Applying the phase space cuts, 
\beq[eq:cuts]
	Q^2 > 10~\GeV^2, 
	\quad
	W^2 \equiv (p + q)^2 > 20~\GeV^2, 
	\quad
	0.01 < y < 0.85,
	\quad
	\l_T > 2~\GeV,
    \quad
    x_B < 0.5,
\eeq
to ensure that we are in the DIS region,
we show $\Sigma_{UU}$ in \fig{fig:unp-xsec}(a).
It covers over 6 orders of magnitude, 
with most of the event yield concentrated in the low-$Q^2$ region.
The high-$Q^2$ region, however, gives more sensitivity to large $x_B$ due to the lower $y$ cut. 
We define the scale uncertainty by varying the scale $\mu$ between $Q/2$ and $2Q$
and define the $K$ factor as the cross section ratio of the NLO to the LO, both at $\mu = Q$.
These are shown in \fig{fig:unp-xsec}(b)(c),
indicating that the NLO correction generally causes 
a $10$--$20\%$ change to the LO cross section,
and brings down the scale uncertainty to within $5\%$.
The PDF uncertainty is negligible compared to the scale uncertainty for the unpolarized case.

%----------------------------------------------------------------
% Fig: Unpolarized cross section
%------------------------------------------------
%----------------------------------------------------------------

The term $\Sigma_{LU}$ for the single electron spin contribution is not of interest here
and will be dropped below. 
It can be easily separated from the other three contributions in experiments using the polarization configurations.

The single transverse spin term $\Sigma_{UT}$ contains contributions 
from the interference and the $Z$ channels.
This has been studied at LO in \refcite{Boughezal:2023ooo},
and as there, we find that the $Z$ channel is highly suppressed by the square of the $Z$ propagator, 
and the $Z$ dipole coupling of the $u$ quark dominates in the interference channel.
To see this, we take the results from 
\eqs{eq:summary-UT/LT-int}{eq:summary-UT/LT-12-int} in Appendix~\ref{app:sec:summary-xsec}, 
neglecting the numerically small $Z$ channel,
\begin{align}
	\Sigma_{UT}(x_B, Q^2, \phi)
	&\simeq N_{\rm int} \, \frac{2 \l_T }{v} \cdot 
		\bigg\{
		y \, g_A^e \Big[
			g_V^u \, \C\!\bb{ t_1, \, h_{u + \bar u} } 
				\, \Re\bigpp{ \Gamma_{\gamma}^u \, e^{i \phi} }
			+ g_V^d \, \C\!\bb{ t_1, \, h_{d + \bar d} } 
				\, \Re\bigpp{ \Gamma_{\gamma}^d \, e^{i \phi} }
			\nn\\
        &\hspace{8.5em}
			+ e_u \, \C\!\bb{ t_1, \, h_{u + \bar u} } 
				\, \Re\bigpp{ \Gamma_{Z}^u \, e^{i \phi} }
			+ e_d \, \C\!\bb{ t_1, \, h_{d + \bar d} } 
				\, \Re\bigpp{ \Gamma_{Z}^d \, e^{i \phi} }
			\Big]
			\nn\\
    	&\hspace{2.8em}
			+ (2 - y) \, g_V^e \Big[
				g_A^u \, \C\!\bb{ t_2, \, h_{u - \bar u} }
				\, \Re\bigpp{ \Gamma_{\gamma}^u \, e^{i \phi} }
				+ g_A^d \, \C\!\bb{ t_2, \, h_{d - \bar d} }
				\, \Re\bigpp{ \Gamma_{\gamma}^d \, e^{i \phi} }
			\Big]
		\bigg\},
\label{eq:UT-approx2}
\end{align}
where $\C[t_i, h_{q \pm \bar q}]$ stands for the convolution,
defined in \eq{eq:condef}, 
of the PDFs $h_{q \pm \bar q} \equiv h_q \pm h_{\bar q}$
with the hard coefficients $t_i$ $(i = 1, 2)$,
which are given in \eq{eq:kernels} and both of order 1. 
If $y$ is also not too small, so that it is of the same order as $(2-y)$, 
we can neglect the last line because it is suppressed by the small $g_V^e = -0.054$.
The remaining four terms can be compared through the couplings,
$
	(g_V^u, g_V^d, e_u, e_d) = (0.203, -0.351, 0.667, -0.333).
$
Together with the fact $|h_u| \gg |h_d|$,
it is then obvious that $\Gamma_Z^u$ is the dominant term.

%----------------------------------------------------------------
% Fig: Cross section: UT
%------------------------------------------------
\begin{figure}[htbp]
	\centering
	\includegraphics[scale=0.42]{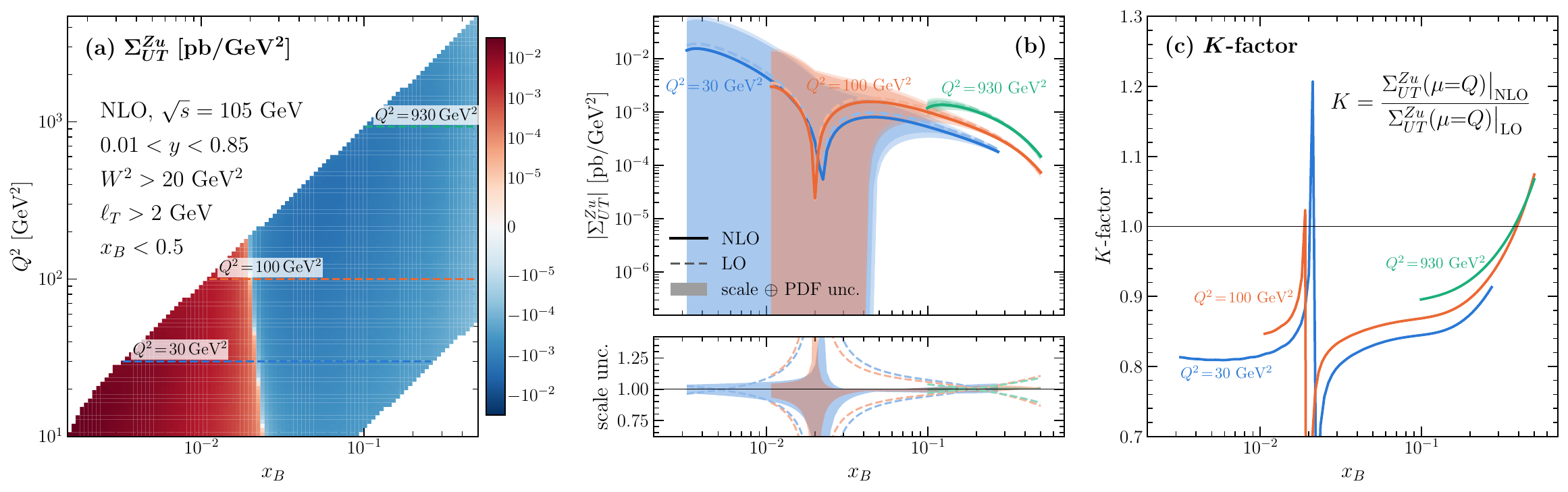}
	\caption{Similar to \fig{fig:unp-xsec}, 
        but for the coefficient $\Sigma_{UT}^{Z u}$ of the $\Gamma_Z^u$ dipole 
		in the single transverse spin cross section term $\Sigma_{UT}$.
		Its absolute value is shown in (b) for three values of $Q^2$, 
        with the sign flips easily identified from (a).}
\label{fig:UT-Zu}
\end{figure}
%----------------------------------------------------------------

In \fig{fig:UT-Zu},
we show the coefficient of $\Gamma_Z^u$ in the single transverse spin term, 
\beq[eq:Sigma-UT-Zu]
	\Sigma_{UT}^{Zu}(x_B, Q^2) \equiv 
	\frac{ \partial \, \Sigma_{UT}(x_B, Q^2, \phi = 0) }{ \partial \Re (\Gamma_Z^u)},
\eeq
which numerically includes both the interference and the $Z$ channels.
It has a sign flip around $x_B = 0.02$ originating from the sign flip of the transversity PDF.
Besides that, both the shapes and magnitude of the $K$ factor in \fig{fig:UT-Zu}(c) 
resemble the unpolarized case in \fig{fig:unp-xsec}(c).
While the scale uncertainty is reduced by the inclusion of the NLO correction, 
it is now the PDF uncertainty,
which is obtained by looping through the set of JAMDiFF transversity PDF replicas,
that dominates, especially at small $x_B \lesssim 0.1$. 

For the double spin term $\Sigma_{LT}$, 
we take the result from \eqs{eq:summary-LT-AA}{eq:summary-UT/LT-12-int},
with the $Z$ channel and the $g_V^e$-suppressed $Z$ dipole terms neglected, 
\begin{align}
	\Sigma_{LT}
	&\simeq N_{\gamma} \, \frac{2 Q \sqrt{1-y}}{v} \bigg\{
		y \bb{
			\frac{2}{3} 
                \, \C\!\bb{ t_1, \, h_{u + \bar u} } 
				\, \Re\bigpp{ \Gamma_{\gamma}^u \, e^{i \phi} }
			- \frac{1}{3} 
                \, \C\!\bb{ t_1, \, h_{d + \bar d} } 
				\, \Re\bigpp{ \Gamma_{\gamma}^d \, e^{i \phi} }
		}
		\nn\\
	&\hspace{2.5em}
		+ \frac{\alpha_Z Q^2}{\alpha_e (Q^2 + m_Z^2)} \, 
		\pp{ 1 - \frac{y}{2} }
		\bb{  
			\frac{1}{2} 
            \, \C\!\bb{ t_2, \, h_{u - \bar u} }
			\, \Re\bigpp{ \Gamma_{\gamma}^u \, e^{i \phi} }
			- \frac{1}{2} 
            \, \C\!\bb{ t_2, \, h_{d - \bar d} }
			\, \Re\bigpp{ \Gamma_{\gamma}^d \, e^{i \phi} }
		}
	\bigg\},
\label{eq:LT-approx2}
\end{align}
where the electric and axial charges of electron and quarks have been explicitly used.
The first line is from the $\gamma$ channel 
and the second line from the interference channel.
The latter is suppressed at low $Q^2$ by the $Z$ propagator, 
but gains an enhancement from the factor $(1 - y / 2)$ at small $y$.
The $\Sigma_{LT}$ contribution is almost exclusively sensitive to the photon dipole coupling, 
orthogonal to that of the $\Sigma_{UT}$ contribution in \eq{eq:UT-approx2},
with stronger dependence on the $u$-quark contribution due to the electric charge and transversity PDF. 

%----------------------------------------------------------------
% Fig: Cross section: LT
%------------------------------------------------
\begin{figure}[htbp]
	\centering
	\includegraphics[scale=0.42]{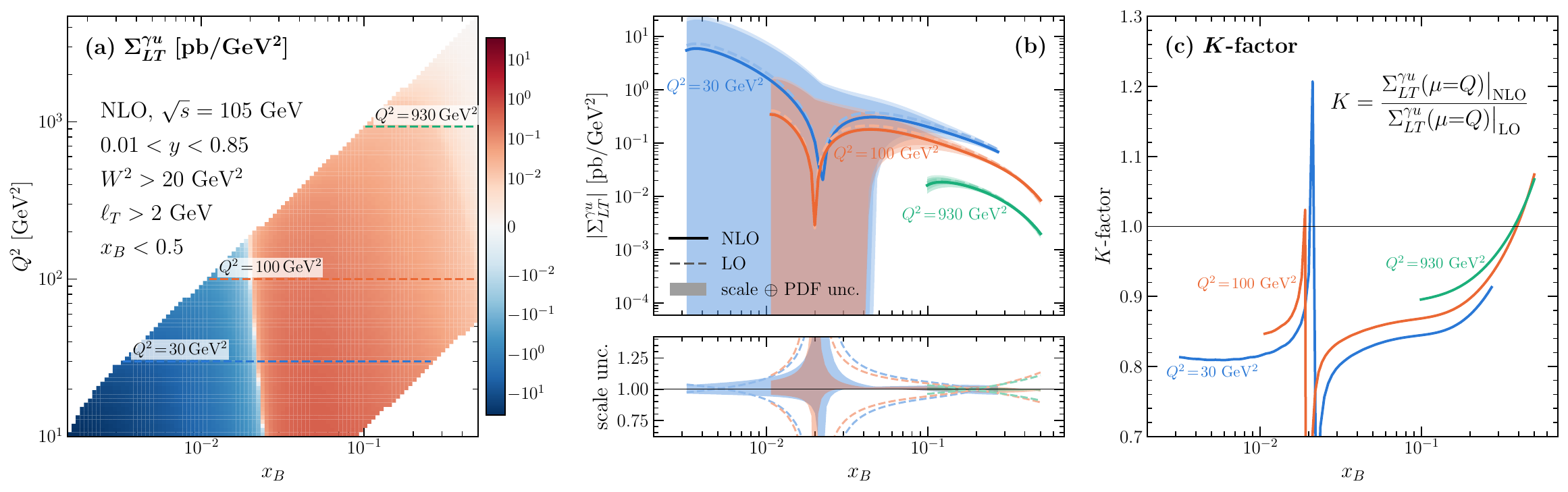}
	\caption{Similar to \fig{fig:UT-Zu}, 
        but for the coefficient $\Sigma_{LT}^{\gamma u}$ of the $\Gamma_{\gamma}^u$ dipole 
        in the double spin cross section term $\Sigma_{LT}$.}
\label{fig:LT-Au}
\end{figure}
%----------------------------------------------------------------

In \fig{fig:LT-Au},
we show the coefficient of $\Gamma_{\gamma}^u$ in the double spin term, 
\beq[eq:Sigma-LT-Au]
	\Sigma_{LT}^{\gamma u}(x_B, Q^2) \equiv 
	\frac{ \partial \, \Sigma_{LT}(x_B, Q^2, \phi = 0) }{ \partial \Re (\Gamma_{\gamma}^u)},
\eeq
which numerically includes both the interference and $\gamma$ channels.
All the features are similar to \fig{fig:UT-Zu}, 
except that the magnitude of $\Sigma_{LT}^{\gamma u}$ is generally bigger,
particularly at small $Q^2$.

\begin{figure}[htbp]
	\centering
	\includegraphics[scale=0.4]{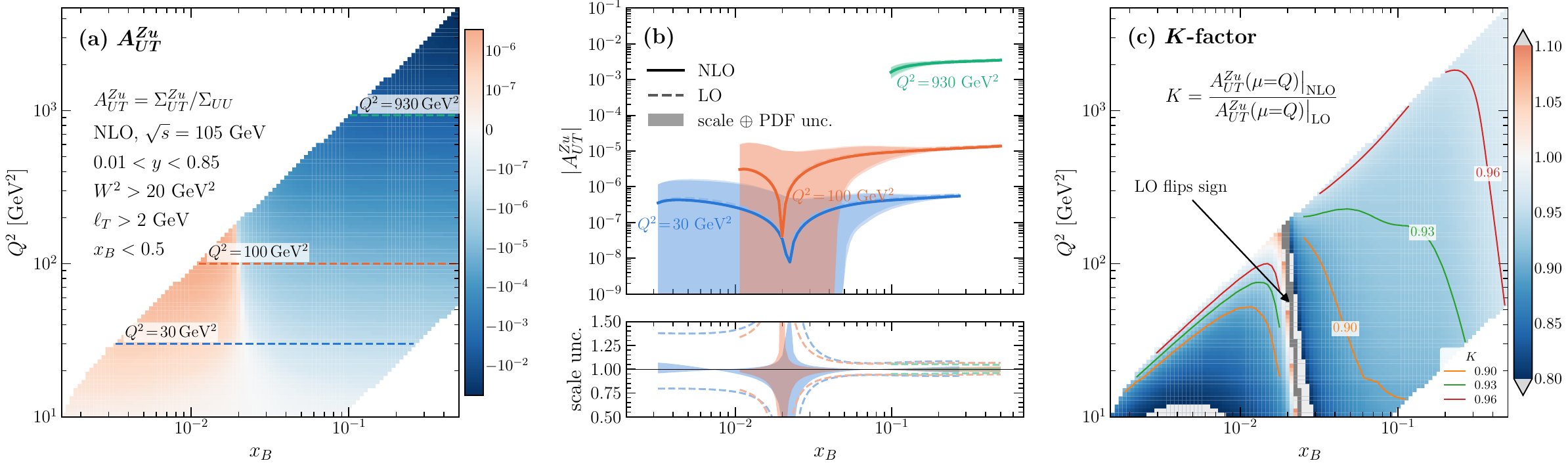}
	\caption{(a) The coefficient $A_{UT}^{Zu}$ of the $\Gamma_Z^u$ dipole in the 
		single transverse spin asymmetry $A_{UT}$ at NLO, 
		in the $(x_B, Q^2)$ space.
		(b) displays its absolute value as functions of $x_B$ for three chosen $Q^2$ values indicated by the dashed lines in (a).
		The upper panel compares LO (dashed lines) and NLO (solid lines), with bands including both scale and PDF uncertainties
		(their differences are hardly visible in the log scale).
		The lower panel gives the relative scale uncertainties for the three $Q^2$ curves, 
		comparing LO (dashed lines) and NLO (color bands).
		(c) gives the $K$ factor distribution of $A_{UT}^{Zu}$ in the $(x_B, Q^2)$ space,
		evaluated at $\mu = Q$ and with the central PDF set. Solid and colored contours mark $K = 0.96$, $0.93$, and $0.90$.}
\label{fig:AUT-Zu}
\end{figure}

Experimentally, it is the polarization asymmetries that are measured,
defined as
\beq[eq:asy-def]
	A_{UT}(x_B, Q^2, \phi)
	\equiv \frac{\Sigma_{UT}(x_B, Q^2, \phi)}{\Sigma_{UU}(x_B, Q^2)},
	\quad
	A_{LT}(x_B, Q^2, \phi)
	\equiv \frac{\Sigma_{LT}(x_B, Q^2, \phi)}{\Sigma_{UU}(x_B, Q^2)}.
\eeq
Dividing \eqs{eq:Sigma-UT-Zu}{eq:Sigma-LT-Au} by $\Sigma_{UU}$,
we get the dominant contributions to these asymmetries, 
\beq[eq:asy-def-flavor]
	A_{UT}^{Z u}(x_B, Q^2)
	\equiv \frac{\Sigma_{UT}^{Z u}(x_B, Q^2)}{\Sigma_{UU}(x_B, Q^2)},
	\quad
	A_{LT}^{\gamma u}(x_B, Q^2)
	\equiv \frac{\Sigma_{LT}^{\gamma u}(x_B, Q^2)}{\Sigma_{UU}(x_B, Q^2)},
\eeq
which are the coefficients of 
$\Re \bigpp{ \Gamma_{Z}^u \, e^{i \phi} }$ and 
$\Re \bigpp{ \Gamma_{\gamma}^u \, e^{i \phi} }$ 
in $A_{UT}(x_B, Q^2, \phi)$ and $A_{LT}(x_B, Q^2, \phi)$, respectively.
They are shown in \figs{fig:AUT-Zu}{fig:ALT-Au}, 
including their magnitude, scale and PDF uncertainties, and $K$ factor distributions.
Interestingly, the $K$ factors of $\Sigma_{UT}^{Zu}$ and $\Sigma_{LT}^{\gamma u}$
largely cancel that of $\Sigma_{UU}$, 
such that the $K$ factors of the asymmetries are close to 1 in a large portion of phase space,
particularly the large-$x_B$ and high-$Q^2$ region where their magnitude is large.
Hence, we conclude that the polarization asymmetry observables are stable 
against radiative QCD corrections. 
They can therefore be used as precision observables for probing new physics effects 
encoded in the dipole operators.

As expected, the value of $A_{LT}^{\gamma u}$ is greater than $A_{UT}^{Zu}$
at low $Q^2$ by a few orders of magnitude,
but they become closer as $Q^2$ grows. 
While we still see a significant reduction of the scale uncertainty from the NLO correction,
it is dominated by the PDF uncertainty.
The impact of that will be studied further in the following.

%----------------------------------------------------------------
% Fig: Asymmetry: UT
%------------------------------------------------
%----------------------------------------------------------------

%----------------------------------------------------------------
% Fig: Asymmetry: LT
%------------------------------------------------
\begin{figure}[htbp]
	\centering
	\includegraphics[scale=0.4]{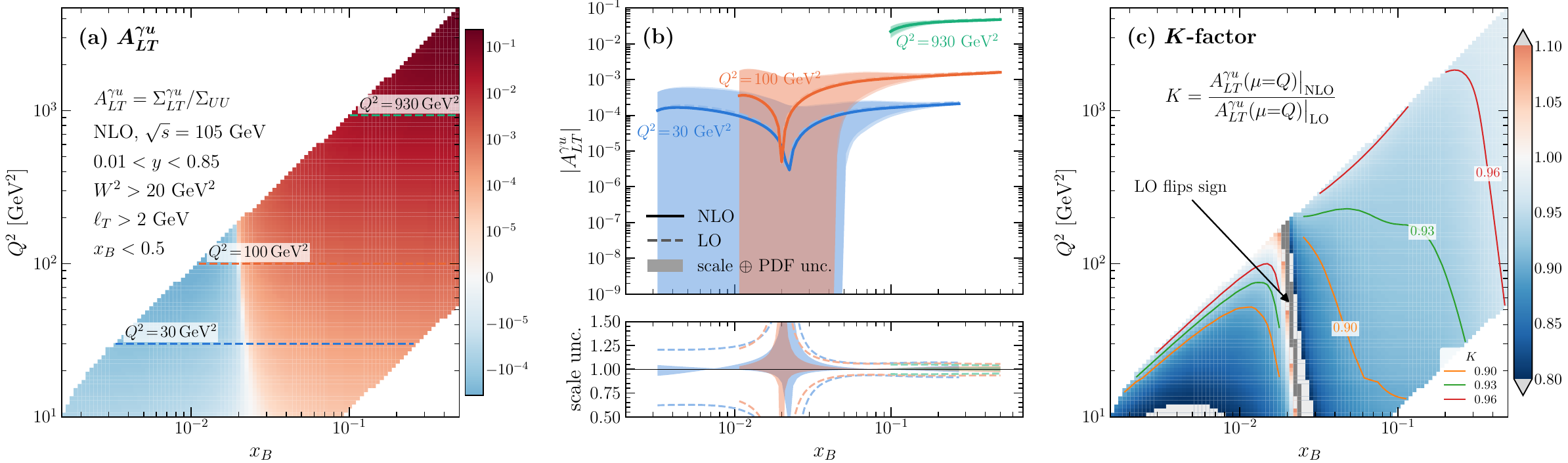}
	\caption{Similar to \fig{fig:AUT-Zu}, but for
		the coefficient $A_{LT}^{\gamma u}$ of the $\Gamma_{\gamma}^u$ dipole in the 
		double spin asymmetry $A_{LT}$.}
\label{fig:ALT-Au}
\end{figure}
%----------------------------------------------------------------

%----------------------------------------------------------------
\subsection{EIC bounds on dipole couplings}
\label{ssec:bound}
%----------------------------------------------------------------
The advantage of the TSA observables, 
as emphasized in the literature~\cite{Boughezal:2023ooo, Wen:2023xxc, Wang:2024zns, Wen:2024cfu, Wen:2024nff, Huang:2025ljp}, 
is that they receive minimal contributions from the SM, since the SM contributions are
either suppressed by light-fermion masses or come only from high-twist QCD dynamics.
This makes them uniquely sensitive to the dipole operators of the SMEFT.
In this subsection, we examine how well the EIC measurements 
can constrain their Wilson coefficients.
We will first use the physical basis in \eq{eq:Lag-AZ},
and then convert to the Warsaw basis using \eq{eq:dip-wcoefs-conver}.

Since the $\phi$ dependence in both $\Sigma_{UT}$ and $\Sigma_{LT}$ of \eq{eq:xsec-pol}
appears in the form $\Re \bigpp{ \Gamma_V^q \, e^{i \phi} }$ 
(for $V = \gamma$ or $Z$ and $q = u$ or $d$),
the real and the imaginary parts of the dipole couplings 
can be separately extracted from experimental signals via $\cos\phi$ and $\sin\phi$ distributions, respectively.
Therefore, let us first assume the dipole couplings are real.
The results we obtain in the following analysis also apply to the imaginary parts.

The two sources of uncertainties that we consider are the scale uncertainty and the PDF uncertainty.
Other systematic uncertainties are assumed to cancel in the polarization asymmetries.
The uncertainties of unpolarized PDFs only affect the total rate,
and are negligible compared to the scale uncertainty.
We thus fix them to the central replica set of JAM22.
On the other hand, the uncertainties of transversity PDFs are numerically important
and will be taken into account via the set of 968 replicas, $\{ h^{(r)} \}_{r = 1}^{968}$, in the JAMDiFF fit.

We perform the statistical analysis using the likelihood method.
Start with the single spin cross section, 
\beq
	\frac{d\sigma_{UT}(\Gamma_{j}, h)}{dx_B \, dQ^2 \, d\phi}
	= \frac{1}{2\pi} \bb{ \Sigma_{UU} (x_B, Q^2) + S_T \, \Gamma_{j} \, \Sigma_{UT}^{j}(x_B, Q^2, h) \cos\phi },
\eeq
where $j$ denotes the four dipole couplings, and is understood to be summed over. 
Besides $\Gamma_{j}$, all other quantities are measurable and calculable, given a choice of the transversity PDF $h$. 
Suppose we divide the measured region of $(x_B, Q^2, \phi)$ into many small bins.
The {\it expected} event yield $\bar{n}_b$ in each bin $b$ is given by 
\begin{align}
	\bar{n}_b \equiv\bar{n}_b(\Gamma_j, h)
	&= \mathcal{L} \, d x_B \, d Q^2 \, d \phi \cdot
		\frac{d\sigma_{UT, b}(\Gamma_j, h)}{dx_B \, dQ^2 \, d\phi}
	\nn\\
	&= \mathcal{L} \, d x_B \, d Q^2 \, d \phi \cdot \frac{1}{2\pi} \bb{ 
		\Sigma_{UU, b} + S_T \, \Gamma_{j} \, \Sigma_{UT, b}^{j}(h) \cos\phi_b
	},
\end{align}
where $\mathcal{L}$ is the integrated luminosity,
and quantities with subscript $b$ refer to their values in bin $b$.
The {\it actual} event number $n_b$ in bin $b$ follows 
a Poisson distribution $P( n_b \mid \bar{n}_b )$ with mean at $\bar{n}_b$,
which satisfies 
\beq[eq:log-Possion]
	\ln P( n_b \mid \bar{n}_b )
	\simeq n_b \pp{ \ln \frac{\bar{n}_b}{n_b} + 1 - \frac{\bar{n}_b}{n_b} } + \order{\ln n_b}.
\eeq
Given a measurement $\cc{ n_b }$ of the event numbers in all the bins, 
we can form the likelihood, 
\begin{align}
	L_{UT} = L_{UT}(\Gamma_j, h)
	= \rho(h) \prod_b P\bigpp{ n_b \mid \bar{n}_b(\Gamma_j, h) }.
\label{eq:likelihood-stat}
\end{align}
The PDF $h$ is treated as a (functional) nuisance parameter,
which follows a probability distribution $\rho(h)$.
The latter is obtained from a global analysis of past experiments
and is practically modeled by the (unweighted) replica set $\{ h^{(r)} \}$.
In terms of Bayesian probability, \eq{eq:likelihood-stat} is understood as 
the conditional probability for the data $\cc{ n_b }$ to happen 
given the values of the parameters $\Gamma_j$ and PDF $h$,
multiplied by the prior probability $\rho(h)$ of $h$ and a flat prior of $\Gamma_j$.

Now, taking the measured event number
$n_b = \bar{n}_b(\Gamma_j = 0) 
	= \mathcal{L} \, d x_B \, d Q^2 \, d \phi \, \Sigma_{UU, b} / (2\pi)$
as the expectation from the SM,
which is independent of $h$,
and using $\prod_b P_b = \exp\bigpp{ \sum_b \ln P_b }$ in \eq{eq:likelihood-stat}
along with \eq{eq:log-Possion},
we can turn the sum over $b$ into an integral over the phase space,
\beq
	L_{UT}(\Gamma_j, h)
	= \rho(h) \, e^{ - \frac{1}{2} G_{UT}(\Gamma_j, \, h)},
\eeq
where the exponent is
\begin{align}
	G_{UT}(\Gamma_j, h)
	&= -\frac{\mathcal{L}}{\pi} \int d x_B \, d Q^2 \, d \phi \, 
		\Sigma_{UU}(x_B, Q^2) \cdot
        \bigg\{ \ln \Bigbb{ 
			1 + S_T \, \Gamma_{j} \, A_{UT}^{j}(x_B, Q^2, h) \cos\phi
		      }
        \nn\\
    &\hspace{16.8em}
            - S_T \, \Gamma_{j} \, A_{UT}^{j}(x_B, Q^2, h) \cos\phi
	    \bigg\}
	\nn\\
	&= -2\mathcal{L} \int d x_B \, d Q^2 \, 
		\Sigma_{UU}(x_B, Q^2) \cdot \ln \biggcc{ \frac{1}{2} \biggbb{
				1 + \sqrt{ 1 - \bigbb{ S_T \, \Gamma_{j} \, A_{UT}^{j}(x_B, Q^2, h) }^2 }
			}
		}
	\nn\\
	&\simeq \frac{\mathcal{L}}{2} \int d x_B \, d Q^2 \, 
		\Sigma_{UU}(x_B, Q^2) \cdot \Bigbb{ S_T \, \Gamma_{j} \, A_{UT}^{j}(x_B, Q^2, h) }^2,
\end{align}
where we used the notation $A_{UT}^{j} = \Sigma_{UT}^{j} / \Sigma_{UU}$ as defined in \eq{eq:asy-def-flavor},
integrated over $\phi$ in the second step, 
and used the approximation in the last step that the value of the asymmetry is generally small, 
as we have seen in \figs{fig:AUT-Zu}{fig:ALT-Au}.
The last step also defines the Fisher information matrix for a given $h$,
\beq[eq:fisher]
	F_{UT}^{jj'}(h) = \frac{1}{2} \, \mathcal{L} \, S_T^2 \int d x_B \, d Q^2 \, 
		\Sigma_{UU}(x_B, Q^2) \, 
		A_{UT}^{j}(x_B, Q^2, h) \, A_{UT}^{j'}(x_B, Q^2, h).
\eeq

It is straightforward to include the double spin asymmetry in the analysis.
We only need to add the corresponding Fisher information $F^{jj'}_{LT}$ to $F^{jj'}_{UT}$,
which then turns the likelihood into
\beq[eq:likelihood-h]
	L(\Gamma_j, h)
	= \rho(h) \, \exp\!\bb{ - \frac{1}{2} \, \Gamma_j \, 
		F^{jj'}(h)
	\, \Gamma_{j'} },
\eeq
with $F^{jj'}(h) = F_{UT}^{jj'}(h) + F_{LT}^{jj'}(h)$ and 
\beq
	F_{LT}^{jj'}(h) 
	= \frac{1}{2} \, \mathcal{L} \, \lambda_e^2 S_T^2 \int d x_B \, d Q^2 \, 
		\Sigma_{UU}(x_B, Q^2) \, 
		A_{LT}^{j}(x_B, Q^2, h) \, A_{LT}^{j'}(x_B, Q^2, h).
\eeq

\begin{figure}[htbp]
	\centering
	\includegraphics[scale=0.63]{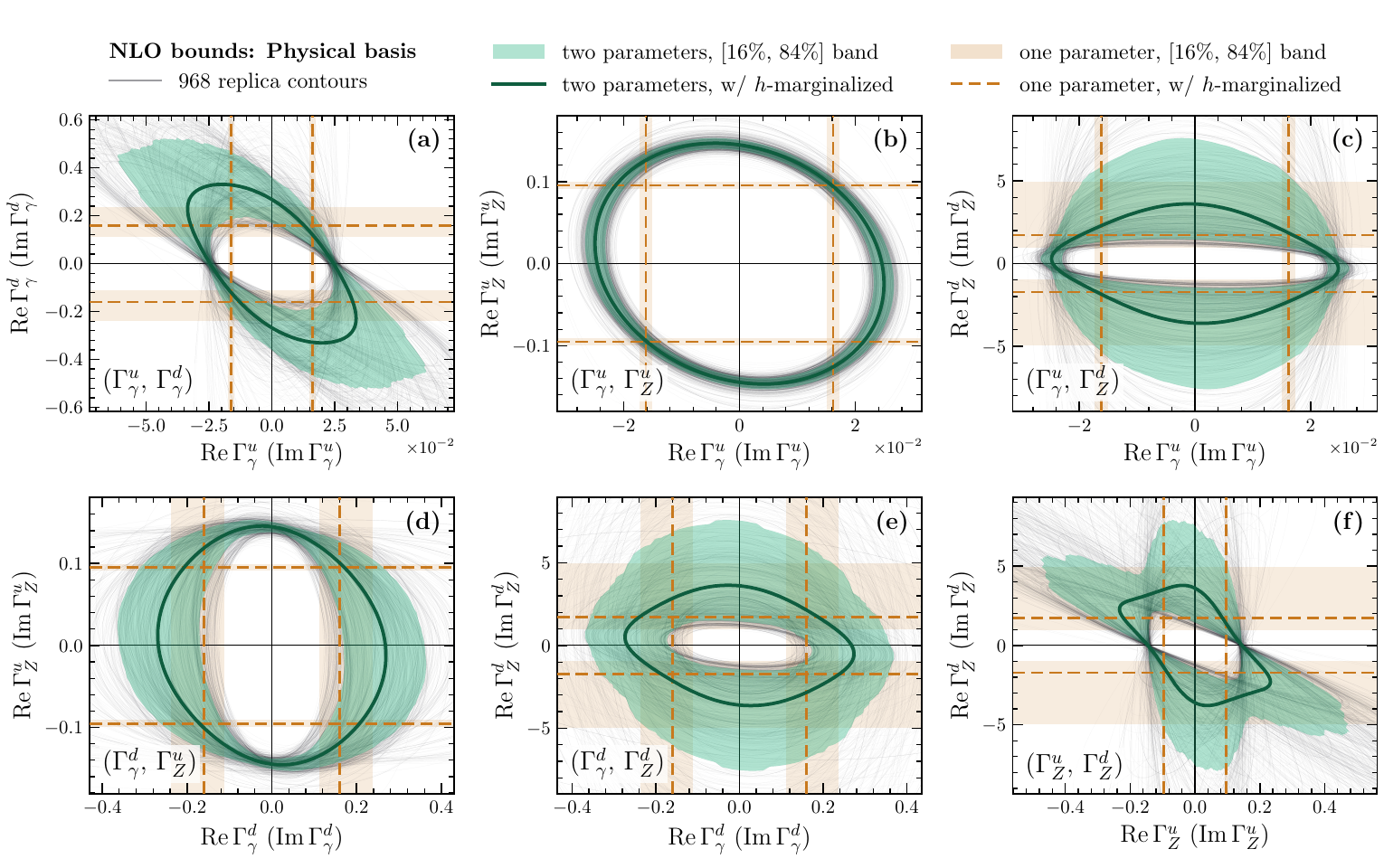}
	\caption{Constraint contour for each pair of the dipole couplings (at scale $m_Z$) at $68\%$ C.L.,
    obtained from the likelihood in \eq{eq:likelihood-h} at NLO and $\mu = Q$, with the cuts in \eq{eq:cuts}. 
	The stack of thin gray lines represents such contours for each replica of the transversity PDF $h$,
	and their $68\%$ central intervals are shaded in light green, surrounding the $h$-marginalized contours.
	Along the horizontal and vertical directions are also shown the single-parameter constraint bands,
	with the dashed lines given by marginalizing the PDF $h$,
	and the fuzzy bands representing the $68\%$ central intervals of the replica stack.
	These constraints apply equally to each (pair) of the real parts of the dipole couplings and imaginary parts.}
\label{fig:bound-az}
\end{figure}

To set the projected constraints at $68\%$ confidence level (C.L.),
we form the log-likelihood ratio (LLR)
\beq[eq:LLR-h]
	R(\Gamma_j, h)
	= -2 \ln\!\biggbb{ \frac{L(\Gamma_j, h)}{L(0, h)} }
	= \Gamma_j \, F^{jj'}(h) \, \Gamma_{j'},
\eeq
which is a quadratic function of the dipole couplings.
We first evaluate this using the NLO cross section formulas at $\mu = Q$ for each $h$ replica, 
with the RG equation \eqref{eq:run-C} of $\Gamma_j$ taken into account to evolve them from $\mu = m_Z$.
For single-parameter constraints, 
turning on one dipole at a time and letting $R(\Gamma_j, h) \leq 1$ returns a band for each $h$,
while for two-parameter constraints, 
turning on two dipoles at a time and letting $R(\Gamma_j, h) \leq 2.3$ returns a contour for each $h$.
These give the $68\%$ C.L. constraint for each $h$, shown in \fig{fig:bound-az}. 
Stacking these bands and contours together, 
we then find the central $[16\%, 84\%]$ interval along each direction of the dipole coupling space
to obtain fuzzy bands and contours to mimic the ``$68\%$'' PDF uncertainty.
These are compared with the ``averaged'' constraints obtained by 
marginalizing the $h$ PDF in the likelihood \eqref{eq:likelihood-h},
\beq[eq:likelihood-h-marg]
	L_m(\Gamma_j)
	= \int [dh] \, L(\Gamma_j, h)
	= \frac{1}{N_r} \sum_{r = 1}^{N_r} 
		\exp\!\bb{ - \frac{1}{2} \, \Gamma_j \, 
			F^{jj'}\!\bigpp{ h^{(r)} }
		\, \Gamma_{j'} },
\eeq
where $[dh]$ represents the functional integral of $h$,
which is practically done by averaging over the $N_r = 968$ replicas.
The resultant marginalized $68\%$ C.L. constraints, obtained from the marginalized LLR 
$R_m(\Gamma_j) = -2 \ln\!\bigbb{ L_m(\Gamma_j) / L_m(0) }$,
are also shown in \fig{fig:bound-az} to lie inside the fuzzy bands or contours.
The constraints apply equally to the real parts and imaginary parts of the dipole couplings.
Since they are respectively associated with orthogonal $\cos\phi$ and $\sin\phi$ signals, 
they can be separately constrained without correlations.

%----------------------------------------------------------------
% Fig: Bound: AZ
%------------------------------------------------
%----------------------------------------------------------------

Notably, the correlations among $Z$ and $\gamma$ dipoles are very small 
as opposed to those of $(\Gamma_{\gamma}^u, \Gamma_{\gamma}^d)$ and $(\Gamma_{Z}^u, \Gamma_{Z}^d)$,
because they are separately probed by the single and the double spin asymmetries,
as we found in \sec{ssec:nlo}.
The $u$ and $d$ dipoles corresponding to the same vector boson are separated 
because of the different $x_B$ distributions encoded in their transversity PDFs.%%%
\footnote{The ability to achieve this separation is due to the binned likelihood analysis method,
which contains all the information of the constraints on the dipole couplings, including their full correlations.
It stands in contrast to the statistical methods in \refs{Wen:2023xxc, Wen:2024cfu, Wen:2024nff, Huang:2025ljp},
where the entire $(x_B, Q^2)$ phase space is integrated over, yielding only two linear constraints. 
Since we have used infinitely fine binning, our results represent the theoretically ideal scenario.}
%%%
Additionally, for the photon dipoles, they have different combinations in \eqs{eq:UT-approx2}{eq:LT-approx2}
due to the charge weights and $y$ dependence,
which makes the contour of $(\Gamma_{\gamma}^u, \Gamma_{\gamma}^d)$ 
 more elliptical than that of $(\Gamma_{Z}^u, \Gamma_{Z}^d)$.
However, we note that a significant limitation is due to the large uncertainty and small magnitude of $h_d$.
This causes the $d$ dipoles to be poorly constrained and the results significantly degraded due to the PDF uncertainties. 

%----------------------------------------------------------------
% Fig: Bound: LO vs NLO
%------------------------------------------------
\begin{figure}[htbp]
	\centering
	\includegraphics[scale=0.63]{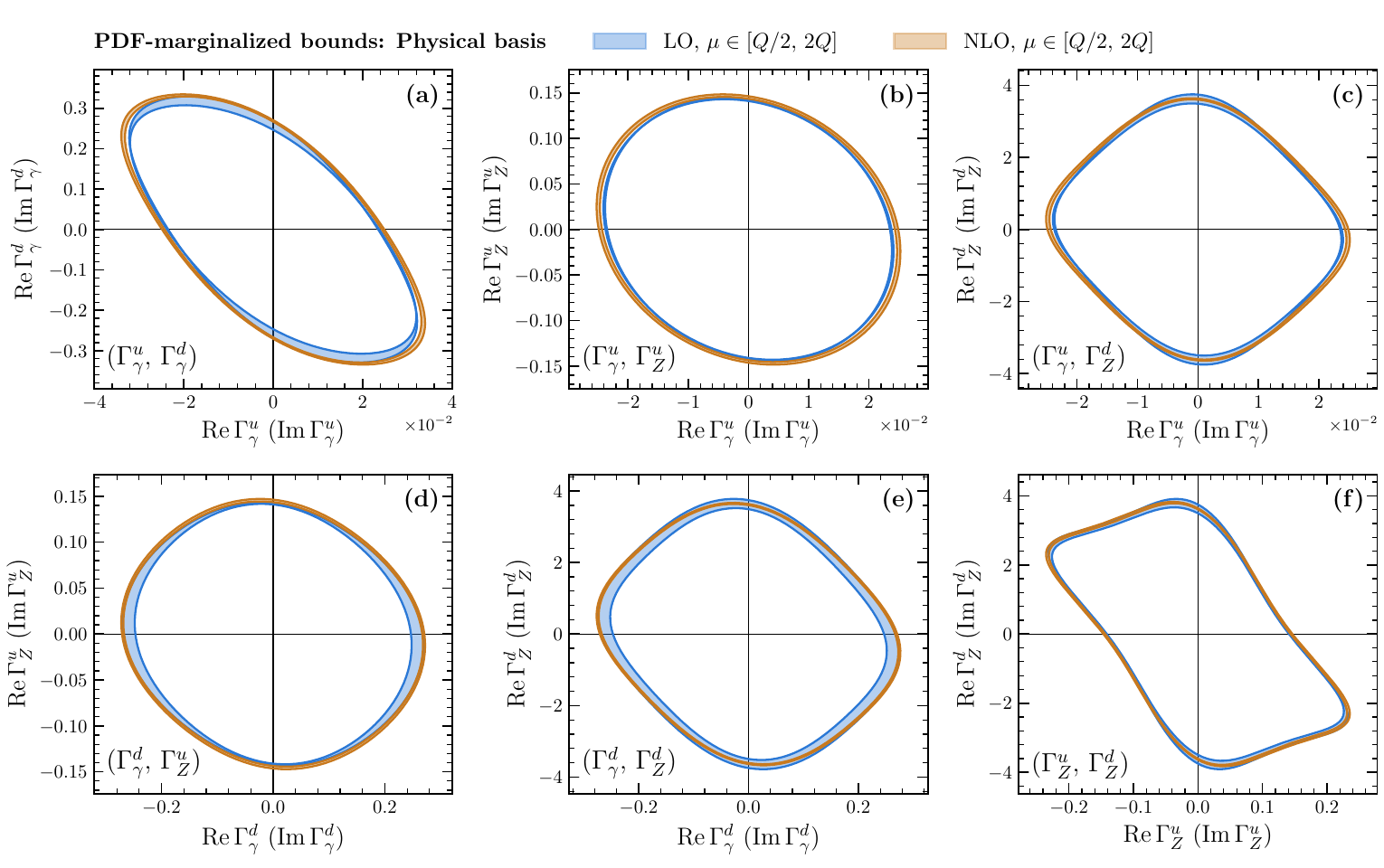}
	\caption{Constraint contour for each pair of the dipole couplings (at scale $\mu = m_Z$) at $68\%$ C.L.,
    	marginalized over the PDF replica set.
    	The scale $\mu$ is varied between $Q / 2$ and $2Q$ 
    	to give a band of the constraint contours for both LO and NLO. 
        Identical constraints apply to both the real and the imaginary contributions.}
\label{fig:lo-nlo-az}
\end{figure}
%----------------------------------------------------------------

Next, we examine the impacts of the NLO corrections and scale uncertainty.
Focusing on the PDF-marginalized likelihood in \eq{eq:likelihood-h-marg}, 
we compare the projected constraints 
with \eq{eq:likelihood-h-marg} evaluated using the LO or NLO expressions,
and with $\mu$ varied within $\{ Q / 2, Q, 2Q \}$.
The results are shown in \fig{fig:lo-nlo-az}.
As expected from \sec{ssec:nlo}, the results of the LO and NLO calculations are very close to each other.
While the scale uncertainty is slightly smaller for the NLO bounds,
it is evident that the PDF uncertainties dominate over scale uncertainties 
and  are much larger than the NLO corrections.

%----------------------------------------------------------------
% Fig: Bound: Warsaw
%------------------------------------------------
\begin{figure}[htbp]
	\centering
	\includegraphics[scale=0.63]{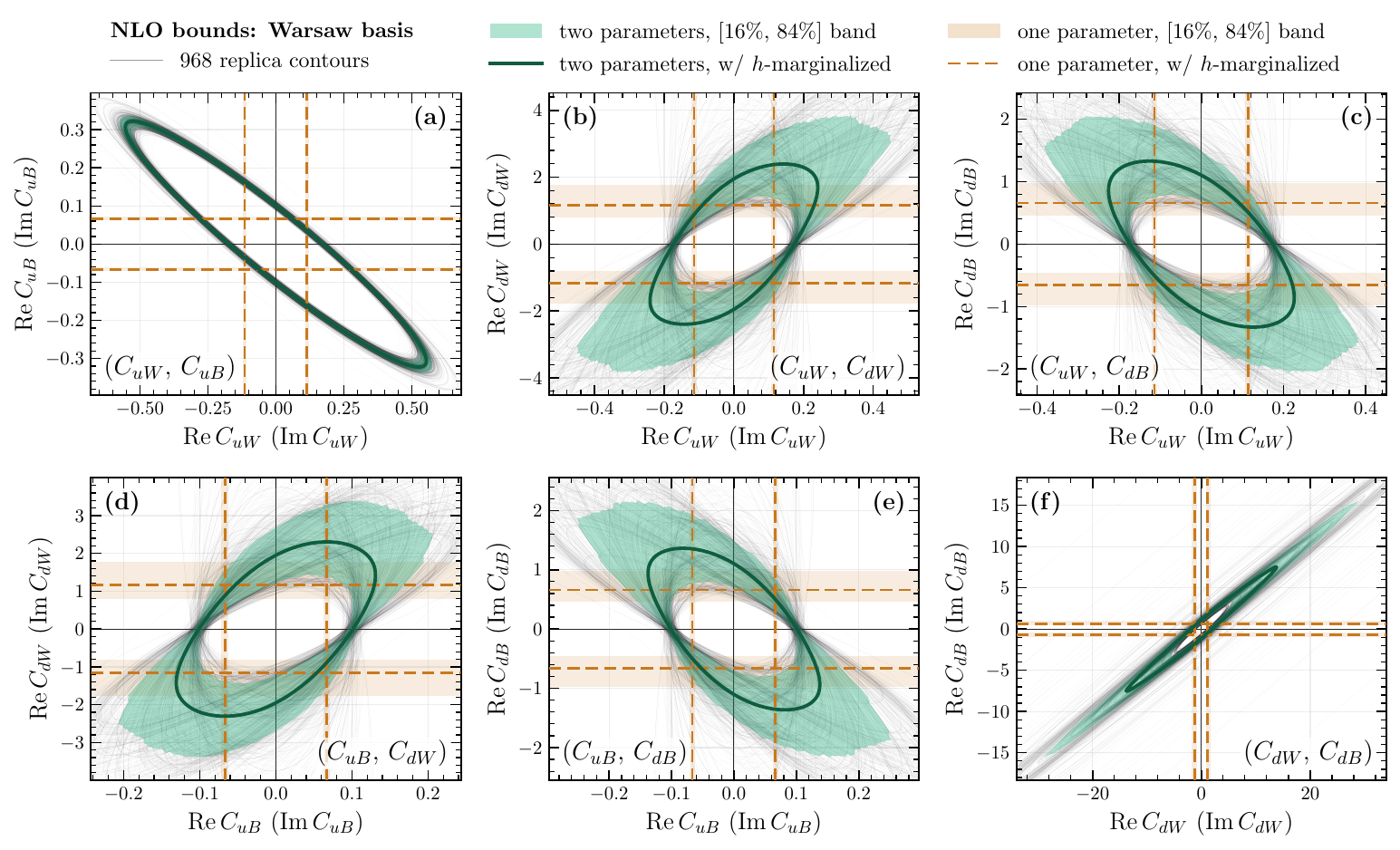}
	\caption{Similar to \fig{fig:bound-az},
    	but for the dipole Wilson coefficients (at scale $\mu = m_Z$) in the Warsaw basis,
    	with $\Lambda = 1~\TeV$. 
        The grey contours correspond to different PDF replicas, 
        while the dashed lines are $68\%$ C.L.\ single parameter fits. 
        The constraints are identical for both the real and the imaginary contributions.}
\label{fig:bound-warsaw}
\end{figure}
%----------------------------------------------------------------

Finally, we present in \fig{fig:bound-warsaw} the analogous constraint contours to those of \fig{fig:bound-az}
for the dipole couplings in the Warsaw basis (with $\Lambda = 1~\TeV$),
which is easily done by replacing the $\Gamma_{j}$ and $\Gamma_{j'}$ in \eqs{eq:LLR-h}{eq:likelihood-h-marg}
with $C$'s using \eq{eq:dip-wcoefs-conver}
before computing the likelihoods.
Most of our conclusions stand. 
We note, however, that the Warsaw basis mixes the photon and $Z$ dipoles,
and the strong correlation of every pair of Warsaw dipole couplings
hides the approximate de-correlation of the photon and $Z$ dipole contributions 
that we observe in \fig{fig:bound-az}.

%----------------------------------------------------------------
\subsection{Comparison to other constraints}
\label{ssec:others}
%----------------------------------------------------------------

We end this section by briefly comparing our results with the dipole constraints from other sources.

\begin{figure}[htbp]
	\centering
	\includegraphics[scale=0.4]{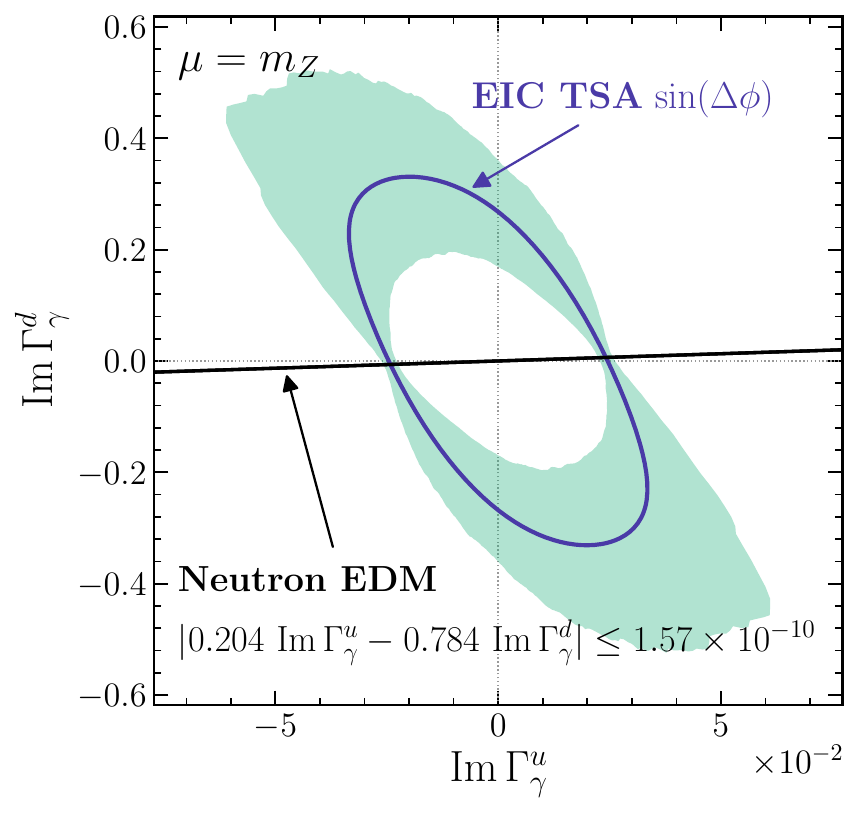}
	\caption{Comparison of the TSA constraint, from \fig{fig:bound-az}(a), 
	with the constraint set by the neutron EDM measurement~\cite{Abel:2020pzs}.
	Both constraints apply to the imaginary parts of photon dipole couplings at scale $\mu = m_Z$.}
\label{fig:edm}
\end{figure}

The first important constraint comes from measurements of low-energy dipole moments.
Unlike the case for leptons, there are no low-energy measurements of the light quarks' ``dipole moments''
because they are confined in hadrons. 
Their dipole couplings can, on the other hand, be matched to the nucleon dipole moments~\cite{Ellis:1996dg, Bhattacharya:2012bf, Pitschmann:2014jxa, Xu:2015kta, Bhattacharya:2015esa, Liu:2017olr, Kley:2021yhn, Kumar:2024yuu}.
For example, the neutron electric dipole moment $d_n$ receives contributions 
from the dipole operators in \eq{eq:Lag-AZ} by
\beq[eq:neutron-edm-dipole]
	d_n = \frac{e}{v} \pp{ g_T^d \, \Im \Gamma_{\gamma}^u + g_T^u \, \Im \Gamma_{\gamma}^d },
\eeq
where $g_T^{u, d}$ are the tensor charges of $u$ or $d$ in the proton target.
They are related to the integrals of transversity PDFs and can be computed using lattice QCD,
which gives~\cite{FlavourLatticeAveragingGroupFLAG:2024oxs, Gupta:2018lvp}
\beq
	g_T^u = 0.784 \pm 0.030,
	\quad
	g_T^d = -0.204 \pm 0.015,
\eeq
at scale $\mu = 2~\GeV$.
Taking the measurement result from \refcite{Abel:2020pzs}, 
$|d_n| < 1.8 \times 10^{-26}~e\cdot {\rm cm}$ at 90\% C.L., 
and converting it to 68\% C.L.,
can give us a constraint of $(\Im \Gamma_{\gamma}^u, \Im \Gamma_{\gamma}^d)$ at $\mu = 2~\GeV$.
Then using the RG equation \eqref{eq:run-C} gives us the bound
\beq
	\left| 0.204 \Im \Gamma_{\gamma}^u -0.784 \Im \Gamma_{\gamma}^d \right| 
	\leq 1.57 \times 10^{-10},
\eeq
at 68\% C.L. and scale $\mu = m_Z$.
This gives a linear constraint band in \fig{fig:edm},
which is extremely narrow and appears simply as a line.
It is displayed together with the EIC projected constraint via the TSA observables, 
extracted from \fig{fig:bound-az}(a).

Although in the TSA observables the photon dipole couplings are weighted by the transversity PDFs 
which have the same signs as the tensor charges,
they have additional weights from the electric charges,
so these two observables give complementary constraints.
While the low-energy measurement can achieve a precision far beyond that of collider-based experiments,
the matching in \eq{eq:neutron-edm-dipole} assumes no participation of other $CP$-violating operators
and it is also sensitive only to the imaginary parts of the photon dipole couplings.
The TSA observables, however, are uniquely sensitive to the dipole operators, 
receiving little contamination from other sources,
and can probe both the real and the imaginary parts of both the photon and $Z$ dipole couplings at the same time.

%----------------------------------------------------------------
% Fig: EDM
%------------------------------------------------
%----------------------------------------------------------------

\begin{figure}[htbp]
	\centering
	\includegraphics[scale=0.45]{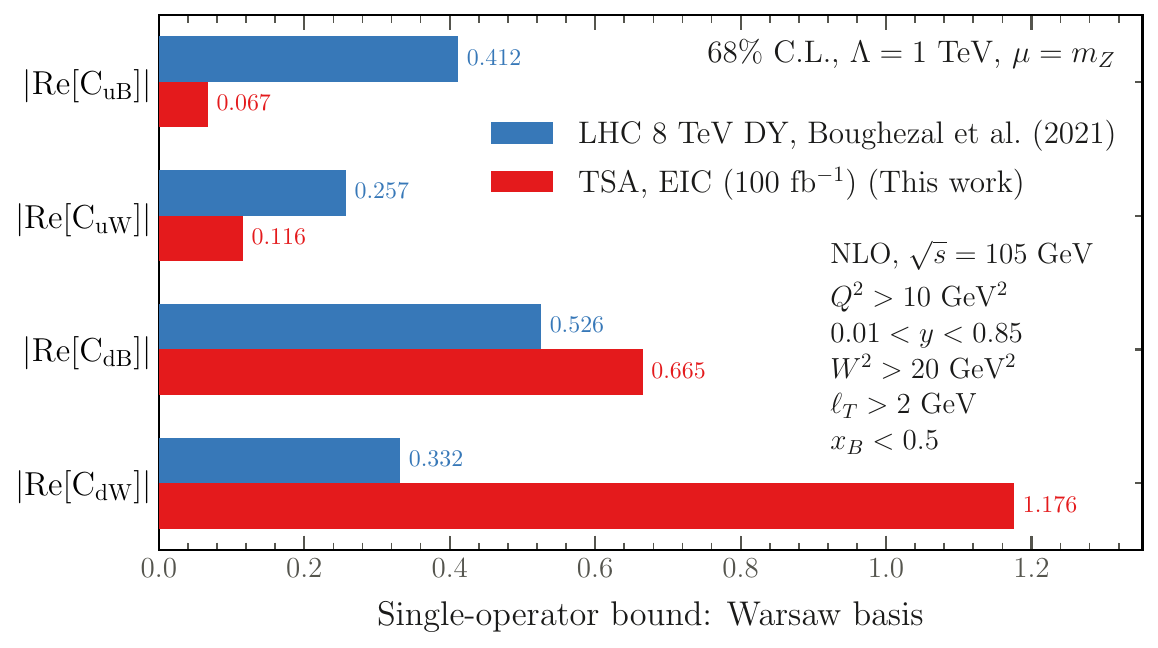}
	\caption{Comparison of the single-parameter constraints 
	from the TSA observables at EIC 
	and from the Drell-Yan measurements at LHC~\cite{ATLAS:2016gic, Boughezal:2021tih}.}
\label{fig:DY}
\end{figure}

The other constraint that we compare with is from the LHC measurement of the Drell-Yan process
in the high invariant mass region of the lepton pair.
As mentioned in the Introduction, due to their chiral-odd nature, 
the dipole operators enter the Drell-Yan cross section quadratically at $\order{1 / \Lambda^4}$,
and need to be analyzed together with dimension-8 operators. 
However, one can take advantage of the high energy to alleviate the suppression.
In \fig{fig:DY}, we compare the single-parameter constraints from the EIC projection 
and from the LHC measurements~\cite{ATLAS:2016gic} analyzed in \refcite{Boughezal:2021tih},
where all the couplings are normalized at scale $\mu = m_Z$ using \eq{eq:run-C}.
Clearly, the constraint on $u$-quark dipoles from EIC greatly outperforms that from the LHC,
but the $d$-quark dipoles are still largely limited by the transversity PDFs.
Also, we note that the quadratic sensitivity from Drell-Yan only constrains the absolute values of the dipole couplings,
while the TSA observables are additionally sensitive to their signs.

%----------------------------------------------------------------
% Fig: DY comparison
%------------------------------------------------
%----------------------------------------------------------------

%================================================================================
\section{Conclusion}
\label{sec:con}
%================================================================================

Making use of the transverse polarizability of the proton beam at the future EIC, 
the transverse spin asymmetry (TSA) observables in inclusive DIS 
provide unique sensitivity to the light-quark dipole operators in the SMEFT. 
In this work, we have extended the theoretical calculation of this observable 
to include one-loop corrections from QCD
and performed a detailed phenomenological study.
The ambiguity of $\gamma_5$ treatment in dimensional regularization is briefly discussed, 
and a general criterion for consistency is proposed based on symmetries and Hermiticity. 
Using Kreimer's scheme with a fixed reading point for fermion traces, 
our results automatically satisfy the consistency condition and gauge invariance.

We found that the TSA observables are remarkably stable against QCD corrections,
which makes them precision observable candidates for probing new physics effects.
A projection of the obtainable constraints on the dipole operators has also been shown.
Using the likelihood analysis method, information about the strengths of all the four light-quark dipoles is separated,
including both real and imaginary parts.
The constraint on the up quark dipoles is significantly better than the current constraint from LHC measurements,
but the down quark dipoles remain largely unconstrained at the EIC due to the large uncertainty 
of the corresponding transversity PDF.

Given that the global fit of transversity PDFs will be further improved 
from EIC measurements of other processes 
such as the semi-inclusive dihadron production process,
we expect that the actual future constraint will be better than the projection in this paper.

%================================================================================================
\begin{acknowledgments}
We thank Christina Cocuzza, Hooman Davoudiasl, Sven-Olaf Moch, Jae Doo Nam, Robert Szafron, Alessandro Tricoli, and Werner Vogelsang for helpful discussions.
This work is supported by the U.S. Department of Energy under Contract No.~DE-SC0012704. 
Digital data are attached as supplemental files.  
We acknowledge the assistance of Claude Opus 5.5 in obtaining the final plots.
\end{acknowledgments}
%================================================================================================

%%%%%%%%%%%%%%%%%%%%%%%%%%%%%%%%%%%%%%%%%%%%%%%%%%%%%%%%
\newpage
\appendix

%================================================================================
\section{Summary of cross section formula}
\label{app:sec:summary-xsec}
%================================================================================

We summarize the analytic calculation results of the hard scattering coefficients and cross section formulas.
We use the notation defined in \eq{eq:ch-notat} for the three EW channels.

%----------------------------------------------------------------
\subsection{Leptonic tensor}
%----------------------------------------------------------------

With the leptonic tensor normalized as \eq{eq:lepton-L-param}, 
the $\gamma$ channel has been calculated in \eq{eq:AA-lepton-trace-tree}
to give $(L^{\gamma}_1, L^{\gamma}_2)
\equiv (L^{\gamma\gamma}_1, L^{\gamma\gamma}_2) = (1, \lambda_e)$.
The $Z$ channel gives
\begin{align}
	L_1^{Z}
	\equiv L_1^{ZZ}
    = (g_V^e)^2 + (g_A^e)^2 - 2 \lambda_e g_V^e g_A^e,
	\qquad
	L_2^{Z}
    \equiv L_2^{ZZ}
	= \lambda_e \bb{ (g_V^e)^2 + (g_A^e)^2 } - 2 g_V^e g_A^e,
\end{align}
and the interference channel,
\begin{align}
	L_1^{\rm int}
    \equiv L_1^{\gamma Z}
	= L_1^{Z\gamma}
	= - \pp{
			g_V^e -  \lambda_e g_A^e
		},
	\qquad
	L_2^{\rm int}
    \equiv L_2^{\gamma Z}
	= L_2^{Z\gamma }
	= \pp{
			g_A^e -  \lambda_e g_V^e
		}.
\end{align}

%----------------------------------------------------------------
\subsection{Hadronic tensor}
%----------------------------------------------------------------

The NLO calculation of the hard coefficients of the factorized hadronic tensor is described 
in \sec{sec:nlores} for the $\gamma$ channel. 
For the other two channels, one just needs to replace the photon vertices. 
Using Kreimer's scheme and sticking to the Dirac reading point at the spin projector, 
the consistency condition in \eq{eq:consistency} is automatically satisfied for all channels. 
Soft and collinear divergences are also canceled in the same way. 
Below, we document only the infrared finite expressions in 4 dimensions.  
Results for $\wt T_{1,j}^{I, a}$ and $\wt T_{2,j}^{I, a}$ are not given separately, with \eq{eq:consistency} understood.

To have concise expressions for the results, we define
\begin{align}
	S(\x) &= - \pp{ \frac{\pi^2}{3} + \frac{9}{2} } \delta(1 - \x)
		+ 2 \pp{ \frac{\ln(1 - \x)}{1 - \x} }_+ 
		- \frac{3}{2} \frac{1}{(1 - \x)_+},
	\nn\\
	R_U(\x) &= - \frac{1 + \x^2}{1 - \x} \ln \x 
		- (1 + \x) \ln (1 - \x),
	\nn\\
	R_T(\x) &= - \frac{2 \x \ln\x}{1 - \x}
        - 2\ln(1 - \x),
\end{align}
and 
\bse[eq:kernels]\begin{align}
	u_{Tq}\pp{ \x, \frac{Q}{\mu} }
	&= \delta(1 - \x)
		+ \frac{\alpha_s C_F}{2\pi} \bb{
    		 P_{qq}(\x) \ln\frac{Q^2}{\mu^2} 
    	   + S(\x)
    	   + R_U(\x) + 3 
		},
	\\
	u_{Tg}\pp{ \x, \frac{Q}{\mu} }
	&= \frac{\alpha_s T_F}{2\pi} \,
		\biggbb{
			P_{qg}(\x) \, \ln\pp{\frac{(1 - \x) Q^2}{\x \, \mu^2}}
			- (1 - 2 \x)^2
		},
	\\
	u_{Lq}\pp{ \x, \frac{Q}{\mu} }
	&= \frac{\alpha_s C_F}{2\pi} \pp{
			2\x 
		},
	\\
	u_{Lg}\pp{ \x, \frac{Q}{\mu} }
	&= \frac{\alpha_s T_F}{2\pi} \,
		\bigbb{ 
			4 \x (1 - \x)
		},
	\\
	u_{3}\pp{ \x, \frac{Q}{\mu} }
	&= \delta(1 - \x)
		+ \frac{\alpha_s C_F}{2\pi} \, \biggbb{
    		P_{qq}(\x) \ln\frac{Q^2}{\mu^2} 
    		+ S(\x)
    		+ R_U(\x)
    		  + 3 + (\x - 1)
		},
	\label{eq:u3-kernel} 
    \\
	t_1\pp{ \x, \frac{Q}{\mu} }
	&= \delta(1 - \x)
		+ \frac{\alpha_s C_F}{2\pi} \, \biggbb{
    		\pp{ \delta P_{qq}(\x) + \frac{1}{2} \delta(1 - \x) } \ln\frac{Q^2}{\mu^2}  
    		+ S(\x)
    		+ R_T(\x) + \frac{5}{2} 
		},
	\\
	t_2\pp{ \x, \frac{Q}{\mu} }
	&= \delta(1 - \x)
		+ \frac{\alpha_s C_F}{2\pi} \, \biggbb{
    		\pp{ \delta P_{qq}(\x) + \frac{1}{2} \delta(1 - \x) } \ln\frac{Q^2}{\mu^2} 
    		+ S(\x)
    		+ R_T(\x) + \frac{9}{2} 
		},
\end{align}\ese
where $(C_F, T_F) = (4/3, 1/2)$ are the SU(3) color factors, 
$P_{qq}(\x)$ and $\delta P_{qq}(\x)$ are the splitting kernels in \eqs{eq:split-Pqq}{eq:split-tPqq}, 
and $P_{qg}(\x)$ is the splitting kernel of $g \to q\bar q$,
\beq
	P_{qg}(\x) = \x^2 + (1 - \x)^2.
\eeq

The partonic tensor is normalized as \eq{eq:w-decomp}. 
The structure functions of the photon channel ($I = \gamma$) are given 
in \eq{eq:AA-Wq0-coefs} at LO and \eq{eq:AA-Uq-Tq-sub} at NLO. 
Summing them up, we have 
\begin{align}
	U^{\gamma, q}_{T}
	&= e_q^2 \, u_{Tq},
	\nn\\
    U^{\gamma, q}_{L}
	&= e_q^2 \, u_{Lq},
    \nn\\
	T^{\gamma, q}_{1, (\gamma q)}
	&= e_q \, t_1,
\label{eq:AA-Wq-coefs}
\end{align}
where the arguments $(\x, Q/\mu)$ are suppressed.
%%%
The $Z$ channel gives 
\begin{align}
	U^{Z, q}_{T}
	&= \bigpp{ (g_V^q)^2 + (g_A^q)^2 } \, u_{Tq},
	\nn\\
    U^{Z, q}_{L}
	&= \bigpp{ (g_V^q)^2 + (g_A^q)^2 } \, u_{Lq},
    \nn\\
    U^{Z, q}_{3}
	&= -2 g_V^q g_A^q \, u_{3},
	\nn\\
	T^{Z, q}_{1, (Z q)}
	&= g_V^q \, t_1,
    \nn\\
	T^{Z, q}_{2, (Z q)}
	&= g_A^q \, t_2,
\label{eq:ZZ-Wq-coefs}
\end{align}
with extra $U_3$ and $T_2$ structures.
%%%
The interference channel gives
\begin{align}
	U^{{\rm int}, q}_{T}
	&= 2 e_q g_V^q \, u_{Tq},
	\nn\\
    U^{{\rm int}, q}_{L}
	&= 2 e_q g_V^q \, u_{Lq},
    \nn\\
    U^{{\rm int}, q}_{3}
	&= -2 e_q g_A^q \, u_3,
	\nn\\
    T^{{\rm int}, q}_{1, (\gamma q)}
	&= g_V^q \, t_1,
    \nn\\
	T^{{\rm int}, q}_{2, (\gamma q)}
	&= g_A^q \, t_2,
    \nn\\
	T^{{\rm int}, q}_{1, (Z q)}
	&= e_q \, t_1,
    \nn\\
	T^{{\rm int}, q}_{2, (Z q)}
	&= 0,
\label{eq:int-Wq-coefs}
\end{align}
where $T^{{\rm int}, q}_{2, (Z q)}$ is forbidden by $\hat P \hat T$ symmetry 
in the same way $T_{2, (\gamma q)}^{\gamma, q}$ is forbidden in \eq{eq:AA-Wq-coefs}. 

The $(U_T, U_L, U_3)$ structures are given by the SM interactions and 
have been checked against the literature~\cite{Altarelli:1978id, Bardeen:1978yd, Collins:2011zzd}. 
A nontrivial issue for the  $U_3$ calculation is that, 
while the reading point at the spin projector works to 
give the correct results for the dipole structures $(T_1, T_2)$, 
it fails to reproduce the $U_3$ result in the literature~\cite{Davies:2016ruz}, 
with the last term in \eq{eq:u3-kernel} being $5(\x - 1)$ rather than $(\x - 1)$. 
The latter result is obtained only if we read the fermion trace from the $Z$ vertex.
Further discussions on this will be given in a future work.

Structure functions for antiquarks are obtained from the above simply by replacing $g_A^q$ with $-g_A^q$.
The gluon structure functions only contain $U_T$ and $U_L$. 
They are the same as the SM and reproduced here as 
\begin{align}
	U_T^{\gamma, g}
	&= \sum_{q} (2 e_q^2) \, u_{Tg}, 
	\\
	U_L^{\gamma, g}
	&= \sum_{q} (2 e_q^2) \, u_{Lg},
\label{eq:AA-Wg-coefs}
\end{align}
for the $\gamma$ channel, 
\begin{align}
	U_T^{Z, g}
	&= \sum_{q} 2 \bigpp{ (g_V^q)^2 + (g_A^q)^2 } \,
		u_{Tg}, 
	\\
	U_L^{Z, g}
	&= \sum_{q} 2 \bigpp{ (g_V^q)^2 + (g_A^q)^2 } \,
		u_{Lg},
\label{eq:ZZ-Wg-coefs}
\end{align}
for the $Z$ channel, and
\begin{align}
	U_T^{{\rm int}, g}
	&= \sum_{q} (4 e_q g_V^q) \,
		u_{Tg}, 
	\\
	U_L^{{\rm int}, g}
	&= \sum_{q} (4 e_q g_V^q) \,
		u_{Lg},
\label{eq:int-Wg-coefs}
\end{align}
for the interference channel.
Here, the sum over $q$ runs over quark flavors, $q = u, d, s, \cdots$.
Loops of quarks and antiquarks contribute the same 
to gluon structure functions by charge conjugation symmetry, 
which also forbids the $U_3$ structure.

%----------------------------------------------------------------
\subsection{Cross section}
%----------------------------------------------------------------

We use the normalization in \eq{eq:xsec-pol} for the cross section, 
and write each term on its right-hand side as a sum of the three channels,
\beq[eq:Sigma-channel-decomp]
    \Sigma_c = \sum_{I = \gamma, Z, {\rm int}} N_I \,  \Sigma_c^I,
    \qquad
    c \in \{ UU, LU, UT, LT \},
\eeq
where $N_I$ is given in \eq{eq:norm-NI}.
We also introduce the shorthand notation for 
the convolution of the kernels in \eq{eq:kernels} with any PDF $f(x, \mu)$,
\begin{align}
	\C[u, f] &= \C[u, f](x_B, Q^2)
	\equiv \int_{x_B}^1 \frac{d x}{x} \, 
		u\pp{ \frac{x_B}{x}, \frac{Q}{\mu} } \, 
		f(x, \mu).
\label{eq:condef}
\end{align}

The $\gamma$ channel contains only an
unpolarized term $\Sigma^{\gamma}_{UU}$
and a double spin term $\Sigma^{\gamma}_{LT}$.
The unpolarized term contains contributions from $q$, $\bar{q}$, and $g$,
\begin{align}
	\Sigma^{\gamma}_{UU}(x_B, Q^2)
	&= \Sigma_{UU, T}^{\gamma}(x_B, Q^2) \, \bigbb{ 1 + (1 - y)^2 } 
		+  \Sigma_{UU, L}^{\gamma}(x_B, Q^2) \, \bigbb{ 2 (1 - y) },
\end{align}
where
\begin{align}
	\Sigma_{UU, T}^{\gamma}(x_B, Q^2)
	&= \sum_{q = u, d, s, \ldots} e_q^2 \, \Bigcc{\, 
			\C\!\bb{ u_{Tq}, f_{q} + f_{\bar q} }
			+ 2 \, \C\!\bb{ u_{Tg}, f_{g} }
		},
	\nn\\
	\Sigma_{UU, L}^{\gamma}(x_B, Q^2)
	&= \sum_{q = u, d, s, \ldots} e_q^2 \, 
		\Bigcc{\, 
			\C\!\bb{ u_{Lq}, f_{q} + f_{\bar q} }
			+ 2 \, \C\!\bb{ u_{L g}, f_{g} }
		}.
\end{align}
%%%
The double spin term
receives contributions only from $u$ and $d$ quarks and antiquarks,
dependent on their photon dipole couplings,
\begin{align}
	\Sigma^{\gamma}_{LT}(x_B, Q^2, \Delta \phi)
	&= \frac{2 Q \sqrt{1 - y} }{v} \cdot 
		y \sum_{q = u, d} e_q 
            \Re\bigpp{ \Gamma_{\gamma}^q \, e^{i \Delta \phi} }
			\, \C\!\bb{ t_1, \, h_q + h_{\bar q} }.
\label{eq:summary-LT-AA}
\end{align}
Single spin asymmetry terms vanish,
$\Sigma^{\gamma}_{LU} = \Sigma^{\gamma}_{UT} = 0$,
to all orders for the $\gamma$ channel.

The $Z$ channel contributes to all the four terms in \eq{eq:Sigma-channel-decomp}.
The unpolarized and beam spin asymmetry terms
both contain contributions from $q$, $\bar{q}$, and $g$,
\begin{align}
	\Sigma^{Z}_{UU}(x_B, Q^2)
	&= \Sigma_{UU, T}^{Z}(x_B, Q^2) \, \bigbb{ 1 + (1 - y)^2 } 
		+ \Sigma_{UU, L}^{Z}(x_B, Q^2) \, \bigbb{ 2 (1 - y) }
        + \Sigma_{UU, 3}^{Z}(x_B, Q^2) \, \bigbb{ y (2 - y) },
    \nn\\
    \Sigma^{Z}_{LU}(x_B, Q^2)
	&= \Sigma_{LU, T}^{Z}(x_B, Q^2) \, \bigbb{ 1 + (1 - y)^2 } 
		+ \Sigma_{LU, L}^{Z}(x_B, Q^2) \, \bigbb{ 2 (1 - y) }
		+ \Sigma_{LU, 3}^{Z}(x_B, Q^2) \, \bigbb{ y (2 - y) },
\end{align}
where
\begin{align}
	\Sigma_{UU, T}^{Z}(x_B, Q^2)
	&= \bigbb{ (g_V^e)^2 + (g_A^e)^2 } 
		\sum_{q = u, d, s, \ldots} \bigbb{ (g_V^q)^2 + (g_A^q)^2 } \cdot 
		\Bigcc{\, 
			\C\!\bb{ u_{Tq}, f_{q} + f_{\bar q} }
			+ 2 \, \C\!\bb{ u_{Tg}, f_{g} }
		},
	\nn\\
	\Sigma_{UU, L}^{Z}(x_B, Q^2)
	&= \bigbb{ (g_V^e)^2 + (g_A^e)^2 } 
		\sum_{q = u, d, s, \ldots} \bigbb{ (g_V^q)^2 + (g_A^q)^2 } \cdot
		\Bigcc{\, 
			\C\!\bb{ u_{Lq}, f_{q} + f_{\bar q} }
			+ 2 \, \C\!\bb{ u_{L g}, f_{g} }
		},
	\nn\\
	\Sigma_{UU, 3}^{Z}(x_B, Q^2)
	&= \pp{ 2g_V^e g_A^e } 
		\sum_{q = u, d, s, \ldots} \pp{ 2 g_V^q g_A^q } \cdot 
		\Bigcc{\, 
			\C\!\bb{ u_{3}, f_{q} - f_{\bar q} }
		},
    \nn\\
	\Sigma_{LU, T}^{Z}(x_B, Q^2)
	&= \pp{ -2g_V^e g_A^e } 
		\sum_{q = u, d, s, \ldots} \bigbb{ (g_V^q)^2 + (g_A^q)^2 } \cdot 
		\Bigcc{\, 
			\C\!\bb{ u_{Tq}, f_{q} + f_{\bar q} }
			+ 2 \, \C\!\bb{ u_{Tg}, f_{g} }
		},
	\nn\\
	\Sigma_{LU, L}^{Z}(x_B, Q^2)
	&= \pp{ -2g_V^e g_A^e } 
		\sum_{q = u, d, s, \ldots} \bigbb{ (g_V^q)^2 + (g_A^q)^2 } \cdot
		\Bigcc{\, 
			\C\!\bb{ u_{Lq}, f_{q} + f_{\bar q} }
			+ 2 \, \C\!\bb{ u_{L g}, f_{g} }
		},
	\nn\\
	\Sigma_{LU, 3}^{Z}(x_B, Q^2)
	&= \bigbb{ (g_V^e)^2 + (g_A^e)^2 } 
		\sum_{q = u, d, s, \ldots} \pp{ -2 g_V^q g_A^q } \cdot 
		\Bigcc{\, 
			\C\!\bb{ u_{3}, f_{q} - f_{\bar q} }
		}.
\end{align}
%%%
For the single transverse spin and double spin asymmetry terms,
we write them as
\begin{align}
	\Sigma^{Z}_{UT}(x_B, Q^2, \Delta \phi)
	&= \frac{2 Q \sqrt{1-y}}{v} \, \Bigbb{
		y \, \Sigma_{UT, 1}^{Z}(x_B, Q^2, \Delta \phi)
		+ (2 - y) \, \Sigma_{UT, 2}^{Z}(x_B, Q^2, \Delta \phi)
	},
	\nn\\
	\Sigma^{Z}_{LT}(x_B, Q^2, \Delta \phi)
	&= \frac{2 Q \sqrt{1-y}}{v} \, \Bigbb{
		y \, \Sigma_{LT, 1}^{Z}(x_B, Q^2, \Delta \phi)
		+ (2 - y) \, \Sigma_{LT, 2}^{Z}(x_B, Q^2, \Delta \phi)
	},
\end{align}
where the four coefficients depend only on $u$ and $d$ quarks via their $Z$ dipole couplings,
\begin{align}
	\Sigma_{UT, 1}^{Z}(x_B, Q^2, \Delta \phi)
	&= -\pp{ 2g_V^e g_A^e } 
		\sum_{q = u, d} g_V^q 
            \Re\bigpp{ \Gamma_{Z}^q \, e^{i \Delta \phi} }
			\, \C\!\bb{ t_1, \, h_q + h_{\bar q} },
	\nn\\
	\Sigma_{UT, 2}^{Z}(x_B, Q^2, \Delta \phi)
	&= - \bigbb{ (g_V^e)^2 + (g_A^e)^2 }  
		\sum_{q = u, d} g_A^q 
            \Re\bigpp{ \Gamma_{Z}^q \, e^{i \Delta \phi} }
			\, \C\!\bb{ t_2, \, h_q - h_{\bar q} },
	\nn\\
	\Sigma_{LT, 1}^{Z}(x_B, Q^2, \Delta \phi)
	&= \bigbb{ (g_V^e)^2 + (g_A^e)^2 }   
		\sum_{q = u, d} g_V^q 
            \Re\bigpp{ \Gamma_{Z}^q \, e^{i \Delta \phi} }
			\, \C\!\bb{ t_1, \, h_q + h_{\bar q} },
	\nn\\
	\Sigma_{LT, 2}^{Z}(x_B, Q^2, \Delta \phi)
	&= \pp{ 2g_V^e g_A^e }
		\sum_{q = u, d} g_A^q 
            \Re\bigpp{ \Gamma_{Z}^q \, e^{i \Delta \phi} }
			\, \C\!\bb{ t_2, \, h_q - h_{\bar q} }.
\end{align}

Similarly, the interference channel
has an unpolarized term and a beam spin asymmetry term
that contain contributions from $q$, $\bar{q}$, and $g$,
\begin{align}
	\Sigma^{\rm int}_{UU}(x_B, Q^2)
	&= \Sigma_{UU, T}^{\rm int}(x_B, Q^2) \, \bigbb{ 1 + (1 - y)^2 } 
		+ \Sigma_{UU, L}^{\rm int}(x_B, Q^2) \, \bigbb{ 2 (1 - y) }
		+ \Sigma_{UU, 3}^{\rm int}(x_B, Q^2) \, \bigbb{ y (2 - y) },
    \nn\\
    \Sigma^{\rm int}_{LU}(x_B, Q^2)
	&= \Sigma_{LU, T}^{\rm int}(x_B, Q^2) \, \bigbb{ 1 + (1 - y)^2 } 
		+ \Sigma_{LU, L}^{\rm int}(x_B, Q^2) \, \bigbb{ 2 (1 - y) }
		+ \Sigma_{LU, 3}^{\rm int}(x_B, Q^2) \, \bigbb{ y (2 - y) },
\end{align}
where
\begin{align}
	\Sigma_{UU, T}^{\rm int}(x_B, Q^2)
	&= -g_V^e
		\sum_{q = u, d, s, \ldots} \pp{ 2 e_q g_V^q } \cdot 
		\Bigcc{\, 
			\C\!\bb{ u_{Tq}, f_{q} + f_{\bar q} }
			+ 2 \, \C\!\bb{ u_{Tg}, f_{g} }
		},
	\nn\\
	\Sigma_{UU, L}^{\rm int}(x_B, Q^2)
	&= -g_V^e
		\sum_{q = u, d, s, \ldots} \pp{ 2 e_q g_V^q } \cdot
		\Bigcc{\, 
			\C\!\bb{ u_{Lq}, f_{q} + f_{\bar q} }
			+ 2 \, \C\!\bb{ u_{L g}, f_{g} }
		},
	\nn\\
	\Sigma_{UU, 3}^{\rm int}(x_B, Q^2)
	&= -g_A^e
		\sum_{q = u, d, s, \ldots} \pp{ 2 e_q g_A^q } \cdot 
		\Bigcc{\, 
			\C\!\bb{ u_{3}, f_{q} - f_{\bar q} }
		},
    \nn\\
    \Sigma_{LU, T}^{\rm int}(x_B, Q^2)
	&= g_A^e
		\sum_{q = u, d, s, \ldots} \pp{ 2 e_q g_V^q } \cdot 
		\Bigcc{\, 
			\C\!\bb{ u_{Tq}, f_{q} + f_{\bar q} }
			+ 2 \, \C\!\bb{ u_{Tg}, f_{g} }
		},
	\nn\\
	\Sigma_{LU, L}^{\rm int}(x_B, Q^2)
	&= g_A^e
		\sum_{q = u, d, s, \ldots} \pp{ 2 e_q g_V^q } \cdot
		\Bigcc{\, 
			\C\!\bb{ u_{Lq}, f_{q} + f_{\bar q} }
			+ 2 \, \C\!\bb{ u_{L g}, f_{g} }
		},
	\nn\\
	\Sigma_{LU, 3}^{\rm int}(x_B, Q^2)
	&= g_V^e
		\sum_{q = u, d, s, \ldots} \pp{ 2 e_q g_A^q } \cdot 
		\Bigcc{\, 
			\C\!\bb{ u_{3}, f_{q} - f_{\bar q} }
		}.
\end{align}
For the single transverse spin and double spin asymmetry terms, 
we write them as
\begin{align}
	\Sigma^{\rm int}_{UT}(x_B, Q^2, \Delta \phi)
	&= \frac{2 Q \sqrt{1-y}}{v} \, \Bigbb{
		y \, \Sigma_{UT, 1}^{\rm int}(x_B, Q^2, \Delta \phi)
		+ (2 - y) \, \Sigma_{UT, 2}^{\rm int}(x_B, Q^2, \Delta \phi)
	},
	\nn\\
	\Sigma^{\rm int}_{LT}(x_B, Q^2, \Delta \phi)
	&= \frac{2 Q \sqrt{1-y}}{v} \, \Bigbb{
		y \, \Sigma_{LT, 1}^{\rm int}(x_B, Q^2, \Delta \phi)
		+ (2 - y) \, \Sigma_{LT, 2}^{\rm int}(x_B, Q^2, \Delta \phi)
	},
\label{eq:summary-UT/LT-int}
\end{align}
where the four coefficients depend only on $u$ and $d$ quarks 
via both of their photon and $Z$ dipole couplings,
\begin{align}
	\Sigma_{UT, 1}^{\rm int}(x_B, Q^2, \Delta \phi)
	&= g_A^e
		\sum_{q = u, d} \Bigbb{ 
			g_V^q \Re\bigpp{ \Gamma_{\gamma}^q \, e^{i \Delta \phi} }
			+ e_q \Re\bigpp{ \Gamma_{Z}^q \, e^{i \Delta \phi} }
		}
			\, \C\!\bb{ t_1, \, h_q + h_{\bar q} },
	\nn\\
	\Sigma_{UT, 2}^{\rm int}(x_B, Q^2, \Delta \phi)
	&= g_V^e
		\sum_{q = u, d} g_A^q 
            \Re\bigpp{ \Gamma_{\gamma}^q \, e^{i \Delta \phi} }
			\, \C\!\bb{ t_2, \, h_q - h_{\bar q} },
	\nn\\
	\Sigma_{LT, 1}^{\rm int}(x_B, Q^2, \Delta \phi)
	&= (- g_V^e)
		\sum_{q = u, d} \Bigbb{ 
			g_V^q \Re\bigpp{ \Gamma_{\gamma}^q \, e^{i \Delta \phi} }
			+ e_q \Re\bigpp{ \Gamma_{Z}^q \, e^{i \Delta \phi} }
		}
			\, \C\!\bb{ t_1, \, h_q + h_{\bar q} },
	\nn\\
	\Sigma_{LT, 2}^{\rm int}(x_B, Q^2, \Delta \phi)
	&= (-g_A^e)
		\sum_{q = u, d} g_A^q 
            \Re\bigpp{ \Gamma_{\gamma}^q \, e^{i \Delta \phi} }
			\, \C\!\bb{ t_2, \, h_q - h_{\bar q} }.
\label{eq:summary-UT/LT-12-int}
\end{align}
%%%%%%%%%%%%%%%%%%%%%%%%%%%%%%%%%%%%%%%%%%%%%%%%%%%%%%%%

%================================================================================
\bibliographystyle{apsrev}
\bibliography{reference}
%================================================================================

%%%%%%%%%%%%%%%%%%%%%%%%%%%%%%%%%%%%%%%%%%%%%%%%%%%%%%%%
\end{document}